\pdfoutput=1

\documentclass[sigconf,authorversion]{acmart}

\usepackage{multirow}
\usepackage[disable]{todonotes}
\usepackage{listings}
\usepackage{caption}
\usepackage{color}
\usepackage{array,graphicx}
\usepackage{booktabs}
\usepackage[utf8]{inputenc}

\definecolor{gray}{rgb}{0.4,0.4,0.4}
\definecolor{darkblue}{rgb}{0.0,0.0,0.6}
\definecolor{cyan}{rgb}{0.0,0.6,0.6}

\lstdefinelanguage{XML}
{
	basicstyle=\ttfamily\small,
	morestring=[b]",
	morestring=[s]{>}{<},
	morecomment=[s]{<?}{?>},
	stringstyle=\color{black},
	identifierstyle=\color{darkblue},
	keywordstyle=\color{cyan},
	morekeywords={xmlns,version,ref,id,xsi,noNamespaceSchemaLocation}
}

\lstdefinelanguage{XQuery}%
{
	basicstyle=\ttfamily\small,
	alsoletter={-},
	morekeywords={
		for, in,
		let,
		as,%
		allowing, empty,
		at,
		tumbling, sliding, window,
		start, when,
		only, end,
		previous, next,
		count,
		where,
		group, by,
		collation,
		order, stable,
		ascending, descending, greatest, least,
		return,
		eq, and, not, or,
		ordered, unordered,
		switch, case, default,
		some, every, satisfies,
		try, catch,
		of,
		typeswitch,
		cast,
		castable,
		treat,
		validate, lax, strict, 
		xquery, encoding, version,
		import, module, namespace,
		declare, boundary-space, preserve, strip, construction,
		base-uri,
		ordering,
		copy-namespaces, no-preserve, inherit, no-inherit,
		decimal-format,
		decimal-separator, grouping-separator, infinity, minus-sign, NaN, percent, per-mille, zero-digit, digit, pattern-separator, exponent-separator,
		schema, element, function,
		variable, external,
		context, item,
		option,
		node,
		ancestor, ancestor-or-self, attribute, child, descendant, descendant-or-self, following, following-sibling, namespace, parent, preceding, preceding-sibling, self
	},
	morestring=[b]',%
	morecomment=[n]{(:}{:)},%
	keywordstyle=\bfseries,
	escapeinside={(*@}{@*)},
	breaklines=true, 
	breakautoindent=true, 
	literate={\-}{}{0\discretionary{-}{}{-}}
	{\\\-}{}{0\discretionary{-}{}{-}}
}

\providecommand{\versionchoice}[4]{#4}

\providecommand{\spaceabove}{
	\setlength\abovecaptionskip{0,5\baselineskip}
	\setlength\belowcaptionskip{0,2\baselineskip}
}
\providecommand{\spacebelow}{\vspace{-2mm}}
\providecommand{\spacebelowb}{\vspace{0mm}}

\copyrightyear{2020}
\acmYear{2020}
\setcopyright{acmcopyright}\acmConference[MODELS '20]{ACM/IEEE 23rd International Conference on Model Driven Engineering Languages and Systems}{October 18--23, 2020}{Montreal, QC, Canada}
\acmBooktitle{ACM/IEEE 23rd International Conference on Model Driven Engineering Languages and Systems (MODELS '20), October 18--23, 2020, Montreal, QC, Canada}
\acmPrice{15.00}
\acmDOI{10.1145/3365438.3410987}
\acmISBN{978-1-4503-7019-6/20/10}

\begin{document}
\title{Detecting Quality Problems in Research Data: A Model-Driven Approach}
\subtitle{Extended Version}

\begin{abstract}

As scientific progress highly depends on the quality of research data, there are strict requirements for data quality coming from the scientific community. 
A major challenge in data quality assurance is to localise quality problems that are inherent to data.	
Due to the dynamic digitalisation in specific scientific fields, especially the humanities, different database technologies and data formats may be used in rather short terms to gain experiences. 
We present a model-driven approach to analyse the quality of research data. 
It allows abstracting from the underlying database technology. 
Based on the observation that many quality problems show anti-patterns, a data engineer formulates analysis patterns that are generic concerning the database format and technology.
A domain expert chooses a pattern that has been adapted to a specific database technology and concretises it for a domain-specific database format.
The resulting concrete patterns are used by data analysts to locate quality problems in their databases. 
As proof of concept, we implemented tool support that realises this approach for XML databases.
We evaluated our approach concerning expressiveness and performance in the domain of cultural heritage
based on a qualitative study on quality problems occurring in cultural heritage data.

\end{abstract}

\author{Arno Kesper}
\orcid{https://orcid.org/0000-0002-5042-1087}
\affiliation{Philipps-Universität Marburg}
\email{arno.kesper@uni-marburg.de}
\author{Viola Wenz}
\orcid{https://orcid.org/0000-0002-2176-9463}
\affiliation{Philipps-Universität Marburg}
\email{viola.wenz@uni-marburg.de}
\author{Gabriele Taentzer}
\orcid{https://orcid.org/0000-0002-3975-5238}
\affiliation{Philipps-Universität Marburg}
\email{taentzer@uni-marburg.de}

\ccsdesc[500]{Information systems~Expert search}
\ccsdesc[500]{Information systems~Query languages}
\ccsdesc[500]{Software and its engineering~Domain specific languages}


\maketitle

\section{Introduction}
\label{sec:introduction}
Research data is defined as data that is ``generated in the course of scientific work'' \cite{rfii}.
The quality of research data is essential for scientific progress. 
Establishing and maintaining data quality throughout the data life cycle is still a challenge \cite{rfii}. 

We present an approach to analyse the quality of research data.
For the requirements elicitation and evaluation we chose the domain of cultural heritage research.
However, we do not see any obstacles to applying our approach to other kinds of research data since the approach itself is not dependent on specific characteristics of the chosen domain.

As common in various research fields, there is a high amount of uncertainty in cultural heritage data, such as data on buildings or artworks. 
It can occur in various forms and often remains implicit. 
Data fields may remain empty, for example, since the requested information is unknown.
Or information that should be unique (e.g. a year of birth) may not as there exist several scientific opinions about that.
Hence, it is very natural for a scientific discourse that there are quality problems in curated databases. 
%
As there is no standard definition for {\em data quality} in a scientific context, we start with a literature survey to investigate multiple quality dimensions, such as consistency, completeness and precision \cite{zaveri2016quality,batini_methodologies_2009,laranjeiro_survey_2015}.
Furthermore, we elicited data {\em quality problems} as they occur in curated cultural heritage databases. 
Identified quality problems, such as imprecise, redundant and semantically incorrect data, are mapped to affected quality dimensions. 

A major step to data quality assurance is to analyse the inherent quality problems. 
%
We can observe that the digitalisation of the humanities has started but this process is not settled yet. 
Dependent on the kinds of research question tackled, the underlying database technologies used may largely differ. 
The digitalisation of texts, for example, has resulted in a standard XML format \cite{xml} called TEI~\cite{tei}.
Recently researchers have started to use graph database technology in the context of digital humanities to interrelate text representations with further information~\cite{Kuc16}.

Due to this dynamic digitalisation process, we present a model-driven approach to data quality analysis based on {\em patterns}. 
In contrast to related approaches \cite{kontokostas2014test,furber2010using,furber_swiqa_2011,bizer_quality-driven_2009,oditis_domain-specific_2017}, 
it allows for specifying anti-patterns for data quality problems that are generic concerning the underlying database technology and format.
Such a generic pattern can be adapted to several database technologies, resulting in several abstract patterns.
A domain expert chooses an abstract pattern as a {\em template} and concretises it to the domain-specific database format and to a concrete quality problem. 
Data analysts can apply the resulting concrete patterns to analyse their databases.

As a proof of concept, we implemented a realisation of our approach for XML databases with the Eclipse Modelling Framework~\cite{emf}.
To apply a pattern to XML data, we translate it to XQuery~\cite{xquery}.
The approach was tested in detail for expressiveness and performance.
In total, 85~\% of the identified data quality problem variants are covered by our approach.
The application to large cultural heritage databases of about 80~\% of 43 patterns took less than 20 seconds.

In summary, this paper makes the following contributions:
\begin{itemize}
	\item A literature survey on quality dimensions for research data and a study of quality problems occurring in cultural heritage data (Sec.~\ref{sec:problems})
	\item A model-driven approach to data quality analysis (Sec.~\ref{sec:concepts})
	\item Tool support for XML data (Sec.~\ref{sec:implementation})
	\item An evaluation wrt. expressiveness and performance (Sec.~\ref{sec:evaluation})
\end{itemize}
We discuss related work in Section~\ref{sec:relatedWork} and conclude in Section~\ref{sec:conclusion}.



\section{Data Quality Problems}
\label{sec:problems}
%
%
Data quality problems can be considered as specific problem instances in one or more data \textit{quality dimensions} \cite{laranjeiro_survey_2015}.
We start by investigating several quality dimensions especially relevant in the context of research data.
Thereafter, we present data quality problems occurring in cultural heritage data collected through a qualitative study and map them to the affected dimensions.
Finally, we introduce a little example to illustrate a few data quality problems. 

\subsection{A Literature Survey}
Since we present an approach for pattern-based quality analysis of research data, we focus on {\em aspects of data quality that are inherent to the data itself} and do not depend on any external impacts, such as technologies employed to provide or secure data.
There are no common definitions of data quality dimensions \cite{zaveri2016quality, laranjeiro_survey_2015, batini_methodologies_2009}.
Thus, based on a brief meta-survey we will define quality dimensions particularly relevant for research data in the following.


%

Zaveri et al. \cite{zaveri2016quality} presented 23 data quality dimensions, based on a review of 21 papers on quality assessment of linked open data.
The identified dimensions are grouped as follows: 
The {\em accessibility} dimensions are related to access and retrieval of data 
by authorized humans or machines.
The {\em intrinsic} dimensions focus on the data itself and in contrast to the {\em contextual} dimensions are independent of the usage context.
{\em Trust} dimensions measure the trustworthiness of data.
Dimensions related to the timeliness 
of data and the frequency of change over time are grouped to {\em dynamicity}.
Dimensions related to the {\em representation} of data are also grouped.
Batini et al.~\cite{batini_methodologies_2009} 
identified the following dimensions to be frequently covered by approaches to classify data quality:
{\em accuracy, completeness, consistency} and {\em timeliness}.
Laranjeiro et al.~\cite{laranjeiro_survey_2015} additionally identified {\em accessibility} to be often defined in the literature.

Since we focus on quality dimensions directly affected by the data itself, we 
exclude the accessibility dimensions described by Laranjeiro et al.~\cite{laranjeiro_survey_2015} and Zaveri et al.~\cite{zaveri2016quality}.
However, we adopt the intrinsic dimensions described by Zaveri et al., namely \emph{accuracy}, \emph{consistency}, \emph{conciseness} (i.e. {\em uniqueness}) and \emph{timeliness}.
We split the dimension that is referred to as {\em accuracy} into \emph{correctness} and \emph{precision} as this enables a finer differentiation, especially concerning the uncertainty often included in research data.
Due to this uncertainty, the \emph{trustworthiness} (called {\em believability} by Zaveri et al.) of data plays an important role as well.
Since lack of knowledge is in the nature of research, we consider \emph{completeness} as a dimension of interest.
Furthermore, the data itself may show characteristics which negatively impact its \emph{understandability}.
We briefly define the selected quality dimensions for research data as follows:

\begin{itemize}
	\item \textbf{Correctness} is the degree to which the data correctly represents the real-world values (semantic correctness)
	and is free of syntactical errors (syntactic correctness).
	\item \textbf{Completeness} is the degree to which all required information is present in the data.
	\item \textbf{Consistency} is the absence of logical or representational contradictions within the data. 
	\item \textbf{Precision} describes how exactly the data represents real-world values.	
	\item \textbf{Uniqueness} is the unambiguous interpretability of data and thus the absence of redundancies. 
	\item \textbf{Understandability} is the ease with which humans can read and interpret the data. 
	\item \textbf{Timeliness} measures how up-to-date the data is.
	\item \textbf{Trustworthiness} is defined as the degree to which the data is accepted to be correct and credible.
\end{itemize}

\subsection{Data Quality Problem Elicitation Considering Cultural Heritage Data}
\label{subsec:problemelicitation}
\versionchoice{This paper emerged in the scope of a funded research project.}{This paper emerged in the scope of the KONDA project\footnote{The project ``Kontinuierliches Qualitäts\-management von dynamischen Forschungsdaten zu Objekten der materiellen Kultur unter Nutzung des LIDO-Standards'' (KONDA) is funded by the German Federal Ministry of Education and Research.}.}{This paper emerged in the scope of a funded research project.}{This paper emerged in the scope of the KONDA project\footnote{The project ``Kontinuierliches Qualitäts\-management von dynamischen Forschungsdaten zu Objekten der materiellen Kultur unter Nutzung des LIDO-Standards'' (KONDA) is funded by the German Federal Ministry of Education and Research.}.}
The goal is to develop a continuous quality management process for cultural heritage data.
We conducted 6 qualitative interviews and a workshop with 19 domain experts that perform acquisition, modelling, management and usage of various kinds of cultural heritage data (e.g. data on technical objects or artworks) to investigate the question: 
{\em What quality problems occur in cultural heritage data?} 

We compiled a comprehensive specification of 94 data quality problems \cite{comprehensive_problem_specifications} through structured capturing of various aspects per problem, such as its impact on data quality and possible causes.
73 of the problems are directly related to the data itself.
The others are beyond the scope of this paper as they depend on external impacts, such as data models. 

By grouping the data-centric problems according to conceptual similarity 
and by further abstracting from the captured problems 
we created a list of rather general quality problems and variants presented in Table~\ref{table:problems}.
For each problem, the table shows the affected quality dimensions.
In the following, we will explain the less obvious problems and discuss relations between certain problems.
\begin{table*}
	\centering
	\spaceabove
	\caption{Overview of identified data quality problems and affected quality dimensions.}
	\begin{tabular}{ p{\dimexpr 0.6\textwidth-2\tabcolsep} p{\dimexpr 0.4\textwidth-2\tabcolsep} } 
		\toprule
		\textbf{Quality Problem} & \textbf{Affected Quality Dimensions}\\
		\midrule 		
		\emph{Illegal values:}
			 wrong datatype,
			 domain violation (interval, set, syntax violation)
		& Syntactic correctness \\
		\midrule		
		\emph{Missing data:} 
			 missing values,
			 missing references,
			 missing records,
			 dummy values
		& Completeness \\
		\midrule
		\emph{Referential integrity violation} & Completeness \\
		\midrule 
		\emph{Unique value violation} & Uniqueness \\
		\midrule
		\emph{Violation of a functional dependency} & Consistency, Semantic correctness \\
		\midrule		
		\emph{Contradictory relationships} & Consistency, Semantic correctness \\
		\midrule 			
		\emph{Imprecise data:} 
			 alternative possible values,
			 imprecise numerical values,
			 abstract terms,
			 ambiguous values,
			 abbreviations
		& Precision, Uniqueness, Understandability \\
		\midrule 
		\emph{Misplaced information:}
		misfielded values,
		extraneous data
		& Understandability, Consistency, Semantic correctness \\
		\midrule
		\emph{Redundant data:} 
			 exact duplicate records,
			 approximate duplicate records,
			 information placed in multiple locations
		& Uniqueness \\
		\midrule
		\emph{Heterogeneous data:} 
			 heterogeneous measure units,
			 heterogeneous value representations,
			 heterogeneous structural representations
		& Consistency, Understandability \\
		\midrule 
		\emph{Misspellings} & Consistency, Understandability \\
		\midrule 
		\emph{Semantically incorrect data:} 
		false values,
		false references,
		doubtful data
		& Semantic correctness, Timeliness, Trustworthiness\\
		\bottomrule
	\end{tabular}	
	\label{table:problems}
	\spacebelowb
\end{table*} 
The identified quality problems are not disjoint.
Hence, a certain characteristic of data may imply multiple of the quality problems listed.
{\em Illegal values}, for example, may be indicators for further problems such as misplaced information or misspellings.
A {\em functional dependency} occurs if the values of a set of fields determine the values of another set of fields. 
{\em Contradictory relationships} are present if constraints concerning multiple references between records (e.g. irreflexivity and asymmetry) are violated.
Regarding {\em imprecise data}, 
multiple possible alternatives may be listed or
imprecise numerical values may be given.
Abstract terms, ambiguous values (e.g. homonyms) and unexplained abbreviations cannot unambiguously be interpreted.
{\em Misplaced information} occurs if values are placed into wrong fields or extraneous data is given in a field (e.g. title and name in a name field).
%
{\em Semantically incorrect data}
may overlap with other quality problems such as \emph{illegal values}, \emph{functional dependency violations} and \emph{contradictory relationships}.

The data quality problems described in the literature \cite{rahm_data_2000, laranjeiro_survey_2015, furber2010using, oliveira_taxonomy_2005, oliveira_formal_2005, kim_taxonomy_2003} and those identified in our study overlap largely. 
However, the strategies for categorizing quality problems and the granularities of subdividing quality problems vary.

Note that most of the identified quality problems could be avoided in theory through constraints ensured during data creation.
However, changes in requirements and constraints and also in the technologies and data formats used must be considered.
\todo[inline,backgroundcolor=yellow]{Inserted advantage of model-driven approach below GT: okay}
Due to these dynamics, methods for retrospective data quality analysis are absolutely required and the model-driven (i.e. generic) approach is especially valuable.
Furthermore, in some cases, data is valid with respect to the schema constraints but is still considered problematic and thus is relevant for data quality analysis.
Examples are incomplete or imprecise data.

\subsection{Running Example}
\label{subsec:example}
To demonstrate concrete instances of some of the quality problems presented in Table \ref{table:problems}, we introduce a simple example for XML data depicted in Listing \ref{lst:runningexample}.
It is based on a schema which allows for describing buildings and architects.
Each such element has an ID and includes a \lstinline|name| element.
\lstinline|Building| elements additionally include a \lstinline|city| and a \lstinline|country| element.
\lstinline|Architect| elements further include \emph{potential} \lstinline|birthyear|s.
The following quality problems can be found in the example data:
\begin{enumerate}

\item The \lstinline|architect| record 
includes two specifications of the architect's year of birth. 
This indicates imprecise data, a form of uncertainty, since conflicting alternatives are listed.
\todo[inline,backgroundcolor=yellow]{Inserted sentence below GT: slightly revised}
Hence, this example shows that data may be valid with respect to the schema but needs additional data quality analysis.
\item The \lstinline|building| records show a \emph{violation of a functional dependency} between city and country.
Both buildings are stated to be located in New York City, but their indicated countries differ.
This suggests semantic incorrectness and inconsistency.
\end{enumerate}


\begin{lstlisting}[language=XML, caption={Running example including two quality problems}, label={lst:runningexample}, basicstyle=\ttfamily\footnotesize]
<data>
	<building id="1">
		<name>Empire State Building</name>
		<city>New York City</city>
		<country>USA</country>	
	</building>		
	<building id="2">
		<name>Chrysler Building</name>
		<city>New York City</city>
		<country>unknown</country>
	</building>	
	<architect id="3">
		<name>William F. Lamb</name>
		<birthyear>1883</birthyear>
		<birthyear>1884</birthyear>
	</architect>	
</data>
\end{lstlisting}
%

\section{A Model-Driven Approach to Data Quality Analysis}
\label{sec:concepts}



Given a large variety of data quality problems which may have various concrete forms, 
a {\em model-driven approach} is promising to develop long-lasting concepts and tooling for data quality analysis independent of concrete technologies and formats.
In the following, we will start with an overview of our approach and present some example patterns.
Thereafter, we will introduce the metamodel for representing patterns. 

\subsection{The Overall Approach}\label{subsec:overallapproach}
\begin{figure}	
	\centering
	\spaceabove
	\includegraphics[width=\linewidth]{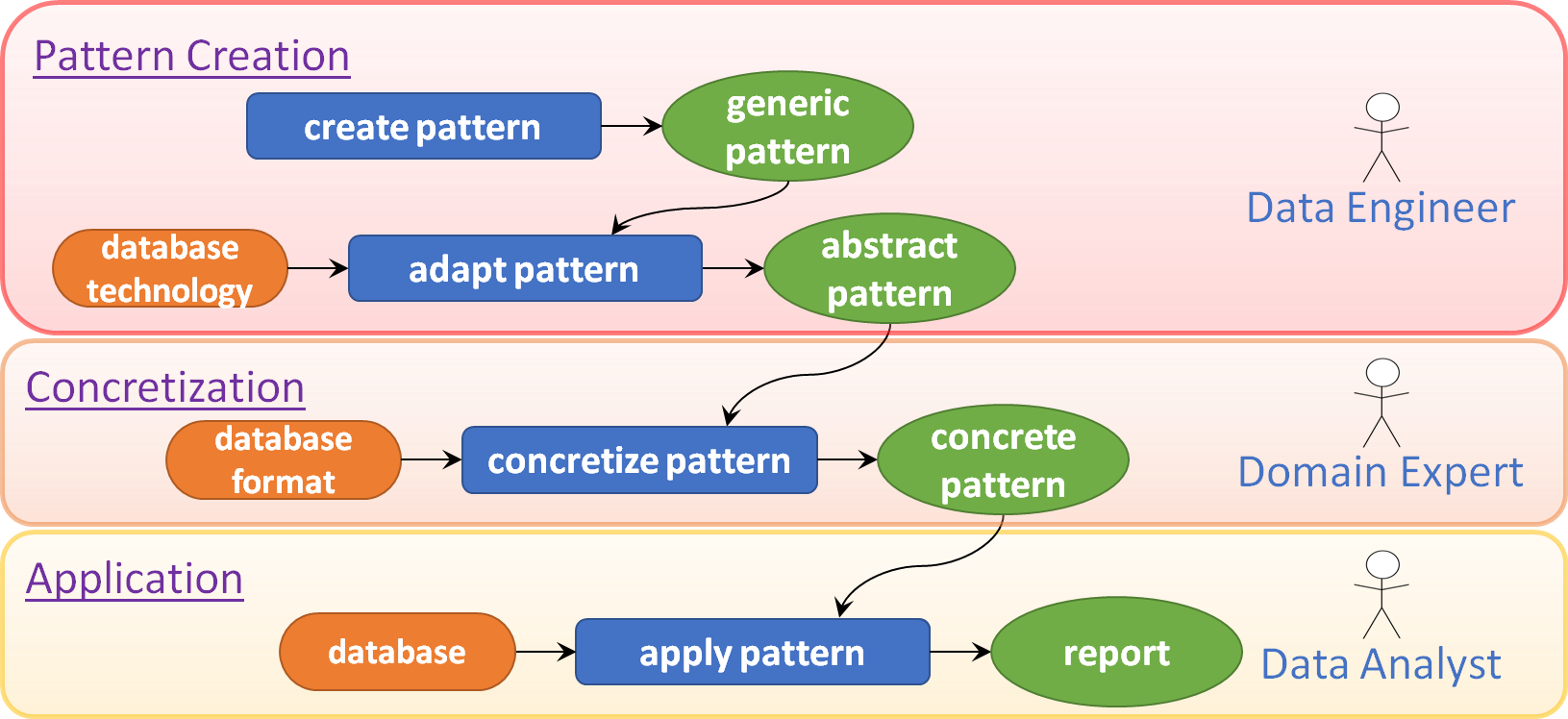}
	\caption{Workflow of pattern creation and application}
	\label{fig:workflow}
	\spacebelow
\end{figure}

Our proposed workflow for data quality analysis is visualised in Fig.~\ref{fig:workflow}.
%
It starts with a {\em data engineer} creating a {\em generic pattern} for identifying a general and often domain-independent quality problem (that is not yet covered by existing patterns).
Since the pattern detects phenomena that negatively impact data quality, it is actually an anti-pattern.
Generic patterns are completely independent of the database technology and database format.
A pattern is defined as a first-order logic expression over graph structures.
Hence, data engineers must have a comprehensive understanding of both first-order logic and graphs.
Furthermore, the ability to think abstractly, structurally and analytically is required.
However, as we plan to provide a graphical modelling workbench, programming skills will not be required to define a pattern.

In the next step, an {\em abstract pattern} is created by adapting a generic pattern to a specific database technology (e.g. XML).
Depending on the database technology this can be done automatically or semi-automatically with input from a data engineer.
Abstract patterns are still independent of any database format (e.g. of a specific XML database).

In the next phase of the workflow, a {\em domain expert} (e.g. culture historian) chooses an abstract pattern fitting for the problem of interest and concretises it for the domain-specific database format.
The result is a {\em concrete pattern}.
To choose the right abstract pattern, 
at least a superficial understanding of first-order logic and graphs is necessary.
To concretise a pattern, both an understanding of the database format and domain knowledge are required.

A {\em data analyst} can apply a concrete pattern (i.e. the generated query) to any database which conforms to the format that the pattern was concretised for.
The result consists of all the data items that match the pattern and thus show some quality problem.
The data analyst gets an overview of located quality problems and can decide then how to handle these problems, thus, how to initiate the improvement of data quality.

\subsection{Example Patterns}\label{subsec:examplepatterns}

To get a first idea of how abstract and concrete patterns may look like, we reconsider the example quality problems of Section~\ref{subsec:example}. 
For each example, we specify an abstract pattern first and show it in a graphical form. 
Thereafter, the abstract pattern is concretised to the XML database format used in Listing~\ref{lst:runningexample} by setting several parameters.
We omit the presentation of the corresponding generic patterns, which
differ from the presented abstract patterns only in that they
do not specify the types of relations and do not contain root elements as they are an XML-specific phenomenon.
\versionchoice{Examples of generic patterns are presented in \cite{supplementary_material}.}{Examples of generic patterns are presented in \cite{supplementary_material}.}{Examples of generic patterns are presented in Section~\ref{appendix:generic} of the appendix.}{Examples of generic patterns are presented in Section~\ref{appendix:generic} of the appendix.}

Each pattern consists of two parts: a graph and a condition.
The graph indicates the elements that are to be selected if they meet the condition.
In the following diagrams it is shown on the left-hand side.
The 
condition is visualised hierarchically on the right-hand side.
It is a first-order logic expression over graph structures, which
entails the following advantages.
Graphs allow a comprehensible representation and
expressing all kinds of data structures (e.g. trees).
Over time, first-order logic has proven to be ``adequate to the axiomatization of all ordinary mathematics'' \cite{folemergence} and to be a good compromise between expressiveness and efficiency \cite{FMORT19}.
The number of graphs in the condition depends on its structure.
Elements are bound from left to right.
Properties are just shown as necessary.

\paragraph{Problem 1}
\begin{figure}
	\centering
	\spaceabove
	\includegraphics[width=\linewidth]{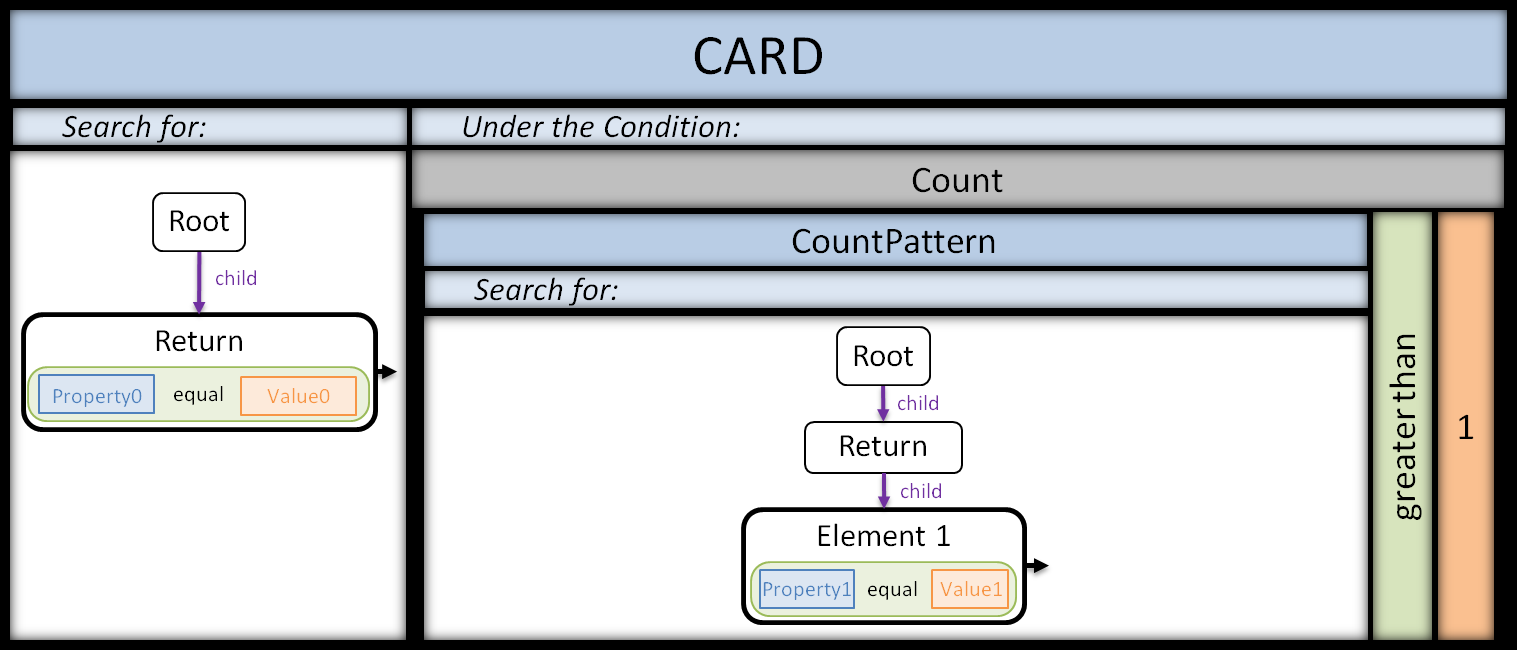}
	\caption{Abstract pattern CARD for detecting violations of a cardinality restriction; XML-adaption}
	\label{fig:card0}
	\spacebelow
\end{figure}
The \emph{abstract pattern} \texttt{CARD} which is visualised in Fig.~\ref{fig:card0} detects field repetitions that, for example, hint at imprecise data.
The right part of this pattern specifies a count condition
which is fulfilled if there is more than one element contained in the return element whose Property1 satisfies the given comparison relation.

To \emph{concretise} the abstract pattern, a \emph{domain expert} specifies properties and values by \emph{setting several parameters}.
They are indicated by the blue and orange boxes in the diagrams.
For simplicity, relations and comparison operators are already predefined in the presented abstract patterns.
To detect problem (1), the parameters need to be specified as shown in Listing \ref{lst:card}.
Since the abstract pattern is concretised for XML data, a property is defined by choosing \lstinline|name|, \lstinline|attribute| (plus specifying the name) or \lstinline|content|.
\begin{lstlisting}[label=lst:card,caption={Concretisation of the CARD pattern (Fig. \ref{fig:card0})},basicstyle=\ttfamily\footnotesize]
Property0 and 1 = name, Value0 = "architect", Value1 = "birthyear" 
\end{lstlisting}
Once all parameters are specified, the \emph{concrete pattern} is automatically translated into a \emph{query}.
It can then be applied to a database by a \emph{data analyst} to detect concrete problem occurrences.
\emph{The concrete pattern returns elements of type \lstinline|architect| that contain more than one element of type \lstinline|birthyear|.}
Hence, when applied to the example data in Listing \ref{lst:runningexample}, it returns the \lstinline|architect| element with \lstinline|ID 3|.

\paragraph{Problem 2}
\begin{figure}	
	\includegraphics[width=\linewidth]{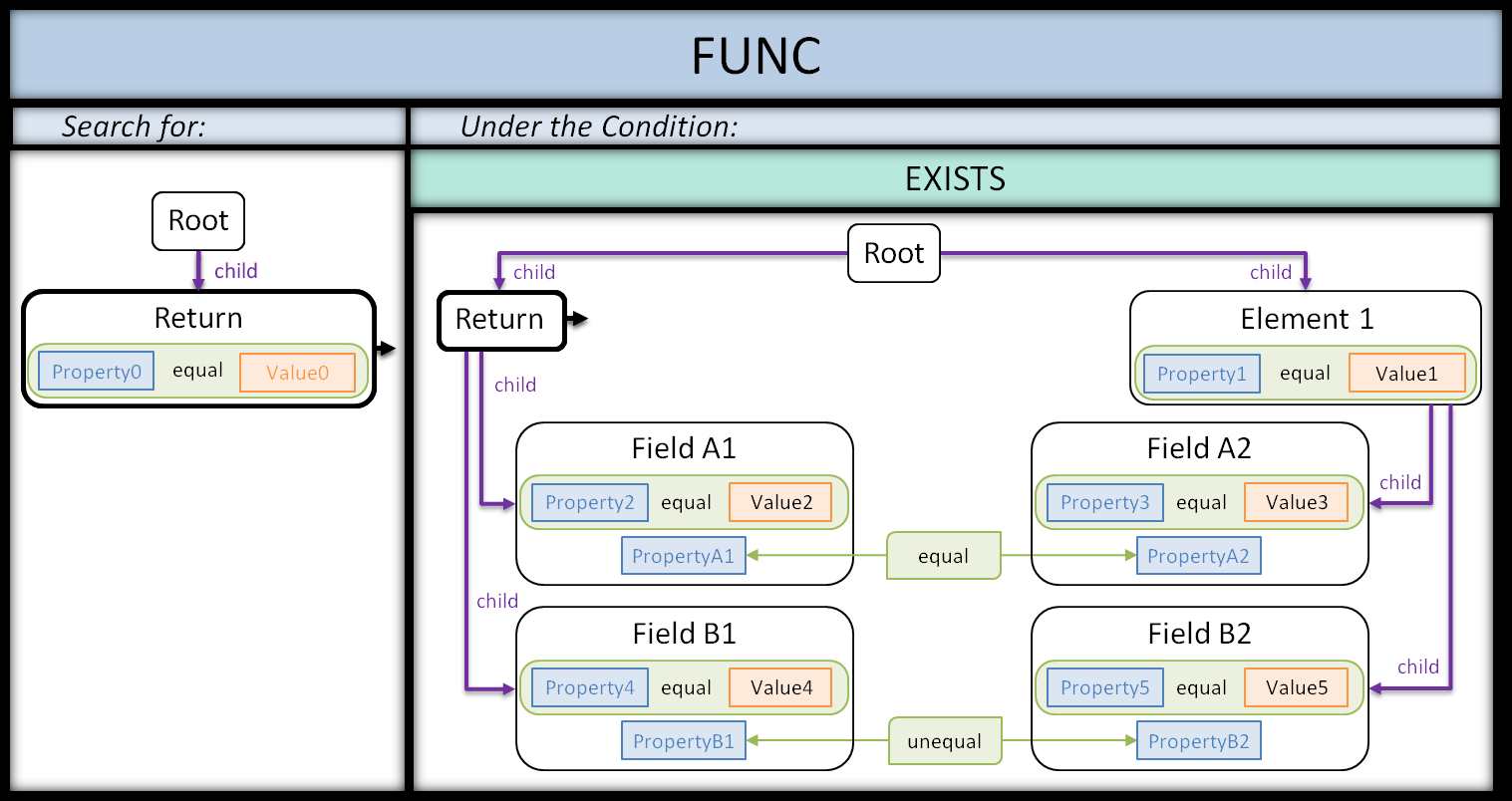}
	\caption{Abstract pattern FUNC for detecting functional dependency violations; XML-adaption}
	\label{fig:func1}
	\centering
\end{figure}
The abstract pattern \texttt{FUNC} which is depicted in Fig.~\ref{fig:func1} finds violations of a functional dependency by detecting two container elements with two subelements each that are in specific comparison relations. One of the container elements is returned.
The pattern can be concretised for the running example by specifying the input values presented in Listing \ref{lst:func}.

\begin{lstlisting}[label=lst:func,caption={Concretisation of the FUNC pattern (Fig. \ref{fig:func1})},basicstyle=\ttfamily\footnotesize]
Property0 to 5 = name, PropertyA and B = content,
Value0 and 1 = "building", 
Value2 and 3 = "city", Value4 and 5 = "country"
\end{lstlisting}

\emph{The concrete pattern finds elements of type \lstinline|building| for which there exists another \lstinline|building| stated to be located in the same \lstinline|city| but in a different \lstinline|country|.}
This pattern is satisfied for the \lstinline|building| elements with \lstinline|ID 1| and \lstinline|ID 2| in the example data in Listing \ref{lst:runningexample}.

\subsection{A Metamodel for Data Quality Analysis}\label{subsec:metamodel}
The core of our model-driven approach to data quality analysis is a metamodel for specifying patterns to localise quality problems in data.
The metamodel allows expressing \emph{generic, XML-adapted abstract and concrete patterns} as first-order logic expressions over graph structures.
In Section \ref{subsubsec:metamodeloverview}, we will explain it in more detail.
In Sections \ref{subsubsec:adaption} and \ref{subsub:concretisation}, we will discuss which parts of the metamodel are affected by the adaption of a \emph{generic pattern} to a specific database technology 
and the concretisation of an \emph{abstract pattern}.
\versionchoice{For an example abstract pattern presented as an instance model we refer to \cite{supplementary_material}.}{For an example abstract pattern presented as an instance model we refer to \cite{supplementary_material}.}{An example abstract pattern presented as an instance model can be found in Section \ref{subsubsec:instance} of the appendix.}{An example abstract pattern presented as an instance model can be found in Section \ref{subsubsec:instance} of the appendix.}
%
%
%
%
\subsubsection{Metamodel Overview}\label{subsubsec:metamodeloverview}
\begin{figure*}	
	\centering
	\spaceabove
	\includegraphics[width=\textwidth]{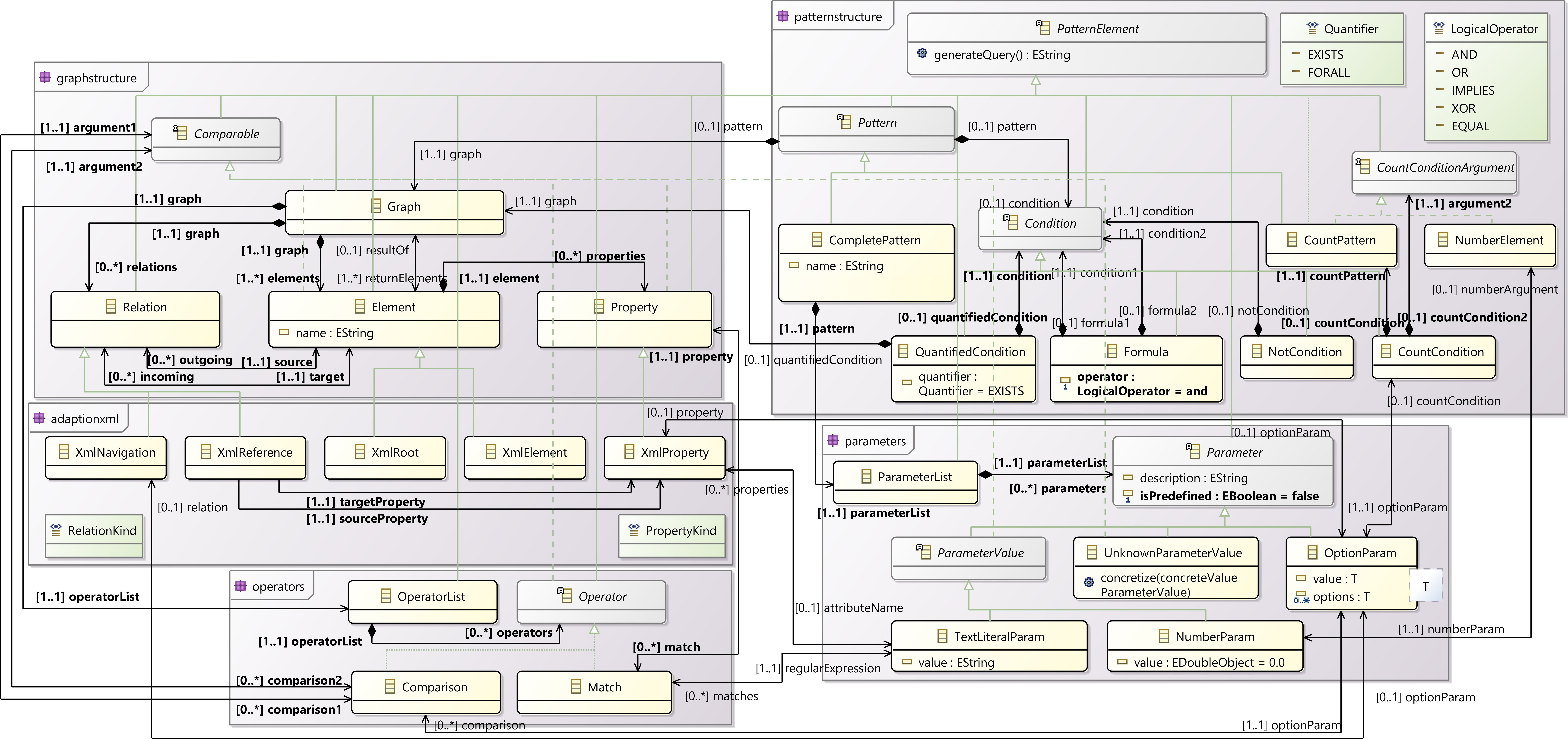}
	\caption{Condensed metamodel for patterns which allow localising quality problems in data}
	\label{fig:metamodel}
	\spacebelow
\end{figure*}

Fig.~\ref{fig:metamodel} shows the most relevant part of our metamodel;
\versionchoice{the complete version and well-formedness rules are available at \cite{supplementary_material}.}{the complete version and well-formedness rules are available at \cite{supplementary_material}.}{the complete version and well-formedness rules are given in Section~\ref{appendix:metamodel} of the appendix.}{the complete version and well-formedness rules are given in Section~\ref{appendix:metamodel} of the appendix.}
The metamodel has five packages: \lstinline|patternstructure| is concerned with the logical structure of a pattern, \lstinline|graphstructure| is used to specify pattern graphs, \lstinline|operators| contains operators to define conditions on pattern elements, \lstinline|parameters| holds placeholders that are not specified in generic or abstract patterns but are to be set in concrete patterns. Package \lstinline|adaptionxml| holds classes which enable the adaption of a generic pattern to XML; 
it will be discussed in Section \ref{subsubsec:adaption}.
It is not required for representing \emph{generic} patterns. 
The design of the packages \lstinline|patternstructure| and \lstinline|graphstructure| is inspired by the implementation design of nested graph conditions presented by Nassar in \cite{Nas20}.

The package \textbf{\lstinline|patternstructure|} includes classes for specifying the logical structure of a pattern.
Each \emph{generic, abstract or concrete pattern} is represented as an instance of \lstinline|CompletePattern|, which extends \lstinline|Pattern|.
A \lstinline|Pattern| always contains a \lstinline|Graph| and a \lstinline|Condition|.
The \lstinline|Graph| determines which elements should be returned if the \lstinline|Condition| is satisfied.
By nesting \lstinline|Condition|s we can express first-order logic formulas over graph structures.
There are four types of conditions.
%
A \lstinline|QuantifiedCondition| is specified via a \lstinline|quantifier|.
It contains a \lstinline|Graph| specifying the domain of discourse as well as a further \lstinline|Condition|.
%
A \lstinline|Formula| is specified by a logical \lstinline|operator| and two further conditions which serve as arguments.
The logical operator $not$ is modelled separately in a \lstinline|NotCondition| as it has only one argument.
A \lstinline|CountCondition| allows expressing a cardinality constraint.
It compares a \lstinline|CountPattern| with another \lstinline|CountPattern| or a \lstinline|NumberElement|, thus a primitive number, via a specified operator.
Just like a \lstinline|CompletePattern|, the \lstinline|CountPattern| consists of a graph and a nested condition.
The \lstinline|CountPattern|, however, does not represent the matches themselves but instead represents the \emph{number} of matches.
These occurrences of the pattern in the data can partly overlap concerning the matched elements.

The package \textbf{\lstinline|graphstructure|} allows specifying \lstinline|Graph|s in \emph{generic patterns} as compositions of named \lstinline|Element|s and arbitrary many directed \lstinline|Relation|s in between.
Elements of different graphs of a pattern correspond to each other if they have the same name.
Each \lstinline|Element| may contain arbitrary many \lstinline|Properties| which may be subject to conditions.
If a \lstinline|Graph| is contained in a \lstinline|QuantifiedCondition|, each included \lstinline|Element| that does not correspond to an \lstinline|Element| in a previous \lstinline|Graph| is bound by the condition\rq{}s \lstinline|quantifier|.
The elements returned by the pattern are specified by the \lstinline|returnElements| association between \lstinline|Graph| and \lstinline|Element|.

The package \textbf{\lstinline|operators|} allows defining conditions on elements.
Each \lstinline|Operator| is assigned to exactly one \lstinline|Graph| by being contained in its \lstinline|operatorList|.
The arguments of an operator must be components of this graph.
A \lstinline|Comparison| has two \lstinline|Comparable| items (of type \lstinline|Element|, \lstinline|Operator|, \lstinline|Property|, \lstinline|ParameterValue| or \lstinline|UnknownParameter-Value|) as arguments.
The concrete operator is specified as a parameter via the association \lstinline|optionParam|, which will be explained later on.
A \lstinline|Match| operator checks a \lstinline|Property| for a regular expression that is given as a parameter. 

The package \textbf{\lstinline|parameters|} holds \lstinline|Parameter|s for indicating which information is not yet given in a generic or abstract pattern but shall be available in a \emph{concretisation} of that pattern.
Each parameter is contained in the \lstinline|parameterList| of the \lstinline|CompletePattern|.
Parameter values can be predefined in a generic or abstract pattern or a description can be provided.
Subclasses of \lstinline|ParameterValue| (only partly shown in Fig.~\ref{fig:metamodel}) each represent concrete literal values of a specific type.
In a generic or abstract pattern, the argument types of a \lstinline|Comparison| do not have to be set.
Thus, an \lstinline|UnknownParameterValue| can be used as a placeholder for a concrete \lstinline|ParameterValue|.
The class \lstinline|OptionParam| allows specifying parameters whose domain is defined via an enumeration of values of type \lstinline|T| given in the \lstinline|options| attribute.
In a concrete pattern, the chosen value for this parameter is specified in the \lstinline|value| attribute. 
This class is used, for example, to specify the concrete operator of a \lstinline|Comparison|, e.g. \lstinline|less than|.
\subsubsection{Adaption to a Specific Database Technology}\label{subsubsec:adaption}
The packages discussed above allow the definition of \emph{generic patterns} independent of any database technology.
To represent \emph{abstract patterns} adapted to a specific database technology, \lstinline|Element|, \lstinline|Relation| and \lstinline|Property| must be subclassed correspondingly. 
For the adaption of a \emph{generic pattern}, instances of \lstinline|Element|, \lstinline|Relation| and \lstinline|Property| in the pattern must be replaced by instances of corresponding subclasses via a (semi-) automatic algorithm.

\paragraph{Adaption to XML}\label{par:xmladaption}
Our implementation provides classes for representing XML-adapted abstract patterns in the package \textbf{\lstinline|adaptionxml|}.
In the course of the adaption, the graphs of the generic pattern are transformed into trees spanned by relations based on XPath~\cite{xpath} axes (represented by \lstinline|XmlNavigation|).
Edges between the branches of these trees represent identifier-based references within the XML data (represented by \lstinline|XmlReference|).

The class \lstinline|XmlRoot| represents the root of the XML document and serves as the root of the trees in the pattern.
The class \lstinline|XmlElement| represents XML elements in general.
Each \lstinline|XmlElement| has exactly one incoming \lstinline|XmlNavigation|.
The axis is specified by the association \lstinline|optionParam<RelationKind>|.
The enumeration \lstinline|RelationKind| comprises XPath axes.
Besides the axes \lstinline|child| and \lstinline|descendant| for representing containment relations, further axes are also practical.
The axes \lstinline|self| and \lstinline|descendant-or-self| are useful if multiple elements of a pattern graph may correspond to the same XML element.
The axis \lstinline|following|, for example, is relevant for XML data in which the order of elements has an impact on the meaning of that data.
%
An XML element may have the following three kinds of properties that we encoded in the enumeration \lstinline|PropertyKind|: 
an element name, a named attribute and some content between the start and end tag.
The kind of an \lstinline|XmlProperty| is specified by \lstinline|optionParam<PropertyKind>|.
If a named attribute is addressed, its name is specified by the \lstinline|attributeName| of type \lstinline|TextLiteralParam|.

For the adaption of a generic pattern to XML, all instances of \lstinline|Element| and \lstinline|Property| are automatically replaced by instances of \lstinline|XmlElement| and \lstinline|XmlProperty|, respectively.
Next, the data engineer decides for each \lstinline|Relation| whether it represents an XPath axis (\lstinline|XmlNavigation|) or a reference (\lstinline|XmlReference|).
The instance of \lstinline|Relation| is replaced correspondingly.
In the latter case, \lstinline|Properties| are automatically inserted into the \lstinline|source| and \lstinline|target| elements. 
In the last step, the \lstinline|XmlRoot| is automatically inserted in each graph of the pattern.
For each \lstinline|XmlElement| which has no incoming \lstinline|XmlNaviga-tion|, an incoming \lstinline|XmlNavigation| from \lstinline|XmlRoot| is inserted.
Thereby the tree structure is completed.





\paragraph{Adaption to other database technologies}\label{par:otherdatabases}
To support other database technologies, further adaption packages with subclasses of \lstinline|Element|, \lstinline|Relation| and \lstinline|Property| as well as the corresponding parameter kinds 
need to be implemented.

In a \emph{relational database}, for example, an \lstinline|Element| corresponds to a row of a specific table while a \lstinline|Property| represents a column of the table. 
To specify a relation between elements, the names of the columns holding the foreign and primary key are required.
Thus, the new subclass of \lstinline|Relation| compares properties of both elements analogously to the \lstinline|XmlReference| class.

In \emph{graph databases} which are specified via subject-predicate-object triples, an \lstinline|Element| corresponds to a node.
A primitive predicate of an \lstinline|Element| may be represented as a \lstinline|Property| 
and a relation between two nodes as a \lstinline|Relation|.
The new subclasses of \lstinline|Property| and \lstinline|Relation| must be implemented correspondingly.

To increase the generality of generic patterns, relations expressing an identity (like the \lstinline|self| axis mentioned above) should be supported for other database technologies as well.

\subsubsection{Concretisation}\label{subsub:concretisation}
As discussed above, all variable parts of an abstract pattern are modelled as \lstinline|Parameter|s contained in the \lstinline|parameter\-List| of a \lstinline|CompletePattern|.
To concretise an abstract pattern, these parameters must be modified as follows.
For each contained \lstinline|Option\-Param| one of the options must be chosen as the \lstinline|value|.
Each \lstinline|Unknown\-ParameterValue| must be replaced by a concrete \lstinline|ParameterValue| via the method \lstinline|concretize|.
Furthermore, each \lstinline|value| of a \lstinline|ParameterValue| must be specified.

\section{Tool Support}
\label{sec:implementation}
%
In the following, we report on a proof-of-concept implementation, in which we have instantiated our conceptual approach to XML.

\subsection{Tool architecture}
\begin{figure}
	\centering
	\spaceabove
	\includegraphics[width=\linewidth]{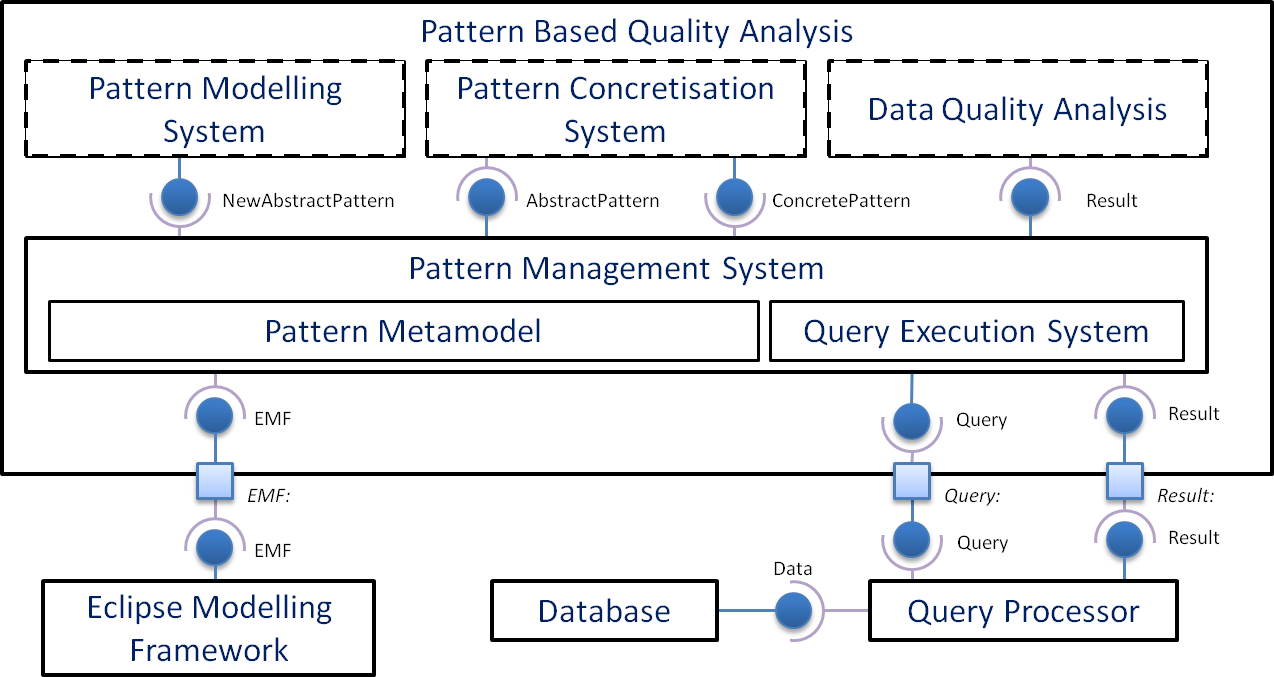}
	\caption{Component diagram of the implementation}
	\label{fig:components}
	\spacebelow
\end{figure}

Fig.~\ref{fig:components} shows the architecture of our tool implementation.
The \lstinline|Pattern Management System| stores predefined and user-defined ge\-ner\-ic, abstract and concrete patterns.
It encompasses an implementation of the metamodel discussed in Section \ref{subsec:metamodel} based on the Eclipse Modelling Framework (EMF)~\cite{emf}. 
The implementation includes an algorithm for adapting a generic pattern to XML (see Section~\ref{subsubsec:adaption})
and an algorithm for translating concrete patterns to XQuery~\cite{xquery}, which will be shortly presented in Section~\ref{subsec:mapping}.

The \lstinline|Query Execution System| applies a chosen set of patterns to a selected database by evaluating the generated queries.
In our implementation this is done via BaseX~\cite{basex}, an open-source XQuery processor.
The result is passed 
to the \lstinline|Data Quality Analysis| component, which
visualises the matched data regions together with metadata on the query execution.

The dashed components in Fig.~\ref{fig:components} form the front end of our implementation and are currently under development.
We will use Eclipse Sirius \cite{sirius} to develop a graphical modelling workbench for specifying generic and abstract patterns.
For the concretisation we will provide a form-based view for entering parameter values.




\subsection{Mapping Patterns to Queries}\label{subsec:mapping}
To detect quality problems by concrete patterns, they must be translated into queries.
The query language and thus the algorithm for translation depends on the database technology chosen. 
\subsubsection{Mapping Patterns to XQuery}\label{subsec:xquerymapping}
We implemented a systematic translation of any concrete pattern represented via the metamodel to XQuery~\cite{xquery} by means of the \lstinline|generateQuery()| method of each class.
Each query consists of a single or multiple \lstinline|for| clauses, a \lstinline|where| clause and a \lstinline|return| clause.
Listing \ref{lst:queryFunc} shows the generated XQuery expression for the concrete \texttt{FUNC} pattern presented in Section \ref{subsec:examplepatterns}.
\versionchoice{Further generated queries are available at \cite{supplementary_material}.}{Further generated queries are available at \cite{supplementary_material}.}{Further generated queries are shown in Section~\ref{appendix:queries} of the appendix.}{Further generated queries are shown in Section~\ref{appendix:queries} of the appendix.}

For the translation of a \lstinline|CompletePattern|, first its \lstinline|Graph| is traversed by
navigating along outgoing \lstinline|XmlNavigation|s and starting at the \lstinline|XmlRoot|.
For each such relation, a \lstinline|for| clause is appended to the query.
Its path expression specifies the navigation from the \lstinline|source| element to the \lstinline|target| element via the given XPath axis.
Conditions at the \lstinline|target| element as well as incoming or outgoing \lstinline|XmlReference|s are translated to XPath predicates inside square brackets following the path expression.

A pattern's nested condition is translated step-by-step to a nested expression being of the kinds \lstinline|some|, \lstinline|every|, \lstinline|and|, \lstinline|or|, \lstinline|not| or \lstinline|count|; 
this expression constitutes the query's \lstinline|where| clause.  
The graphs contained in a \lstinline|QuantifiedCondition| are traversed as described above.
Each \lstinline|XmlElement| which does not correspond to an element in a previous graph or is subject to a condition is translated to a \lstinline|some| or \lstinline|every| expression depending on the given quantifier.
The path expressions and predicates are created as discussed above.
A \lstinline|CountPattern| is translated as follows.
Its graph and the nested condition are translated as those of a \lstinline|CompletePattern|.
The resulting expression serves as the argument to a \lstinline|count| expression.

The \lstinline|return| clause determines that the pattern's \lstinline|returnElements| are returned by the query. 


\begin{lstlisting}[label=lst:queryFunc,caption={XQuery for the concrete FUNC pattern},language=XQuery,basicstyle=\ttfamily\footnotesize]
for $var1 in /child::*[./name()="building"]
where some $var2 in $var1/child::*[./name()="city"]
 satisfies some $var3 in $var1/child::*[./name()="country"]
 satisfies some $var4 in /child::*[./name()="building"]
 satisfies some $var5 in $var4/child::*[./name()="city"]
  [$var2/data()=./data()]
 satisfies some $var6 in $var4/child::*
  [./name()="country"][$var3/data()!=./data()]
satisfies true()
return $var1
\end{lstlisting}

\subsubsection{Mapping Patterns to other Query Languages}\label{subsec:othermapping}
To enable the application of patterns adapted to other database technologies, analogous algorithms for translating concrete patterns into corresponding query languages must be implemented.
%
For example, patterns adapted for relational databases may be translated into SQL \cite{sql}.
%
Patterns adapted for RDF~\cite{rdf} databases may be translated to SPARQL \cite{sparql}.
There already exist approaches \cite{kontokostas2014test,furber2010using,furber_swiqa_2011} for pattern-based quality analysis of linked data based on SPARQL templates.
We will compare with those in detail in Section \ref{sec:relatedWork}.


\section{Evaluation}
\label{sec:evaluation}
For the evaluation we will investigate two research questions: 
\begin{enumerate}
	\item[\textbf{RQ1:}] How far does our approach enable the detection of quality problems in research data?
	\item[\textbf{RQ2:}] What is the query response time of the problem detection?
\end{enumerate}
RQ1 will be discussed based on the application of our approach to two cultural heritage databases in Section~\ref{subsec:application} and on a more general level in Section~\ref{subsec:expressiveness}.
RQ2 will be answered in Section~\ref{subsec:scalability}.

\subsection{Application to Cultural Heritage Databases}\label{subsec:application}

To evaluate our approach, we compiled a representative selection of quality problems, 
created patterns to detect these problems and applied them to two databases.
The first database is present in \versionchoice{MIDAS~\cite{bove_marburger_2001_anonym}}{MIDAS~\cite{bove_marburger_2001}}{MIDAS~\cite{bove_marburger_2001_anonym}}{MIDAS~\cite{bove_marburger_2001}}, an XML-format for the description of art-historical objects.
This database holds information on more than 700,000 such objects plus related entities in more than 47 million XML elements. 
The second database is present in LIDO \cite{coburn_lido_nodate}, a CIDOC-CRM \cite{cidoc} application and XML format for harvesting and exchanging metadata of collectibles.
It describes more than 300,000 cultural heritage objects in more than 48 million XML elements.

Remember that the list of quality problems shown in Table~\ref{table:problems} is the result of our elicitation of quality problems occurring in cultural heritage databases.
For the evaluation, our goal was to cover each problem variant that is listed in Table \ref{table:problems} by a pattern.
We selected concrete quality problems that were either explicitly mentioned by domain experts or which violate the LIDO schema specification \cite{coburn_lido_nodate} or the MIDAS manual  \versionchoice{\cite{bove_marburger_2001_anonym}}{\cite{bove_marburger_2001}}{\cite{bove_marburger_2001_anonym}}{\cite{bove_marburger_2001}}.
We created a set of patterns that cover 85~\% of the problem variants listed in Table~\ref{table:problems}.
However, \emph{false values and references}, \emph{missing records} and \emph{misspellings} could not be covered.
We will elaborate on this in the next section. 

We created altogether \emph{21 generic patterns} and adapted them to XML, resulting in \emph{21 abstract patterns}.
Some of the patterns are closely related to those presented by Kontokostas et al. \cite{kontokostas2014test} and Fürber et al. \cite{furber2010using} \cite{furber_swiqa_2011}.
Table \ref{table:potentials} covers 27 problem variants.
As mentioned above, four variants could not be covered.
Hence, to detect all other variants in both databases, 46 \emph{concrete patterns} would be needed.
Because internal references are not intended in the LIDO format, LIDO data cannot be checked for \emph{referential integrity violations} and \emph{contradictory relationships}.
The integrity of references to external resources such as published controlled vocabulary cannot be checked with our current implementation.
As the LIDO schema specification explicitly allows multiple different measure units, we do not consider heterogeneous units in LIDO databases to be a problem and thus did not create a corresponding pattern.
Hence, 
we created a set of \emph{43 representative concrete patterns} in total.
We applied them to our selected databases and checked, for each pattern, a random sample of the matched elements for correctness.
\versionchoice{The generic, abstract and concrete patterns are presented in \cite{supplementary_material}.}{The generic, abstract and concrete patterns are presented in \cite{supplementary_material}.}{The generic, abstract and concrete patterns are presented in Section~\ref{appendix:generic} of the appendix.}{The generic, abstract and concrete patterns are presented in Section~\ref{appendix:generic} of the appendix.}
%

\emph{We conclude that, for cultural heritage data, our approach is effective since it enables 
the detection of 85\% of the quality problem variants elicited in our study.}
This answers RQ1 in the context of the application of our approach to cultural heritage databases.
We will discuss RQ1 on a more general level in the following section.
The main \emph{threat to validity} is that we applied our approach only to research data of one domain (i.e. cultural heritage) and one database technology (i.e. XML).


\subsection{Expressiveness}\label{subsec:expressiveness}
In the following, we will discuss the expressiveness of our approach with regard to the data quality problems that can be detected.
Table~\ref{table:potentials} gives an overview of \emph{potentials} enabling and \emph{limitations} preventing the detection of certain quality problems.
Problem variants are listed only if the affected potentials and limitations differ significantly.

\begin{table*}
	\centering
	\spaceabove
	\caption{Potentials allowing and limitations preventing the detection of quality problems.}
	\label{table:potentials}
	
	\begin{tabular}{ l l c c c c c c c c c c c c } 
		\toprule
		\multirow{2}{*}{\textbf{Quality Problem}} & \multirow{2}{*}{\textbf{Variants}} & \multicolumn{6}{c}{\emph{\textbf{Potentials}}} & & \multicolumn{5}{c}{\emph{\textbf{Limitations}}}\\
		\cmidrule{3-8} \cmidrule{10-14}
		& & \textbf{RE} & \textbf{C} & \textbf{E} & \textbf{GS} & \textbf{OC} & \textbf{FOL} & & \textbf{M} & \textbf{O} & \textbf{CVO} & \textbf{RWV} & \textbf{HOL}\\
		\midrule 
		\multirow{2}{*}{\emph{Illegal values}}		
		& wrong datatype &\checkmark&&&\checkmark&&&&&&&& \\
		& domain violation (interval, set, syntax) &\checkmark&\checkmark&\checkmark&&&&&&&&& \\
		\midrule		
		\multirow{3}{*}{\emph{Missing data}}	
		& missing values &\checkmark&\checkmark&&\checkmark&&\checkmark&&&&&& \\
		& missing references and records &&&&\checkmark&&\checkmark&&&&&$\times$& \\
		& dummy values &\checkmark&\checkmark&\checkmark&&&&&&$\times$&&& \\
		\midrule
		\multicolumn{2}{l}{\emph{Referential integrity violation}} &&&&\checkmark&&\checkmark&&&&&& \\
		\midrule
		\multicolumn{2}{l}{\emph{Unique value violation}} &&\checkmark&&\checkmark&\checkmark&&&&&&& \\
		\midrule 
		\multicolumn{2}{l}{\emph{Violation of a functional dependency}} &&\checkmark&&\checkmark&&&&&&&& \\
		\midrule 		
		\multicolumn{2}{l}{\emph{Contradictory relationships}}  &&&&\checkmark&&&&&&&&$\times$ \\
		\midrule			
		\multirow{4}{*}{\emph{Imprecise data}}
		& alternative possible values &\checkmark&&&\checkmark&\checkmark&&&&&&& \\
		& imprecise numerical values &\checkmark&&&\checkmark&&&&&&&& \\
		& abstract terms and ambiguous values &&&\checkmark&&&&&&$\times$&&& \\
		& abbreviations &\checkmark&&&&&&&&&&& \\
		\midrule
		\multicolumn{2}{l}{\emph{Misplaced information}} &\checkmark&&&&&&&&&&$\times$& \\
		\midrule
		\multicolumn{2}{l}{\emph{Redundant data}} &&\checkmark&&\checkmark&\checkmark&&&$\times$&&&& \\
		\midrule
		\multirow{3}{*}{\emph{Heterogeneous data}}
		& heterogeneous measure units &\checkmark&&&&&&&&&&& \\
		& heterogeneous value representations &\checkmark&&\checkmark&&&&&&$\times$&&& \\
		& heterogeneous structural representations &&&&\checkmark&&\checkmark&&&&&& \\
		\midrule
		\multicolumn{2}{l}{\emph{Misspellings}} &\checkmark&&&&&&&$\times$&&&& \\
		\midrule
		\multicolumn{2}{l}{\emph{Semantically incorrect data}} &\checkmark&\checkmark&&\checkmark&\checkmark&\checkmark&&&&$\times$&$\times$& \\
		\bottomrule	
	\end{tabular}
	\vspace{1mm}
	\captionsetup{width=\textwidth-1cm}
	\caption*{\normalfont		
		\textbf{RE}: regular expressions, 
		\textbf{C}: comparisons, 
		\textbf{E}: enumerations, 
		\textbf{GS}: graph structures, 
		\textbf{OC}: occurrence count (metric), 
		\textbf{FOL}: first-order logic, 
		\textbf{M}: further metrics, 
		\textbf{O}: ontology, 
		\textbf{CVO}: complex value operations, 
		\textbf{RWV}: real-world values, 
		\textbf{HOL}: higher-order logic
	}	
	\spacebelow
\end{table*} 

\subsubsection{Potentials}
The strength of our approach lies in the detection of problems that reveal themselves through the syntax and especially, the structure of the data.
In the following, we will discuss important features for enabling this functionality.

Our approach allows checking values concerning {\em regular expressions}.
This enables the detection of syntactical patterns that hint at certain quality problems, e.g. implicitly encoded \emph{imprecise numerical values}.
Further, \emph{misplaced information}, \emph{misspellings} and \emph{semantically incorrect data} can be detected via regular expressions if they are accompanied by violations of syntactical rules.

Moreover, our approach allows casting values in the data via predefined operations and \emph{comparing} these values to other values in the data or constant literals, e.g. to detect \emph{interval violations}.

By \emph{enumerations}, we can check whether a value is (not) contained in a list of constant literals and thereby detect problems such as \emph{set violations} and \emph{dummy values}. 
This could alternatively be checked using regular expressions or disjunction of multiple graphs.

Our approach allows checking the {\em existence of complex graph structures}, which may include \emph{comparisons} between elements and properties.
For example, this enables the detection of \emph{contradictory relationships} of a fixed size.


A further feature is the ability to express conditions concerning the \emph{number of occurrences} of a pattern in the data. 
For example, this allows detecting \emph{unique value violations}. 
Restricted forms of \emph{approximate duplicate records} can be detected by specifying a fixed set of properties whose equalities across records indicate that the records are duplicates.

A particular strength of our approach lies in {\em first-order logic expressions over graph structures}.
For example, this allows checking the non-existence of certain graph structures.
Hence, \emph{referential integrity violations} can be detected, which additionally hint at \emph{missing records}.
Via first-order logic expressions and comparisons, our approach allows for checking complex domain-specific dependencies between values of one or multiple related records.
For example, an artist's date of birth must precede the dates of creation of his or her artworks.
Violations of such plausibility rules hint at \emph{semantically incorrect data}.

\subsubsection{Limitations}
In the following, we will discuss the limitations of our approach and affected quality problems.

Besides calculating the number of occurrences of a pattern in the data, our approach does not include any {\em metrics} currently, e.g. for similarity checks \cite{sun2015comparative,elmagarmid2006duplicate}.
This prevents the detection of \emph{misspellings} and \emph{approximate duplicates} in general. 
We plan to investigate the integration of such metrics in the future.

We detected occurrences of \emph{dummy values}, \emph{ambiguous values}, \emph{abstract terms} and \emph{heterogeneous value representations} (e.g. synonyms) in the cultural heritage data by the enumerations feature discussed above.
The corresponding patterns scan the data for keywords manually specified by domain experts.
Integrating a fuzzy search mechanism \cite{fuzzysearch2,ontologysearch,fuzzysearch} based on domain-specific {\em ontologies}
would enable the universal detection of such problems. 


As discussed above, our approach supports the comparison of values within the data, e.g. as part of plausibility rules.
For the selected cultural heritage data, there hardly exist any syntactical rules (e.g. on how to record a date).
Therefore, the syntax is very heterogeneous.
Furthermore, uncertainties are often encoded implicitly.
Thus, the comparison of such values without applying \emph{complex operations on these values} beforehand is not possible.
We will investigate this aspect in future work.


Some data quality problems require \emph{external knowledge about the real world} to be detected.
For instance, our approach does not allow detecting any \emph{semantically incorrect values} that, however, satisfy the specified plausibility rules and do not show through illegal values, functional dependency violations or contradictory relationships.
The required real world values, however, might be present in other databases.
In future work, we will investigate the comparison of records from different databases (possibly with different formats and underlying technologies) to detect semantically incorrect values.
If the data does not contain any hints for a relation or record to be missing (e.g. a referential integrity violation), then the \emph{absence of relations and records} can also not be detected. 


As we do not support {\em higher-order logic}, we cannot express \emph{contradictory relationships} of variable size.


\subsubsection{Summary}
Concerning RQ1, we found that, in general, the strength of our approach lies in expressing first-order logic conditions over arbitrarily complex graph structures, thereby detecting problems that reveal themselves through the structure of the data.
Problems that require the ability to assess the meaning of data, instead, or the knowledge of real-world values to be detected, are not covered by our approach yet.

\subsection{Performance}\label{subsec:scalability}
We measured the runtimes of the 43 applied patterns via BaseX, which is a "main memory application"~\cite{basex}, on a Windows 10 PC with a 4.2 GHz CPU and 16 GB RAM.
Table~\ref{table:runtime} gives a rough overview. 

\begin{table}
	\centering
	\spaceabove
	\caption{Runtimes of patterns applied to cultural heritage databases}
	\begin{tabular}{l c c c c}
		Percentage of patterns & 70 \% & 80 \% & 90 \% & 95 \% \\
		\midrule
		Runtime & < 10 s & < 20 s & < 4 min & < 6 h \\
	\end{tabular}	
	\label{table:runtime}
	\spacebelow
\end{table}
To analyse the factors that impact the runtime, we take a closer look at the generated queries.
For each \lstinline|Element| in the pattern a nested loop expression is created.
We observed that in practice, 
these expressions typically select only elements on one specific level in the XML hierarchy (via the \lstinline|child| axis) or all elements below a certain level (via the \lstinline|descendant| axis) that is not the top level.
Nevertheless, some of the patterns are translated into multiple nested loops that when applied to the large cultural heritage databases iterate through very large sets of XML elements.
%
The patterns that ran longer than 4 minutes are concerned with the following quality problems: 
{\em unique value violation}, {\em violation of a functional dependency}, {\em exact and approximate duplicate records}.
For example, the pattern that detects functional dependency violations
in the LIDO data, is translated into a query consisting of 7 nested loops.
Its application results in more than 77 billion element comparisons and a runtime of almost six hours.

\emph{Concerning RQ2 we conclude that 
80\% of the patterns can easily be executed each time the data is modified without causing major limitations in day-to-day work since they run less then 20 seconds.}  
Others may run multiple hours or days and thus, in practice, should be executed less often.
The main \emph{threat to validity} is that we considered only two XML databases of a single domain.

The runtime can be tweaked by designing patterns such that the number of visited XML elements is minimised, e.g.,
by using a fixed number of \lstinline|child| axes instead of the \lstinline|descendant| axis
or by applying patterns only to parts of the database.
In future work we will investigate how to support users in designing efficient patterns.

\section{Related Work}
\label{sec:relatedWork}


In our literature study, we focused on pattern-based approaches to data quality analysis presented in research papers.
For each approach, we determine the level of abstraction,
measure the expressiveness by means of the features discussed in Section~\ref{subsec:expressiveness} and consider the pattern language.
Table~\ref{table:relatedWork} gives an overview.
\begin{table}
	\centering
	\spaceabove
	\caption{Comparison of related work with our approach}
	\begin{tabular}{ 
			p{\dimexpr 0.22\textwidth/2-3\tabcolsep} 
			p{\dimexpr 0.29\textwidth/2-3\tabcolsep} 
			p{\dimexpr 0.25\textwidth/2-3\tabcolsep} 
			p{\dimexpr 0.24\textwidth/2-3\tabcolsep} } 
		\toprule
		\textbf{Approach} 
		& \textbf{Generality} 
		& \textbf{Expressive\-ness} 
		& \textbf{Pattern Language} \\
		\midrule 
		Kontokostas \newline et al. \cite{kontokostas2014test} 
		& abstract \newline (RDF) 
		& RE, C, GS, OC, FOL 
		& SPARQL \\
		\midrule 
		Fürber et al. \cite{furber2010using, furber_swiqa_2011} 
		& abstract \newline (RDF) 
		& RE, C, E, GS, OC, FOL 
		& SPIN \\
		\midrule
		Bizer et al. \cite{bizer_quality-driven_2009} 
		& concrete (Named Graphs) 
		& RE, C, GS, PL 
		& WIQA-PL \\
		\midrule
		Bicevska et al. \cite{oditis_domain-specific_2017} 
		& concrete \newline (relational) 
		& C, GS, FOL, CS, GR 
		& graphical DSL, SQL \\
		\midrule
		\textbf{Our \newline approach} 
		& generic \newline (PoC: XML) 
		& RE, C, E, GS, OC, FOL 
		& metamodel \newline (+ GUI)\\
		\bottomrule 
	\end{tabular}
	\vspace{1mm}
	\captionsetup{width=\textwidth/2-2cm}
	\caption*{\normalfont
		\textbf{RE}: regular expressions, 
		\textbf{C}: comparisons,
		\textbf{GS}: graph structures,
		\textbf{OC}: occurrence count,
		\textbf{E}: enumerations,
		\textbf{PL}: propositional logic,
		\textbf{CS}: column sum,
		\textbf{GR}: group rows
	}
	\label{table:relatedWork}
	\spacebelow
\end{table}

Kontokostas et al. \cite{kontokostas2014test} and Fürber et al. \cite{furber2010using, furber_swiqa_2011} presented similar approaches to detect quality problems in Linked Data based on parameterized SPARQL or SPIN query templates.
Hence, they do not consider patterns on the generic level.
The approaches have similar potentials and limitations as ours.
In the evaluation part of their paper, Kontokostas et al. 
identified the necessity of creating further abstract patterns to cover a wider range of quality problems.
However, both approaches require abstract patterns (i.e. queries) to be written in SPARQL or SPIN containing more implementation details than in our model-driven approach. 


Bizer et al. \cite{bizer_quality-driven_2009} presented a policy framework for quality-driven information filtering.
They consider patterns on the concrete level only.
The approach is limited to propositional logic and does not support the counting of occurrences of a pattern in the data.
Patterns are expressed with a custom SPARQL-based language. 

Bicevska et al. \cite{oditis_domain-specific_2017} 
proposed to specify data quality requirements for specific database technologies and formats using a domain-specific language (DSL).
Thus, concrete patterns are considered only.
The approach differs slightly from the others with regard to the supported features.
The pattern language, i.e., the DSL, is not presented in the paper. 
Informal explanations are used instead to describe examples for quality specifications. 
The authors proposed to translate them into a query language, such as SQL, but did not present an algorithm. 
To motivate their approach, Bicevska et al. discussed the use of the object constraint language (OCL) to define data quality.
OCL is powerful enough to specify constraints in first-order logic and beyond.
However, it is not well-suited for domain experts without good skills in object-oriented programming.
Furthermore, OCL is fully typed and therefore, well-suited for structured data, whereas research data is often semi-structured~\cite{semistructured_data}.
\emph{In summary, our approach is the only one that supports generic patterns and thus, is flexible concerning the underlying database technology.}
The expressiveness of most of the approaches considered seems to be similar to ours.
Existing pattern languages often require the specification of details that may require skills that domain experts usually do not have.
By being model-driven,
our approach abstracts from query languages.
This lays the foundation for a potentially user-friendly and universal representation of patterns. 

\todo[inline,backgroundcolor=yellow]{Inserted short paragraph on future work below}
In future work we will compare our approach also to non-pattern-based approaches and industrial tools \cite{ERW19} to detect quality problems, for example via machine learning \cite{AbedjanAOPS15} or metric-based methods \cite{mendes_sieve:_2012}.
Furthermore, we will compare our work to approaches for generic data model definitions \cite{rennau}.

\section{Conclusion}
\label{sec:conclusion}
\todo[inline,backgroundcolor=yellow]{Reworked}
To support a dynamic digitalisation of specific scientific fields, especially the humanities, we presented a model-driven approach to analyse the quality of research data.
It supports the specification of patterns to identify data quality problems, independent of the underlying database technology and format.
A proof-of-concept implementation shows how this approach can be used for XML databases.
We evaluated it for expressiveness and performance by applying it to two cultural heritage databases.
While its expressiveness is comprehensive for pattern-based approaches, it has to be integrated with further techniques for quality analysis such as metrics and ontologies, to cover the wide range of data quality problems.
\todo[inline,backgroundcolor=yellow]{Moved humanities discussion here GT: erased here and moved up}

In future work, we will empirically evaluate the usability of the proposed pattern notation.
As we currently see no obstacle to applying our approach to research data from other scientific fields such as biodiversity and also non-research data, we will investigate those applications in the future.
To evaluate the overall concept of our approach we will implement it for further database technologies.
\todo{This sentence changed.}
Further ahead, we plan to combine our approach with further analysis techniques such as machine learning.

Our overall goal is to develop a framework for quality assurance of research data, where the detection of quality problems is the first essential step.

\begin{acks}
	\todo[inline,backgroundcolor=yellow]{Use commands grantsponsor and grantnum? BMBF Förderkennzeichen?}
	This work has been developed in the project KONDA.
	It is partly funded by the \grantsponsor{bmbf}{German Federal Ministry of Education and Research}{https://www.bmbf.de}
	(Grant No.: \grantnum{bmbf}{16QK06A}). 
	We would like to thank the members of the project for the excellent collaboration and support.
	Many thanks to Oguzhan Balandi, Michalis Famelis and Regine Stein for the extensive reviews and their constructive criticism.
	Further we would like to thank the anonymous reviewers for their valuable suggestions.
\end{acks}

\clearpage

\bibliographystyle{ACM-Reference-Format}
\bibliography{document}


\begin{thebibliography}{42}


\ifx \showCODEN    \undefined \def \showCODEN     #1{\unskip}     \fi
\ifx \showDOI      \undefined \def \showDOI       #1{#1}\fi
\ifx \showISBNx    \undefined \def \showISBNx     #1{\unskip}     \fi
\ifx \showISBNxiii \undefined \def \showISBNxiii  #1{\unskip}     \fi
\ifx \showISSN     \undefined \def \showISSN      #1{\unskip}     \fi
\ifx \showLCCN     \undefined \def \showLCCN      #1{\unskip}     \fi
\ifx \shownote     \undefined \def \shownote      #1{#1}          \fi
\ifx \showarticletitle \undefined \def \showarticletitle #1{#1}   \fi
\ifx \showURL      \undefined \def \showURL       {\relax}        \fi
\providecommand\bibfield[2]{#2}
\providecommand\bibinfo[2]{#2}
\providecommand\natexlab[1]{#1}
\providecommand\showeprint[2][]{arXiv:#2}

\bibitem[\protect\citeauthoryear{Abedjan, Akcora, Ouzzani, Papotti, and
  Stonebraker}{Abedjan et~al\mbox{.}}{2015}]%
        {AbedjanAOPS15}
\bibfield{author}{\bibinfo{person}{Ziawasch Abedjan},
  \bibinfo{person}{Cuneyt~Gurcan Akcora}, \bibinfo{person}{Mourad Ouzzani},
  \bibinfo{person}{Paolo Papotti}, {and} \bibinfo{person}{Michael
  Stonebraker}.} \bibinfo{year}{2015}\natexlab{}.
\newblock \showarticletitle{Temporal Rules Discovery for Web Data Cleaning}.
\newblock \bibinfo{journal}{\emph{Proc. {VLDB} Endow.}} \bibinfo{volume}{9},
  \bibinfo{number}{4} (\bibinfo{year}{2015}), \bibinfo{pages}{336--347}.
\newblock
\urldef\tempurl%
\url{https://doi.org/10.14778/2856318.2856328}
\showDOI{\tempurl}


\bibitem[\protect\citeauthoryear{Abiteboul}{Abiteboul}{1997}]%
        {semistructured_data}
\bibfield{author}{\bibinfo{person}{Serge Abiteboul}.}
  \bibinfo{year}{1997}\natexlab{}.
\newblock \showarticletitle{Querying Semi-Structured Data}. In
  \bibinfo{booktitle}{\emph{Database Theory - {ICDT} '97, 6th International
  Conference, Delphi, Greece, January 8-10, 1997, Proceedings}}
  \emph{(\bibinfo{series}{Lecture Notes in Computer Science})},
  \bibfield{editor}{\bibinfo{person}{Foto~N. Afrati} {and}
  \bibinfo{person}{Phokion~G. Kolaitis}} (Eds.), Vol.~\bibinfo{volume}{1186}.
  \bibinfo{publisher}{Springer}, \bibinfo{pages}{1--18}.
\newblock
\urldef\tempurl%
\url{https://doi.org/10.1007/3-540-62222-5\_33}
\showDOI{\tempurl}


\bibitem[\protect\citeauthoryear{Batini, Cappiello, Francalanci, and
  Maurino}{Batini et~al\mbox{.}}{2009}]%
        {batini_methodologies_2009}
\bibfield{author}{\bibinfo{person}{Carlo Batini}, \bibinfo{person}{Cinzia
  Cappiello}, \bibinfo{person}{Chiara Francalanci}, {and}
  \bibinfo{person}{Andrea Maurino}.} \bibinfo{year}{2009}\natexlab{}.
\newblock \showarticletitle{Methodologies for data quality assessment and
  improvement}.
\newblock \bibinfo{journal}{\emph{{ACM} Comput. Surv.}} \bibinfo{volume}{41},
  \bibinfo{number}{3} (\bibinfo{year}{2009}), \bibinfo{pages}{16:1--16:52}.
\newblock
\urldef\tempurl%
\url{https://doi.org/10.1145/1541880.1541883}
\showDOI{\tempurl}


\bibitem[\protect\citeauthoryear{Bijral and Mukhopadhyay}{Bijral and
  Mukhopadhyay}{2014}]%
        {fuzzysearch2}
\bibfield{author}{\bibinfo{person}{Simran Bijral} {and}
  \bibinfo{person}{Debajyoti Mukhopadhyay}.} \bibinfo{year}{2014}\natexlab{}.
\newblock \showarticletitle{Efficient Fuzzy Search Engine with {B} -Tree Search
  Mechanism}. In \bibinfo{booktitle}{\emph{2014 International Conference on
  Information Technology, {ICIT} 2014, Bhubaneswar, India, December 22-24,
  2014}}. \bibinfo{publisher}{{IEEE}}, \bibinfo{pages}{118--122}.
\newblock
\urldef\tempurl%
\url{https://doi.org/10.1109/ICIT.2014.19}
\showDOI{\tempurl}


\bibitem[\protect\citeauthoryear{Biron and Malhotra}{Biron and
  Malhotra}{2004}]%
        {xmlschema}
\bibfield{author}{\bibinfo{person}{Paul~V. Biron} {and} \bibinfo{person}{Ashok
  Malhotra}.} \bibinfo{year}{2004}\natexlab{}.
\newblock \bibinfo{booktitle}{\emph{{XML} Schema Part 2: Datatypes Second
  Edition}}.
\newblock \bibinfo{type}{{W3C} Recommendation}. \bibinfo{institution}{W3C}.
\newblock
\newblock
\shownote{http://www.w3.org/TR/2004/REC-xmlschema-2-20041028/.}


\bibitem[\protect\citeauthoryear{Bizer and Cyganiak}{Bizer and
  Cyganiak}{2009}]%
        {bizer_quality-driven_2009}
\bibfield{author}{\bibinfo{person}{Christian Bizer} {and}
  \bibinfo{person}{Richard Cyganiak}.} \bibinfo{year}{2009}\natexlab{}.
\newblock \showarticletitle{Quality-driven information filtering using the
  {WIQA} policy framework}.
\newblock \bibinfo{journal}{\emph{J. Web Semant.}} \bibinfo{volume}{7},
  \bibinfo{number}{1} (\bibinfo{year}{2009}), \bibinfo{pages}{1--10}.
\newblock
\urldef\tempurl%
\url{https://doi.org/10.1016/j.websem.2008.02.005}
\showDOI{\tempurl}


\bibitem[\protect\citeauthoryear{Bonino, Corno, Farinetti, and Bosca}{Bonino
  et~al\mbox{.}}{2004}]%
        {ontologysearch}
\bibfield{author}{\bibinfo{person}{Dario Bonino}, \bibinfo{person}{Fulvio
  Corno}, \bibinfo{person}{Laura Farinetti}, {and} \bibinfo{person}{Alessio
  Bosca}.} \bibinfo{year}{2004}\natexlab{}.
\newblock \showarticletitle{Ontology driven semantic search}.
\newblock \bibinfo{journal}{\emph{WSEAS Transaction on Information Science and
  Application}} \bibinfo{volume}{1}, \bibinfo{number}{6}
  (\bibinfo{year}{2004}), \bibinfo{pages}{1597--1605}.
\newblock


\bibitem[\protect\citeauthoryear{Bove, Heusinger, and Kailus}{Bove
  et~al\mbox{.}}{2001}]%
        {bove_marburger_2001}
\bibfield{author}{\bibinfo{person}{Jens Bove}, \bibinfo{person}{Lutz
  Heusinger}, {and} \bibinfo{person}{Angela Kailus}.}
  \bibinfo{year}{2001}\natexlab{}.
\newblock \bibinfo{booktitle}{\emph{Marburger {Informations}-,
  {Dokumentations}- und {Administrations}-{System} ({MIDAS}): {Handbuch} und
  {CD} ({Literatur} und {Archiv}; 4). - 4. überarbeitete {Auflage}}}.
\newblock
\urldef\tempurl%
\url{https://archiv.ub.uni-heidelberg.de/artdok/3770/}
\showURL{%
\tempurl}


\bibitem[\protect\citeauthoryear{Coburn, Light, McKenna, Stein, and
  Vitzthum}{Coburn et~al\mbox{.}}{line}]%
        {coburn_lido_nodate}
\bibfield{author}{\bibinfo{person}{Erin Coburn}, \bibinfo{person}{Richard
  Light}, \bibinfo{person}{Gordon McKenna}, \bibinfo{person}{Regine Stein},
  {and} \bibinfo{person}{Axel Vitzthum}.} \bibinfo{year}{[Online]}\natexlab{}.
\newblock \bibinfo{title}{{LIDO} ({Lightweight} {Information} {Describing}
  {Objects})}.
\newblock
  \bibinfo{howpublished}{\url{http://network.icom.museum/cidoc/working-groups/lido/}}.
\newblock
\urldef\tempurl%
\url{http://network.icom.museum/cidoc/working-groups/lido/}
\showURL{%
\tempurl}


\bibitem[\protect\citeauthoryear{Date and Darwen}{Date and Darwen}{1997}]%
        {sql}
\bibfield{author}{\bibinfo{person}{C.~J. Date} {and} \bibinfo{person}{Hugh
  Darwen}.} \bibinfo{year}{1997}\natexlab{}.
\newblock \bibinfo{booktitle}{\emph{A Guide to {SQL} Standard, 4th Edition}}.
\newblock \bibinfo{publisher}{Addison-Wesley}.
\newblock
\showISBNx{0-201-96426-0}


\bibitem[\protect\citeauthoryear{Doerr, Bruseker, Bekiari, Ore, Velios, and
  Stead}{Doerr et~al\mbox{.}}{2020}]%
        {cidoc}
\bibfield{author}{\bibinfo{person}{Martin Doerr}, \bibinfo{person}{George
  Bruseker}, \bibinfo{person}{Chryssoula Bekiari},
  \bibinfo{person}{Christian~Emil Ore}, \bibinfo{person}{Thanasis Velios},
  {and} \bibinfo{person}{Stephen Stead}.} \bibinfo{year}{2020}\natexlab{}.
\newblock \bibinfo{title}{Definition of the CIDOC Conceptual Reference Model}.
\newblock
  \bibinfo{howpublished}{\url{http://www.cidoc-crm.org/sites/default/files/CIDOC
  CRM_v6.2.9 30-4-2020 .pdf}}.
\newblock
\newblock
\shownote{(Accessed on 05/15/2020).}


\bibitem[\protect\citeauthoryear{Dyck, Robie, and Spiegel}{Dyck
  et~al\mbox{.}}{2017}]%
        {xquery}
\bibfield{author}{\bibinfo{person}{Michael Dyck}, \bibinfo{person}{Jonathan
  Robie}, {and} \bibinfo{person}{Josh Spiegel}.}
  \bibinfo{year}{2017}\natexlab{}.
\newblock \bibinfo{booktitle}{\emph{{XQuery} 3.1: An {XML} Query Language}}.
\newblock \bibinfo{type}{{W3C} Recommendation}. \bibinfo{institution}{W3C}.
\newblock
\newblock
\shownote{https://www.w3.org/TR/2017/REC-xquery-31-20170321/.}


\bibitem[\protect\citeauthoryear{Ehrlinger, Rusz, and W{\"{o}}{\ss}}{Ehrlinger
  et~al\mbox{.}}{2019}]%
        {ERW19}
\bibfield{author}{\bibinfo{person}{Lisa Ehrlinger}, \bibinfo{person}{Elisa
  Rusz}, {and} \bibinfo{person}{Wolfram W{\"{o}}{\ss}}.}
  \bibinfo{year}{2019}\natexlab{}.
\newblock \showarticletitle{A Survey of Data Quality Measurement and Monitoring
  Tools}.
\newblock \bibinfo{journal}{\emph{CoRR}}  \bibinfo{volume}{abs/1907.08138}
  (\bibinfo{year}{2019}).
\newblock
\showeprint[arxiv]{1907.08138}
\urldef\tempurl%
\url{http://arxiv.org/abs/1907.08138}
\showURL{%
\tempurl}


\bibitem[\protect\citeauthoryear{Elmagarmid, Ipeirotis, and
  Verykios}{Elmagarmid et~al\mbox{.}}{2007}]%
        {elmagarmid2006duplicate}
\bibfield{author}{\bibinfo{person}{Ahmed~K. Elmagarmid},
  \bibinfo{person}{Panagiotis~G. Ipeirotis}, {and}
  \bibinfo{person}{Vassilios~S. Verykios}.} \bibinfo{year}{2007}\natexlab{}.
\newblock \showarticletitle{Duplicate Record Detection: {A} Survey}.
\newblock \bibinfo{journal}{\emph{{IEEE} Trans. Knowl. Data Eng.}}
  \bibinfo{volume}{19}, \bibinfo{number}{1} (\bibinfo{year}{2007}),
  \bibinfo{pages}{1--16}.
\newblock
\urldef\tempurl%
\url{https://doi.org/10.1109/TKDE.2007.250581}
\showDOI{\tempurl}


\bibitem[\protect\citeauthoryear{Ewald}{Ewald}{2019}]%
        {folemergence}
\bibfield{author}{\bibinfo{person}{William Ewald}.}
  \bibinfo{year}{2019}\natexlab{}.
\newblock \showarticletitle{The Emergence of First-Order Logic}.
\newblock In \bibinfo{booktitle}{\emph{The Stanford Encyclopedia of Philosophy}
  (\bibinfo{edition}{spring 2019} ed.)},
  \bibfield{editor}{\bibinfo{person}{Edward~N. Zalta}} (Ed.).
  \bibinfo{publisher}{Metaphysics Research Lab, Stanford University}.
\newblock


\bibitem[\protect\citeauthoryear{Foundation}{Foundation}{linea}]%
        {emf}
\bibfield{author}{\bibinfo{person}{Eclipse Foundation}.}
  \bibinfo{year}{[Online]}\natexlab{a}.
\newblock \bibinfo{title}{Eclipse {Modeling} {Project}}.
\newblock
\newblock
\urldef\tempurl%
\url{https://www.eclipse.org/modeling/emf/}
\showURL{%
\tempurl}


\bibitem[\protect\citeauthoryear{Foundation}{Foundation}{lineb}]%
        {sirius}
\bibfield{author}{\bibinfo{person}{Eclipse Foundation}.}
  \bibinfo{year}{[Online]}\natexlab{b}.
\newblock \bibinfo{title}{Sirius}.
\newblock
\newblock
\urldef\tempurl%
\url{https://www.eclipse.org/sirius/}
\showURL{%
\tempurl}


\bibitem[\protect\citeauthoryear{Franconi, Mosca, Oriol, Rull, and
  Teniente}{Franconi et~al\mbox{.}}{2019}]%
        {FMORT19}
\bibfield{author}{\bibinfo{person}{Enrico Franconi},
  \bibinfo{person}{Alessandro Mosca}, \bibinfo{person}{Xavier Oriol},
  \bibinfo{person}{Guillem Rull}, {and} \bibinfo{person}{Ernest Teniente}.}
  \bibinfo{year}{2019}\natexlab{}.
\newblock \showarticletitle{$\hbox {OCL}_\textsf {FO}$: First-Order Expressive
  OCL Constraints for Efficient Integrity Checking}.
\newblock \bibinfo{journal}{\emph{Softw. Syst. Model.}} \bibinfo{volume}{18},
  \bibinfo{number}{4} (\bibinfo{date}{Aug.} \bibinfo{year}{2019}),
  \bibinfo{pages}{2655 -- 2678}.
\newblock
\showISSN{1619-1366}
\urldef\tempurl%
\url{https://doi.org/10.1007/s10270-018-0688-z}
\showDOI{\tempurl}


\bibitem[\protect\citeauthoryear{F{\"{u}}rber and Hepp}{F{\"{u}}rber and
  Hepp}{2010}]%
        {furber2010using}
\bibfield{author}{\bibinfo{person}{Christian F{\"{u}}rber} {and}
  \bibinfo{person}{Martin Hepp}.} \bibinfo{year}{2010}\natexlab{}.
\newblock \showarticletitle{Using {SPARQL} and {SPIN} for Data Quality
  Management on the Semantic Web}. In \bibinfo{booktitle}{\emph{Business
  Information Systems, 13th International Conference, {BIS} 2010, Berlin,
  Germany, May 3-5, 2010. Proceedings}} \emph{(\bibinfo{series}{Lecture Notes
  in Business Information Processing})},
  \bibfield{editor}{\bibinfo{person}{Witold Abramowicz} {and}
  \bibinfo{person}{Robert Tolksdorf}} (Eds.), Vol.~\bibinfo{volume}{47}.
  \bibinfo{publisher}{Springer}, \bibinfo{pages}{35--46}.
\newblock
\urldef\tempurl%
\url{https://doi.org/10.1007/978-3-642-12814-1\_4}
\showDOI{\tempurl}


\bibitem[\protect\citeauthoryear{F{\"{u}}rber and Hepp}{F{\"{u}}rber and
  Hepp}{2011}]%
        {furber_swiqa_2011}
\bibfield{author}{\bibinfo{person}{Christian F{\"{u}}rber} {and}
  \bibinfo{person}{Martin Hepp}.} \bibinfo{year}{2011}\natexlab{}.
\newblock \showarticletitle{Swiqa - a semantic web information quality
  assessment framework}. In \bibinfo{booktitle}{\emph{19th European Conference
  on Information Systems, {ECIS} 2011, Helsinki, Finland, June 9-11, 2011}},
  \bibfield{editor}{\bibinfo{person}{Virpi~Kristiina Tuunainen},
  \bibinfo{person}{Matti Rossi}, {and} \bibinfo{person}{Joe Nandhakumar}}
  (Eds.). \bibinfo{pages}{76}.
\newblock
\urldef\tempurl%
\url{http://aisel.aisnet.org/ecis2011/76}
\showURL{%
\tempurl}


\bibitem[\protect\citeauthoryear{{German Council for Scientific Information
  Infrastructures (RfII)}}{{German Council for Scientific Information
  Infrastructures (RfII)}}{2020}]%
        {rfii}
\bibfield{author}{\bibinfo{person}{{German Council for Scientific Information
  Infrastructures (RfII)}}.} \bibinfo{year}{2020}\natexlab{}.
\newblock \bibinfo{booktitle}{\emph{The {Data} {Quality} {Challenge}.
  {Recommendations} for {Sustainable} {Research} in the {Digital} {Turn}}}.
\newblock \bibinfo{address}{Göttingen}.
\newblock
\urldef\tempurl%
\url{http://www.rfii.de/?p=4203}
\showURL{%
\tempurl}


\bibitem[\protect\citeauthoryear{Gr{\"{u}}n}{Gr{\"{u}}n}{2006}]%
        {basex}
\bibfield{author}{\bibinfo{person}{Christian Gr{\"{u}}n}.}
  \bibinfo{year}{2006}\natexlab{}.
\newblock \showarticletitle{Pushing {XML} Main Memory Databases to their
  Limits}. In \bibinfo{booktitle}{\emph{Tagungsband zum 18. GI-Workshop
  {\"{u}}ber Grundlagen von Datenbanken (18th GI-Workshop on the Foundations of
  Databases), Wittenberg, Sachsen-Anhalt, Deutschland, 6.-9. Juni 2006}},
  \bibfield{editor}{\bibinfo{person}{Stefan Brass} {and}
  \bibinfo{person}{Alexander Hinneburg}} (Eds.). \bibinfo{publisher}{Institute
  of Computer Science, Martin-Luther-University}, \bibinfo{pages}{60--64}.
\newblock
\urldef\tempurl%
\url{http://dbs.informatik.uni-halle.de/GvD2006/gvd06\_gruen.pdf}
\showURL{%
\tempurl}


\bibitem[\protect\citeauthoryear{Harris and Seaborne}{Harris and
  Seaborne}{2013}]%
        {sparql}
\bibfield{author}{\bibinfo{person}{Steven Harris} {and} \bibinfo{person}{Andy
  Seaborne}.} \bibinfo{year}{2013}\natexlab{}.
\newblock \bibinfo{booktitle}{\emph{{SPARQL} 1.1 Query Language}}.
\newblock \bibinfo{type}{{W3C} Recommendation}. \bibinfo{institution}{W3C}.
\newblock
\newblock
\shownote{http://www.w3.org/TR/2013/REC-sparql11-query-20130321/.}


\bibitem[\protect\citeauthoryear{Kesper, Matoni, Rössel, Weidling, and
  Wenz}{Kesper et~al\mbox{.}}{2020}]%
        {comprehensive_problem_specifications}
\bibfield{author}{\bibinfo{person}{Arno Kesper}, \bibinfo{person}{Markus
  Matoni}, \bibinfo{person}{Julia Rössel}, \bibinfo{person}{Michelle
  Weidling}, {and} \bibinfo{person}{Viola Wenz}.}
  \bibinfo{year}{2020}\natexlab{}.
\newblock \bibinfo{title}{Catalog of Quality Problems for Data, Data Models and
  Data Transformations}.
\newblock
\newblock
\urldef\tempurl%
\url{https://doi.org/10.5281/zenodo.3955500}
\showDOI{\tempurl}


\bibitem[\protect\citeauthoryear{Kim, Choi, Hong, Kim, and Lee}{Kim
  et~al\mbox{.}}{2003}]%
        {kim_taxonomy_2003}
\bibfield{author}{\bibinfo{person}{Won~Y. Kim}, \bibinfo{person}{Byoung{-}Ju
  Choi}, \bibinfo{person}{Eui~Kyeong Hong}, \bibinfo{person}{Soo{-}Kyung Kim},
  {and} \bibinfo{person}{Doheon Lee}.} \bibinfo{year}{2003}\natexlab{}.
\newblock \showarticletitle{A Taxonomy of Dirty Data}.
\newblock \bibinfo{journal}{\emph{Data Min. Knowl. Discov.}}
  \bibinfo{volume}{7}, \bibinfo{number}{1} (\bibinfo{year}{2003}),
  \bibinfo{pages}{81--99}.
\newblock
\urldef\tempurl%
\url{https://doi.org/10.1023/A:1021564703268}
\showDOI{\tempurl}


\bibitem[\protect\citeauthoryear{Kontokostas, Westphal, Auer, Hellmann,
  Lehmann, Cornelissen, and Zaveri}{Kontokostas et~al\mbox{.}}{2014}]%
        {kontokostas2014test}
\bibfield{author}{\bibinfo{person}{Dimitris Kontokostas},
  \bibinfo{person}{Patrick Westphal}, \bibinfo{person}{S{\"{o}}ren Auer},
  \bibinfo{person}{Sebastian Hellmann}, \bibinfo{person}{Jens Lehmann},
  \bibinfo{person}{Roland Cornelissen}, {and} \bibinfo{person}{Amrapali
  Zaveri}.} \bibinfo{year}{2014}\natexlab{}.
\newblock \showarticletitle{Test-driven evaluation of linked data quality}. In
  \bibinfo{booktitle}{\emph{23rd International World Wide Web Conference, {WWW}
  '14, Seoul, Republic of Korea, April 7-11, 2014}},
  \bibfield{editor}{\bibinfo{person}{Chin{-}Wan Chung},
  \bibinfo{person}{Andrei~Z. Broder}, \bibinfo{person}{Kyuseok Shim}, {and}
  \bibinfo{person}{Torsten Suel}} (Eds.). \bibinfo{publisher}{{ACM}},
  \bibinfo{pages}{747--758}.
\newblock
\urldef\tempurl%
\url{https://doi.org/10.1145/2566486.2568002}
\showDOI{\tempurl}


\bibitem[\protect\citeauthoryear{Kuczera}{Kuczera}{2016}]%
        {Kuc16}
\bibfield{author}{\bibinfo{person}{Andreas Kuczera}.}
  \bibinfo{year}{2016}\natexlab{}.
\newblock \showarticletitle{Digital Editions beyond {XML} - Graph-based Digital
  Editions}. In \bibinfo{booktitle}{\emph{Proceedings of the 3rd
  HistoInformatics Workshop on Computational History (HistoInformatics 2016)
  co-located with Digital Humanities 2016 conference {(DH} 2016), Krakow,
  Poland, July 11, 2016}} \emph{(\bibinfo{series}{{CEUR} Workshop
  Proceedings})}, \bibfield{editor}{\bibinfo{person}{Marten D{\"{u}}ring},
  \bibinfo{person}{Adam Jatowt}, \bibinfo{person}{Johannes
  Preiser{-}Kappeller}, {and} \bibinfo{person}{Antal van~den Bosch}} (Eds.),
  Vol.~\bibinfo{volume}{1632}. \bibinfo{publisher}{CEUR-WS.org},
  \bibinfo{pages}{37--46}.
\newblock
\urldef\tempurl%
\url{http://ceur-ws.org/Vol-1632/paper\_5.pdf}
\showURL{%
\tempurl}


\bibitem[\protect\citeauthoryear{Lai, Wu, Lin, and Huang}{Lai
  et~al\mbox{.}}{2011}]%
        {fuzzysearch}
\bibfield{author}{\bibinfo{person}{Lien~Fu Lai}, \bibinfo{person}{Chao{-}Chin
  Wu}, \bibinfo{person}{Pei{-}Ying Lin}, {and} \bibinfo{person}{Liang{-}Tsung
  Huang}.} \bibinfo{year}{2011}\natexlab{}.
\newblock \showarticletitle{Developing a fuzzy search engine based on fuzzy
  ontology and semantic search}. In \bibinfo{booktitle}{\emph{{FUZZ-IEEE} 2011,
  {IEEE} International Conference on Fuzzy Systems, Taipei, Taiwan, 27-30 June,
  2011, Proceedings}}. \bibinfo{publisher}{{IEEE}},
  \bibinfo{pages}{2684--2689}.
\newblock
\urldef\tempurl%
\url{https://doi.org/10.1109/FUZZY.2011.6007378}
\showDOI{\tempurl}


\bibitem[\protect\citeauthoryear{Laranjeiro, Soydemir, and
  Bernardino}{Laranjeiro et~al\mbox{.}}{2015}]%
        {laranjeiro_survey_2015}
\bibfield{author}{\bibinfo{person}{Nuno Laranjeiro}, \bibinfo{person}{Seyma~Nur
  Soydemir}, {and} \bibinfo{person}{Jorge Bernardino}.}
  \bibinfo{year}{2015}\natexlab{}.
\newblock \showarticletitle{A Survey on Data Quality: Classifying Poor Data}.
  In \bibinfo{booktitle}{\emph{21st {IEEE} Pacific Rim International Symposium
  on Dependable Computing, {PRDC} 2015, Zhangjiajie, China, November 18-20,
  2015}}, \bibfield{editor}{\bibinfo{person}{Guojun Wang},
  \bibinfo{person}{Tatsuhiro Tsuchiya}, {and} \bibinfo{person}{Dong Xiang}}
  (Eds.). \bibinfo{publisher}{{IEEE} Computer Society},
  \bibinfo{pages}{179--188}.
\newblock
\urldef\tempurl%
\url{https://doi.org/10.1109/PRDC.2015.41}
\showDOI{\tempurl}


\bibitem[\protect\citeauthoryear{Lassila}{Lassila}{1999}]%
        {rdf}
\bibfield{author}{\bibinfo{person}{Ora Lassila}.}
  \bibinfo{year}{1999}\natexlab{}.
\newblock \bibinfo{booktitle}{\emph{Resource Description Framework ({RDF})
  Model and Syntax Specification}}.
\newblock \bibinfo{type}{{W3C} Recommendation}. \bibinfo{institution}{W3C}.
\newblock
\newblock
\shownote{http://www.w3.org/TR/1999/REC-rdf-syntax-19990222/.}


\bibitem[\protect\citeauthoryear{Mendes, M{\"{u}}hleisen, and Bizer}{Mendes
  et~al\mbox{.}}{2012}]%
        {mendes_sieve:_2012}
\bibfield{author}{\bibinfo{person}{Pablo~N. Mendes}, \bibinfo{person}{Hannes
  M{\"{u}}hleisen}, {and} \bibinfo{person}{Christian Bizer}.}
  \bibinfo{year}{2012}\natexlab{}.
\newblock \showarticletitle{Sieve: linked data quality assessment and fusion}.
  In \bibinfo{booktitle}{\emph{Proceedings of the 2012 Joint {EDBT/ICDT}
  Workshops, Berlin, Germany, March 30, 2012}},
  \bibfield{editor}{\bibinfo{person}{Divesh Srivastava} {and}
  \bibinfo{person}{Ismail Ari}} (Eds.). \bibinfo{publisher}{{ACM}},
  \bibinfo{pages}{116--123}.
\newblock
\urldef\tempurl%
\url{https://doi.org/10.1145/2320765.2320803}
\showDOI{\tempurl}


\bibitem[\protect\citeauthoryear{Nassar}{Nassar}{2020}]%
        {Nas20}
\bibfield{author}{\bibinfo{person}{Nebras Nassar}.}
  \bibinfo{year}{2020}\natexlab{}.
\newblock \emph{\bibinfo{title}{Consistency-by-Construction Techniques for
  Software Models and Model Transformations}}.
\newblock \bibinfo{thesistype}{Ph.D. Dissertation}.
  \bibinfo{school}{Philipps-Universit{\"a}t Marburg, Germany}.
\newblock


\bibitem[\protect\citeauthoryear{Oditis, Bicevskis, and Bicevska}{Oditis
  et~al\mbox{.}}{2017}]%
        {oditis_domain-specific_2017}
\bibfield{author}{\bibinfo{person}{Ivo Oditis}, \bibinfo{person}{Janis
  Bicevskis}, {and} \bibinfo{person}{Zane Bicevska}.}
  \bibinfo{year}{2017}\natexlab{}.
\newblock \showarticletitle{Domain-Specific Characteristics of Data Quality}.
  In \bibinfo{booktitle}{\emph{Proceedings of the 2017 Federated Conference on
  Computer Science and Information Systems, FedCSIS 2017, Prague, Czech
  Republic, September 3-6, 2017}} \emph{(\bibinfo{series}{Annals of Computer
  Science and Information Systems})}, \bibfield{editor}{\bibinfo{person}{Maria
  Ganzha}, \bibinfo{person}{Leszek~A. Maciaszek}, {and} \bibinfo{person}{Marcin
  Paprzycki}} (Eds.), Vol.~\bibinfo{volume}{11}. \bibinfo{pages}{999--1003}.
\newblock
\urldef\tempurl%
\url{https://doi.org/10.15439/2017F279}
\showDOI{\tempurl}


\bibitem[\protect\citeauthoryear{Oliveira, Rodrigues, Henriques, and
  Galhardas}{Oliveira et~al\mbox{.}}{2005b}]%
        {oliveira_taxonomy_2005}
\bibfield{author}{\bibinfo{person}{Paulo Oliveira}, \bibinfo{person}{Fátima
  Rodrigues}, \bibinfo{person}{Pedro Henriques}, {and} \bibinfo{person}{Helena
  Galhardas}.} \bibinfo{year}{2005}\natexlab{b}.
\newblock \showarticletitle{A taxonomy of data quality problems}. In
  \bibinfo{booktitle}{\emph{2nd {Int}. {Workshop} on {Data} and {Information}
  {Quality}}}. \bibinfo{pages}{219--233}.
\newblock


\bibitem[\protect\citeauthoryear{Oliveira, Rodrigues, and Henriques}{Oliveira
  et~al\mbox{.}}{2005a}]%
        {oliveira_formal_2005}
\bibfield{author}{\bibinfo{person}{Paulo Oliveira},
  \bibinfo{person}{F{\'{a}}tima Rodrigues}, {and} \bibinfo{person}{Pedro~Rangel
  Henriques}.} \bibinfo{year}{2005}\natexlab{a}.
\newblock \showarticletitle{A Formal Definition of Data Quality Problems}. In
  \bibinfo{booktitle}{\emph{Proceedings of the 2005 International Conference on
  Information Quality {(MIT} {ICIQ} Conference), Sponsored by Lockheed Martin,
  MIT, Cambridge, MA, USA, November 10-12, 2006}},
  \bibfield{editor}{\bibinfo{person}{Felix Naumann}, \bibinfo{person}{Michael
  Gertz}, {and} \bibinfo{person}{Stuart~E. Madnick}} (Eds.).
  \bibinfo{publisher}{{MIT}}.
\newblock
\urldef\tempurl%
\url{http://mitiq.mit.edu/iciq/iqdownload.aspx?ICIQYear=2005\&File=AFormalDefinitionofDQProblems.pdf}
\showURL{%
\tempurl}


\bibitem[\protect\citeauthoryear{Rahm and Do}{Rahm and Do}{2000}]%
        {rahm_data_2000}
\bibfield{author}{\bibinfo{person}{Erhard Rahm} {and} \bibinfo{person}{Hong~Hai
  Do}.} \bibinfo{year}{2000}\natexlab{}.
\newblock \showarticletitle{Data Cleaning: Problems and Current Approaches}.
\newblock \bibinfo{journal}{\emph{{IEEE} Data Eng. Bull.}}
  \bibinfo{volume}{23}, \bibinfo{number}{4} (\bibinfo{year}{2000}),
  \bibinfo{pages}{3--13}.
\newblock
\urldef\tempurl%
\url{http://sites.computer.org/debull/A00DEC-CD.pdf}
\showURL{%
\tempurl}


\bibitem[\protect\citeauthoryear{Rennau}{Rennau}{line}]%
        {rennau}
\bibfield{author}{\bibinfo{person}{Hans-Juergen Rennau}.}
  \bibinfo{year}{[Online]}\natexlab{}.
\newblock \bibinfo{title}{Combining graph and tree: writing {SHAX}, obtaining
  {SHACL}, {XSD} and more}.
\newblock
\newblock
\urldef\tempurl%
\url{https://www.parsqube.de/publikationen/combining-graph-and-tree-writing-shax-obtaining-shacl-xsd-and-more/}
\showURL{%
\tempurl}


\bibitem[\protect\citeauthoryear{Sperberg-McQueen, Paoli, Bray, Yergeau, and
  Maler}{Sperberg-McQueen et~al\mbox{.}}{2008}]%
        {xml}
\bibfield{author}{\bibinfo{person}{Michael Sperberg-McQueen},
  \bibinfo{person}{Jean Paoli}, \bibinfo{person}{Tim Bray},
  \bibinfo{person}{Fran\c{c}ois Yergeau}, {and} \bibinfo{person}{Eve Maler}.}
  \bibinfo{year}{2008}\natexlab{}.
\newblock \bibinfo{booktitle}{\emph{Extensible Markup Language ({XML}) 1.0
  (Fifth Edition)}}.
\newblock \bibinfo{type}{{W3C} Recommendation}. \bibinfo{institution}{W3C}.
\newblock
\newblock
\shownote{http://www.w3.org/TR/2008/REC-xml-20081126/.}


\bibitem[\protect\citeauthoryear{Spiegel, Robie, and Dyck}{Spiegel
  et~al\mbox{.}}{2017}]%
        {xpath}
\bibfield{author}{\bibinfo{person}{Josh Spiegel}, \bibinfo{person}{Jonathan
  Robie}, {and} \bibinfo{person}{Michael Dyck}.}
  \bibinfo{year}{2017}\natexlab{}.
\newblock \bibinfo{booktitle}{\emph{{XML} Path Language ({XPath}) {3.1}}}.
\newblock \bibinfo{type}{{W3C} Recommendation}. \bibinfo{institution}{W3C}.
\newblock
\newblock
\shownote{https://www.w3.org/TR/2017/REC-xpath-31-20170321/.}


\bibitem[\protect\citeauthoryear{Sun, Ma, and Wang}{Sun et~al\mbox{.}}{2015}]%
        {sun2015comparative}
\bibfield{author}{\bibinfo{person}{Yufei Sun}, \bibinfo{person}{Liangli Ma},
  {and} \bibinfo{person}{Shuang Wang}.} \bibinfo{year}{2015}\natexlab{}.
\newblock \showarticletitle{A comparative evaluation of string similarity
  metrics for ontology alignment}.
\newblock \bibinfo{journal}{\emph{Journal of Information \&Computational
  Science}} \bibinfo{volume}{12}, \bibinfo{number}{3} (\bibinfo{year}{2015}),
  \bibinfo{pages}{957--964}.
\newblock


\bibitem[\protect\citeauthoryear{{TEI Consortium}}{{TEI Consortium}}{line}]%
        {tei}
\bibfield{author}{\bibinfo{person}{{TEI Consortium}}.}
  \bibinfo{year}{[Online]}\natexlab{}.
\newblock \bibinfo{title}{{TEI} {P5}: {Guidelines} for {Electronic} {Text}
  {Encoding} and {Interchange}}.
\newblock
\newblock
\urldef\tempurl%
\url{http://www.tei-c.org/Guidelines/P5/}
\showURL{%
\tempurl}


\bibitem[\protect\citeauthoryear{Zaveri, Rula, Maurino, Pietrobon, Lehmann, and
  Auer}{Zaveri et~al\mbox{.}}{2016}]%
        {zaveri2016quality}
\bibfield{author}{\bibinfo{person}{Amrapali Zaveri}, \bibinfo{person}{Anisa
  Rula}, \bibinfo{person}{Andrea Maurino}, \bibinfo{person}{Ricardo Pietrobon},
  \bibinfo{person}{Jens Lehmann}, {and} \bibinfo{person}{S{\"{o}}ren Auer}.}
  \bibinfo{year}{2016}\natexlab{}.
\newblock \showarticletitle{Quality assessment for Linked Data: {A} Survey}.
\newblock \bibinfo{journal}{\emph{Semantic Web}} \bibinfo{volume}{7},
  \bibinfo{number}{1} (\bibinfo{year}{2016}), \bibinfo{pages}{63--93}.
\newblock
\urldef\tempurl%
\url{https://doi.org/10.3233/SW-150175}
\showDOI{\tempurl}


\end{thebibliography}

\versionchoice{}{}{
\clearpage
\appendix

\section{Metamodel}\label{appendix:metamodel}

The complete metamodel for representing patterns is shown in Fig.~\ref{fig:metamodeltotal}.
In the following, we will briefly explain its differences compared to the condensed version depicted in Fig.~\ref{fig:metamodel}.


The class \lstinline|TrueElement| is used to indicate the most inner levels of the nested condition.
Hence, it does not contain further conditions.

Instead of considering \lstinline|Element|s contained in different \lstinline|Graph|s as equivalent if their names are equal, we actually express this equivalency via \lstinline|ElementMapping|s contained in \lstinline|Morphism|s.
An \lstinline|ElementMap-ping| is specified by a \lstinline|source| and a \lstinline|target| \lstinline|Element|.
Analogously, the class \lstinline|RelationMapping| allows defining two \lstinline|Relation|s that are contained in different graphs as equivalent.
A \lstinline|Morphism| is specified by a \lstinline|source| and a \lstinline|target| \lstinline|Graph| and is contained in a \lstinline|MorphismContainer|, which is either a \lstinline|CountPattern| or a \lstinline|QuantifiedCondition|.

The interface \lstinline|Adaptable| represents items in generic patterns that can be adapted to a specific database technology by being replaced by an instance of a corresponding subclass.

The class \lstinline|BooleanOperator| represents operators that return a boolean value.
The included list \lstinline|elements| holds the \lstinline|Element|s whom the \lstinline|BooleanOperator| serves as a predicate.
This association simplifies the translation of nested operators into queries.
Only \lstinline|BooleanOp-erator|s that are not an argument of a \lstinline|Comparison| can serve as predicates.
Such a \lstinline|BooleanOperator| is a predicate of those \lstinline|Element|s that directly or indirectly (via nested operators) serve as its arguments or that contain the \lstinline|Properties| that directly or indirectly serve as its arguments. 
The \lstinline|elements| association is updated automatically when the arguments of an operator are modified.

The class \lstinline|Comparison| includes an attribute \lstinline|type| which specifies the types of the values that are compared.
It determines which casting functions are applied to the values in the data prior to the comparison.

Instead of implementing the generic class \lstinline|OptionParam<T>| and passing \lstinline|ComparisonOperator|, \lstinline|RelationKind| and \lstinline|PropertyKind| as the type argument \lstinline|T|, we implemented \lstinline|ComparisonOptionParam|, \lstinline|Rela-tionOptionParam| and \lstinline|PropertyOptionParam| as we consider these the only cases in which we need to express parameters whose domain is defined via an enumeration of allowed values.

Fig.~\ref{fig:metamodeltotal} shows more subclasses of \lstinline|ParameterValue| than Fig.~\ref{fig:metamodel}. 
They allow expressing boolean values, lists of string values as well as some of the XML Schema data types used in XQuery (i.e. \lstinline|date|, \lstinline|time| and \lstinline|dateTime|).





\begin{figure*}	
	\includegraphics[width=\textwidth]{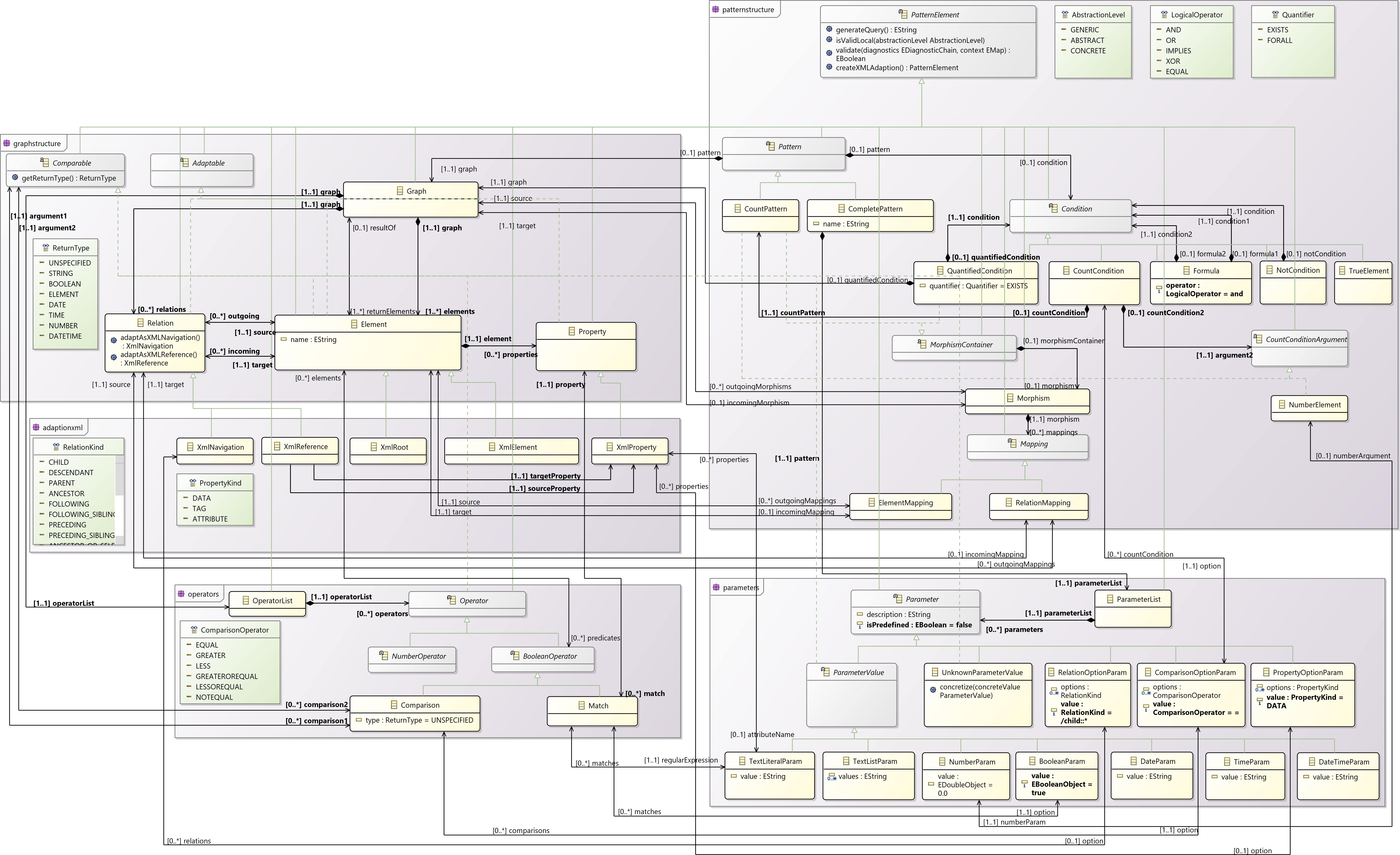}
	\caption{Complete metamodel}
	\label{fig:metamodeltotal}
	\centering
\end{figure*}

\subsection{Constraints}
Besides the \textit{multiplicity constraints} depicted in Fig.~\ref{fig:metamodeltotal}, our metamodel includes further constraints.
They are implemented in the \lstinline|isValidLocal| method of the corresponding \lstinline|PatternElement|. 
Since some constraints may only apply to patterns of a certain level of abstraction (i.e. generic, abstract or concrete), the method has a parameter of type \lstinline|AbstractionLevel|. 
In the following, we will specify the constraints in natural language grouped by package.

\subsubsection{Constraints of the Package \lstinline|patternstructure|}
\begin{itemize}	
	\item Each \lstinline|Morphism| must reference the preceding \lstinline|Graph| in the nested condition via the association \lstinline|source| and the \lstinline|Graph| of its \lstinline|MorphismContainer| via the association \lstinline|target|.
	
	\item The \lstinline|source| \lstinline|Element| of an \lstinline|ElementMapping| must be contained in the \lstinline|source| \lstinline|Graph| of the \lstinline|Morphism| that contains the \lstinline|Element-Mapping|. 
	The \lstinline|target| \lstinline|Element| must be contained in the \lstinline|target| \lstinline|Graph| of this \lstinline|Morphism|. 
	Analogously, the \lstinline|source| and \lstinline|target| \lstinline|Relation| of a \lstinline|RelationMapping| must be contained in the \lstinline|source| or \lstinline|target| \lstinline|Graph| of the containing \lstinline|Morphism|.
	
	\item Two \lstinline|Mapping|s contained in the same \lstinline|Morphism| must not have the same \lstinline|source|.
\end{itemize}

\subsubsection{Constraints of the Package \lstinline|graphstructure|}
\begin{itemize}	
	\item The list of \lstinline|returnElements| of a \lstinline|Graph| must not be empty.
	
	\item Each \lstinline|Element| contained in the list of \lstinline|returnElements| of a \lstinline|Graph| must be contained in this \lstinline|Graph|.
	
	\item The \lstinline|returnElements| of all \lstinline|Graph|s directly contained in a \lstinline|Pattern| or directly contained in any of its \lstinline|QuantifiedCondi-tion|s must be equivalent (specified via \lstinline|ElementMapping|s).
	
	\item For generic patterns: a graph must not contain items of the package \lstinline|adaptionxml|.
	
	\item For XML-adapted abstract and concrete patterns only: Each \lstinline|Graph| must contain exactly one \lstinline|XmlRoot|.
	
	\item For XML-adapted abstract and concrete patterns: a graph must not contain instances of \lstinline|Element|, \lstinline|Relation| or \lstinline|Property|.
	
	\item The \lstinline|Element|s and \lstinline|Relation|s contained in the \lstinline|graph| of a \lstinline|Com-pletePattern| must not be the target of a \lstinline|Mapping|.
	
	\item For each \lstinline|RelationMapping| there must be two \lstinline|ElementMapping|s which indicate that the \lstinline|source| \lstinline|Element|s of both \lstinline|Relation|s are equivalent and that the \lstinline|target| \lstinline|Element|s of both \lstinline|Relation|s are equivalent.
\end{itemize}

\subsubsection{Constraints of the Package \lstinline|operators|}
\begin{itemize}
	\item For each \lstinline|BooleanOperator| that does not serve as an argument of a comparison the set of \lstinline|Element|s that directly or indirectly serve as its arguments must be equal to the set \lstinline|elements|.
	
	\item For each \lstinline|BooleanOperator| that serves as an argument of a comparison its association \lstinline|elements| must be empty.
	
	\item The return types of both arguments of a \lstinline|Comparison| must be equal to the \lstinline|type| of the \lstinline|Comparison|. 
	For a concrete pattern this type must not be \lstinline|UNSPECIFIED|.
	
	\item A \lstinline|Comparison| that has two \lstinline|Element|s as arguments may be specified by \lstinline|EQUAL| or \lstinline|NOTEQUAL| only.
	
	\item Each \lstinline|ComparisonOperator| must be free of cycles: 
	it must not have another \lstinline|ComparisonOperator| as an argument that it also directly or indirectly serves as an argument.
	
	\item Each \lstinline|Operator| referenced within a \lstinline|Graph| must be contained in this graph's \lstinline|operatorList|. 
	Each \lstinline|Operator| contained in a graph's \lstinline|operatorList| must be referenced from within this \lstinline|Graph|.
\end{itemize}

\subsubsection{Constraints of the Package \lstinline|parameters|}
\begin{itemize}
	\item Each \lstinline|Parameter| referenced within a \lstinline|CompletePattern| must be contained in this pattern's \lstinline|parameterList|. 
	Each \lstinline|Parameter| contained in a pattern's \lstinline|parameterList| must be referenced from within this \lstinline|CompletePattern|.
	
	\item For a concrete pattern each \lstinline|ParameterValue| must be specified. 
	Thus, its \lstinline|value| (or \lstinline|values|) attribute must not be null (or empty). 
	For \lstinline|Date|, \lstinline|Time| and \lstinline|DateTime| the \lstinline|value| must satisfy a specific regular expression given in the specification of the XML Schema language ~\cite{xmlschema}.
	
	\item A concrete pattern must not contain an \lstinline|UnknownParameter\-Value|.
	
	\item For a concrete pattern each \lstinline|ComparisonOptionParam|, \lstinline|Relation-OptionParam| and \lstinline|PropertyOptionParam| must be specified. 
	The \lstinline|option| attribute must give at least one choice.
	Further, the attribute \lstinline|value| must not be null and it must be one of the values listed in the \lstinline|option| attribute.
\end{itemize}

\subsubsection{Constraints of the Package \lstinline|adaptionxml|}
\begin{itemize}
	\item Each \lstinline|XmlElement| must have exactly one incoming \lstinline|XmlNaviga-tion|.

	\item An \lstinline|XmlRoot| must not have any incoming \lstinline|Relation|s.
	
	\item An \lstinline|XmlRoot| must not have any outgoing \lstinline|XmlReference|s.

	\item The association \lstinline|predicates| of an \lstinline|XmlRoot| must be empty.

	\item The \lstinline|option| of an \lstinline|XmlNavigtion| must be non-null if and only if it is not the target of a \lstinline|RelationMapping|.

	\item An \lstinline|XmlNavigation| may only be mapped to \lstinline|Relation|s of type \lstinline|XmlNavigation| and an \lstinline|XmlReference| may only be mapped to \lstinline|Relation|s of type \lstinline|XmlReference|.

	\item The \lstinline|sourceProperty| of an \lstinline|XmlReference| must be contained in its \lstinline|source| \lstinline|Element| and the \lstinline|targetProperty| of an \lstinline|XmlReference| must be contained in its \lstinline|target| \lstinline|Element|.
	These are automatically generated during the adaptation.

	\item If an \lstinline|XmlProperty| is specified to be of type \lstinline|ATTRIBUTE|, its \lstinline|attributeName| of type \lstinline|TextLiteral| must not be null.
	In the concrete pattern, its \lstinline|TextLiteral| must contain a non-empty string as \lstinline|value| attribute.
\end{itemize}

\subsection{An Abstract Pattern as Instance Model}\label{subsubsec:instance}


\begin{figure*}
	\includegraphics[width=\textwidth]{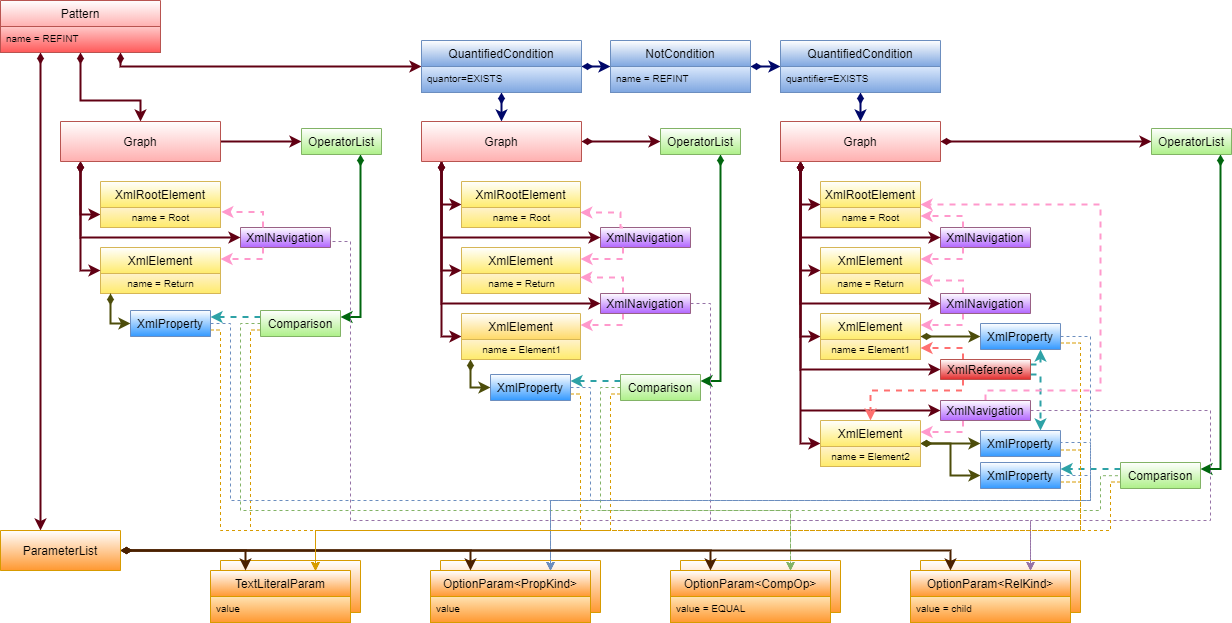}
	\caption{Abstract pattern REFINT  as instance model [solid arrows: containment, dashed arrows: non-containment]}
	\label{fig:refintobjects}
	\centering
\end{figure*}

Fig.~\ref{fig:refintobjects} shows the abstract pattern \texttt{REFINT} (visualised in Fig.~\ref{fig:refint}) as an instance model of our metamodel as depicted in Fig.~\ref{fig:metamodel}.
%
The instance model represents the internal structure of the pattern, which is mostly hidden behind the graphical representation as shown in Fig.~\ref{fig:refint}.
The red objects are the surrounding \lstinline|Pattern| and the \lstinline|Graph|s, which are shown as square white blocks in the graphical representation.
The structure around the blocks represents conditions; 
corresponding objects are represented in light blue.
Each \lstinline|Graph| contains \lstinline|Element|s, which are the nodes of the graph.
When adapted to XML these are represented as \lstinline|XmlElement|s.
The element names in Fig.~\ref{fig:refintobjects} match with those in Fig.~\ref{fig:refint}.
\lstinline|Element|s that are contained in different graphs but have the same name correspond to each other.

The \lstinline|Element|s are connected by \lstinline|Relation|s, which in the XML adaption are represented as \lstinline|XmlNavigation| and \lstinline|XmlReference| contained in \lstinline|Graph|.
It has to be assured, that the \lstinline|Relation|s of elements with the same name occurring in different \lstinline|Graph|s are equivalent.
To guarantee that, only the first \lstinline|XmlNavigation| has an \lstinline|OptionParam| attached.
All corresponding \lstinline|XmlNavigation|s in other graphs (identified via equivalent element names) have implicitly the same specification. 

The \lstinline|Parameter|s are shown in orange at the bottom of the model.
The first group represents 8 \lstinline|TextLiteralParam|s.
3 of them are used in comparisons, while the other 5 are used in an \lstinline|XmlProperty| each;
they are required only in case the option \emph{PropKind = attribute} is chosen.
The other groups consist of 12 \lstinline|OptionParam|s: 3 of them for \lstinline|Relation|s, 5 specify \lstinline|Property|s and the other 4 specify \lstinline|Comparison|s.
Each of them has a variable \lstinline|value|, which represents the chosen value.
In this example, the values of \lstinline|OptionParam<RelKind>| and \lstinline|OptionParam<CompOp>| are predefined.
The values of \lstinline|OptionParam<Prop-Kind>| and \lstinline|TextLiteralParam| have to be specified during the concretisation.
Meanwhile, they are represented as \lstinline|null|.


\clearpage

\section{Generated Queries}\label{appendix:queries}

The following listings show the XQuery expressions generated by our tool for the concrete example pattern \texttt{CARD} presented in Section~\ref{subsec:examplepatterns} and \texttt{REFINT} presented in Section~\ref{appendix:example} of the appendix.

\begin{lstlisting}[label=lst:queryCard,caption={XQuery for the concrete CARD pattern presented in Section~\ref{subsec:examplepatterns}},language=XQuery]
for $var1 in /child::*[./name()="artist"]
where count(
 for $var2 in /child::*[./name()="name"]
 where true()
 return $var2
) > 1.0
return $var1
\end{lstlisting}

\begin{lstlisting}[label=lst:queryRefint,caption={XQuery for the concrete REFINT pattern presented in Section~\ref{appendix:example}},language=XQuery,basicstyle=\ttfamily\footnotesize]
for $var1 in /child::*[./name()="building"]
where some $var2 in $var1/child::*[./name()="creator"]
 satisfies not(
  some $var3 in /child::*[./name()="artist"]
     [$var2/@ref=./@id]
   satisfies true()
 )
return $var25
\end{lstlisting}


\clearpage
\section{Generic and Abstract Patterns}\label{appendix:generic}
In the following, we will present the generic and XML-adapted abstract patterns that were used for the evaluation presented in Section~\ref{sec:evaluation}.
Patterns that have a similar purpose are grouped together.
For each pattern we will briefly explain its purpose, present the diagram of the generic pattern
and finally present the diagram of the XML-adapted abstract version of that pattern.
Concretisations of these patterns will be presented in Section~\ref{appendix:problems}.

The patterns are depicted via diagrams that are structured as those presented in Section~\ref{subsec:examplepatterns}.
Note that there is the possibility to use the same parameter multiple times.
For example, if we want to determine that two elements of a pattern represent the same XML elements
we can use the same properties, values and XPath axes for both pattern elements.
In the presented patterns, the concrete operator of a comparison is often predefined in the generic pattern as ``equal''.
In this case the comparison serves for identifying certain elements.
Some patterns contain further comparisons that are not predefined.
They allow expressing conditions to the values of the selected elements.


\subsection{MATCH}
The MATCH patterns (Figures \ref{fig:match_gen}-\ref{fig:match2_abs}) check a value in the data (PropertyA) against a given regular expression (ValueA).
\begin{figure}
	\centering
	\includegraphics[width=\linewidth]{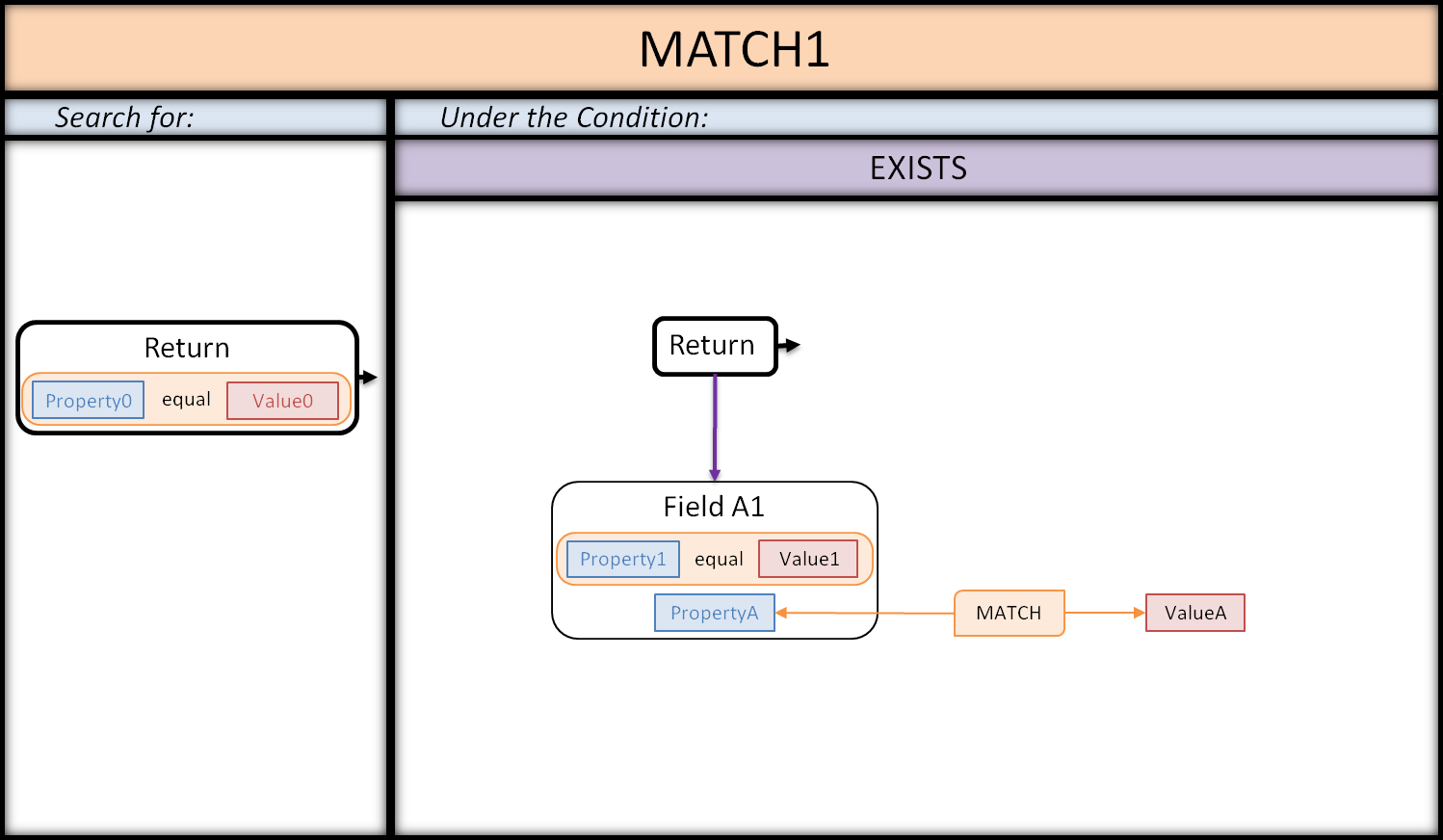}
	\caption{Generic pattern MATCH1}
	\label{fig:match_gen}
\end{figure}
\begin{figure}
	\centering
	\includegraphics[width=\linewidth]{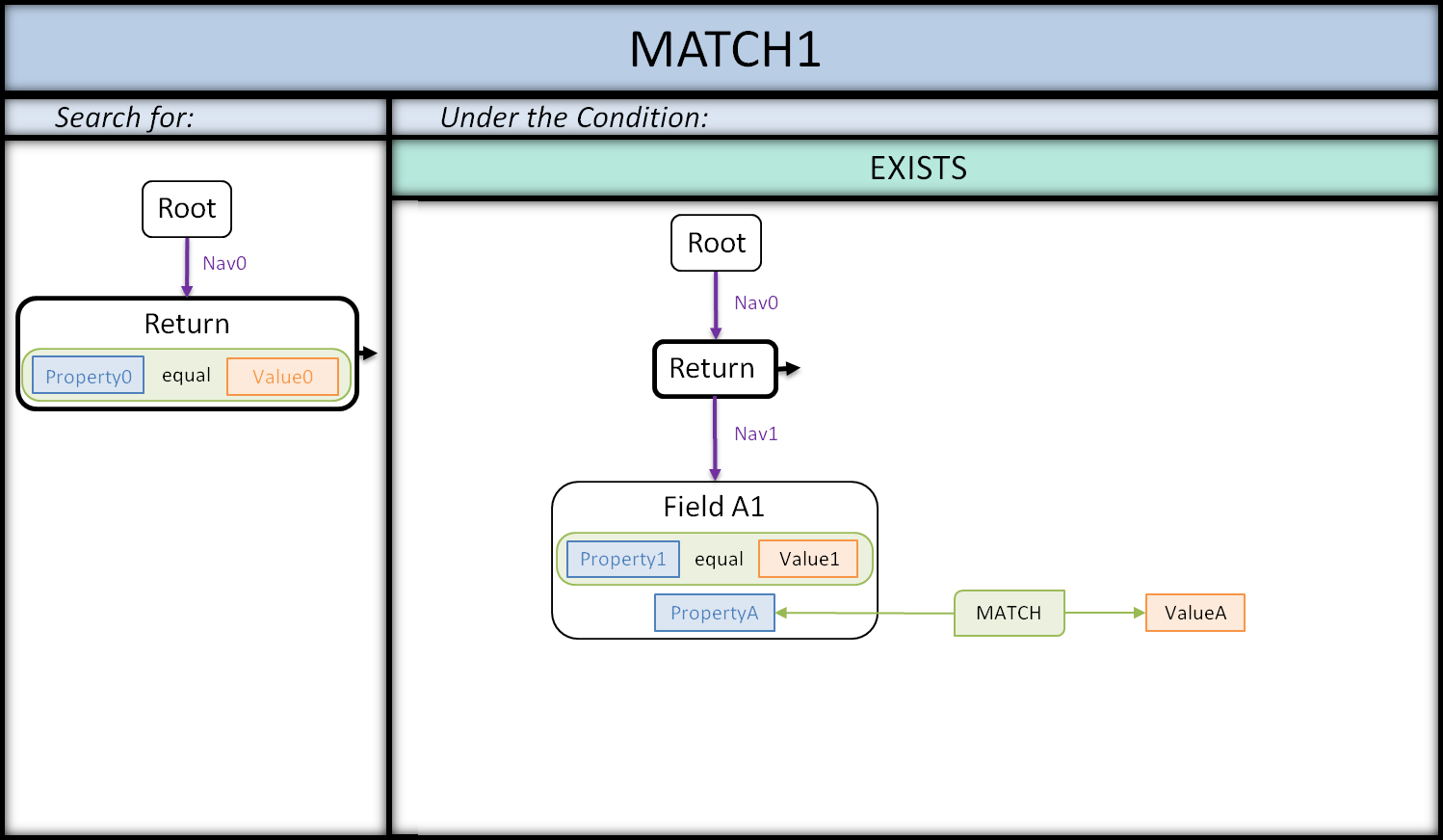}
	\caption{Abstract pattern MATCH1}
	\label{fig:match_abs}
\end{figure}
\begin{figure}
	\centering
	\includegraphics[width=\linewidth]{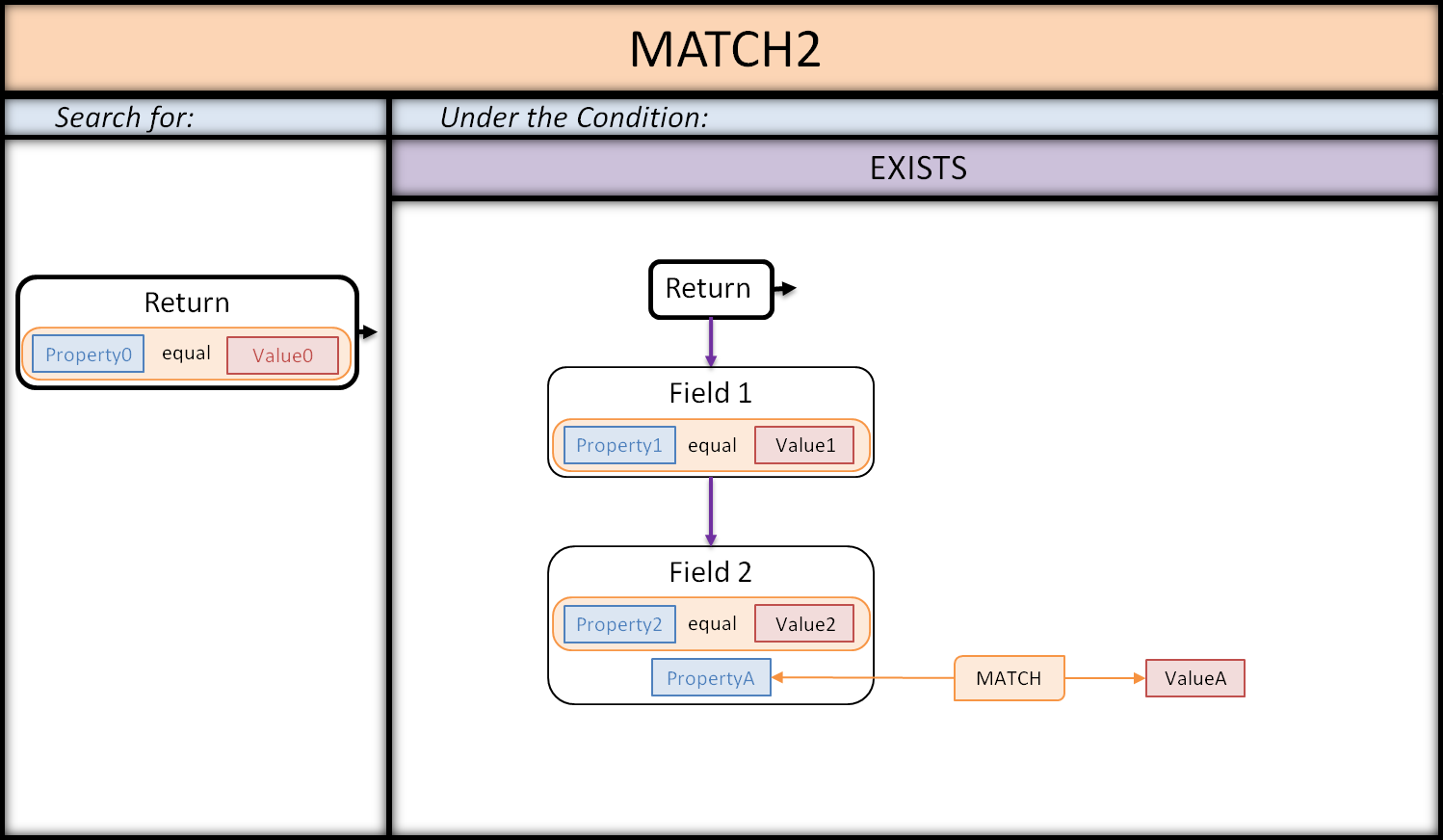}
	\caption{Generic pattern MATCH2}
	\label{fig:match2_gen}
\end{figure}
\begin{figure}
	\centering
	\includegraphics[width=\linewidth]{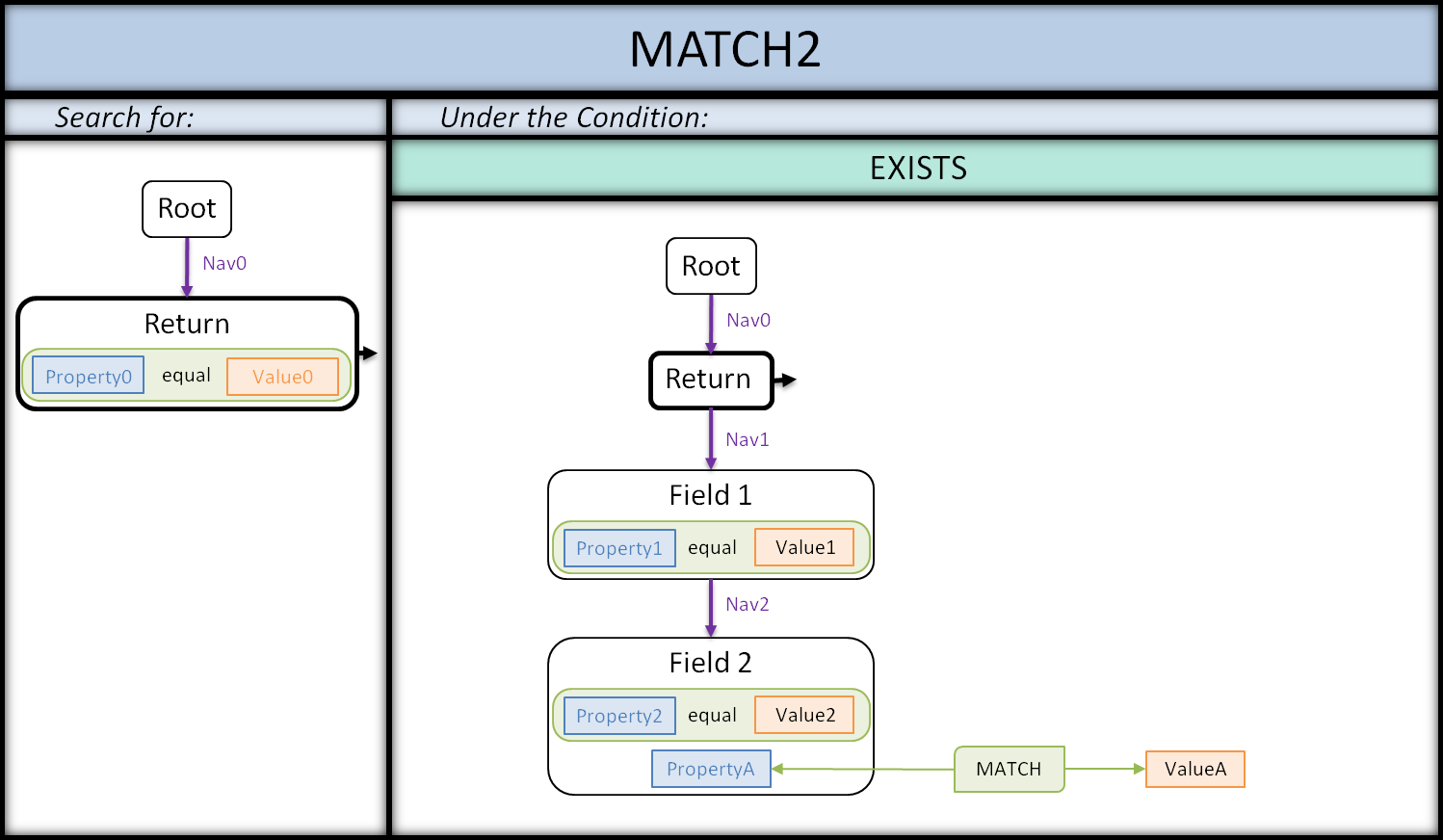}
	\caption{Abstract pattern MATCH2}
	\label{fig:match2_abs}
\end{figure}

\subsection{COMP}
The COMP pattern (Figures \ref{fig:comp_gen}, \ref{fig:comp_abs}) compares two values that are indirectly related to one record.
\begin{figure}
	\centering
	\includegraphics[width=\linewidth]{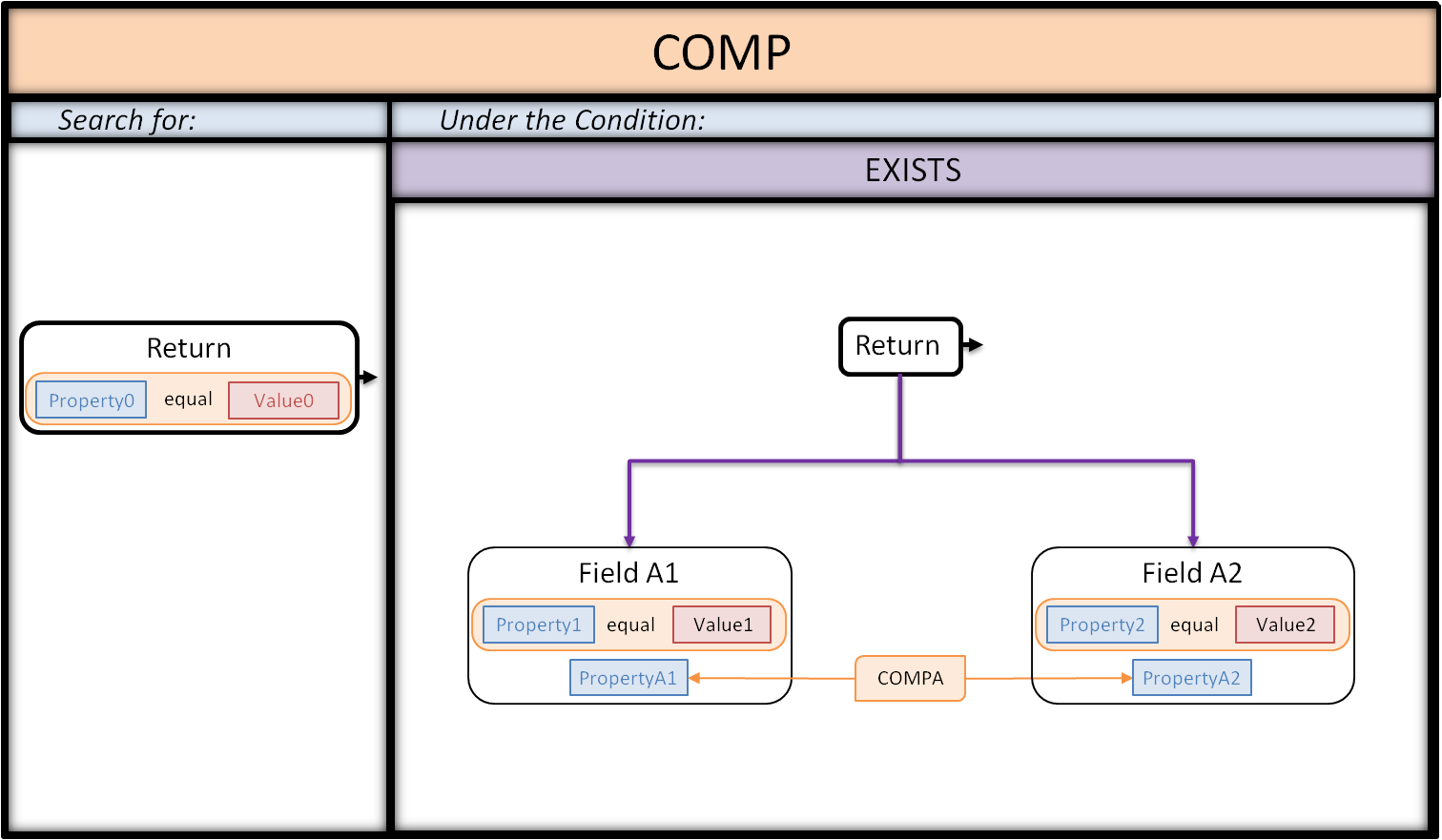}
	\caption{Generic pattern COMP}
	\label{fig:comp_gen}
\end{figure}
\begin{figure}
	\centering
	\includegraphics[width=\linewidth]{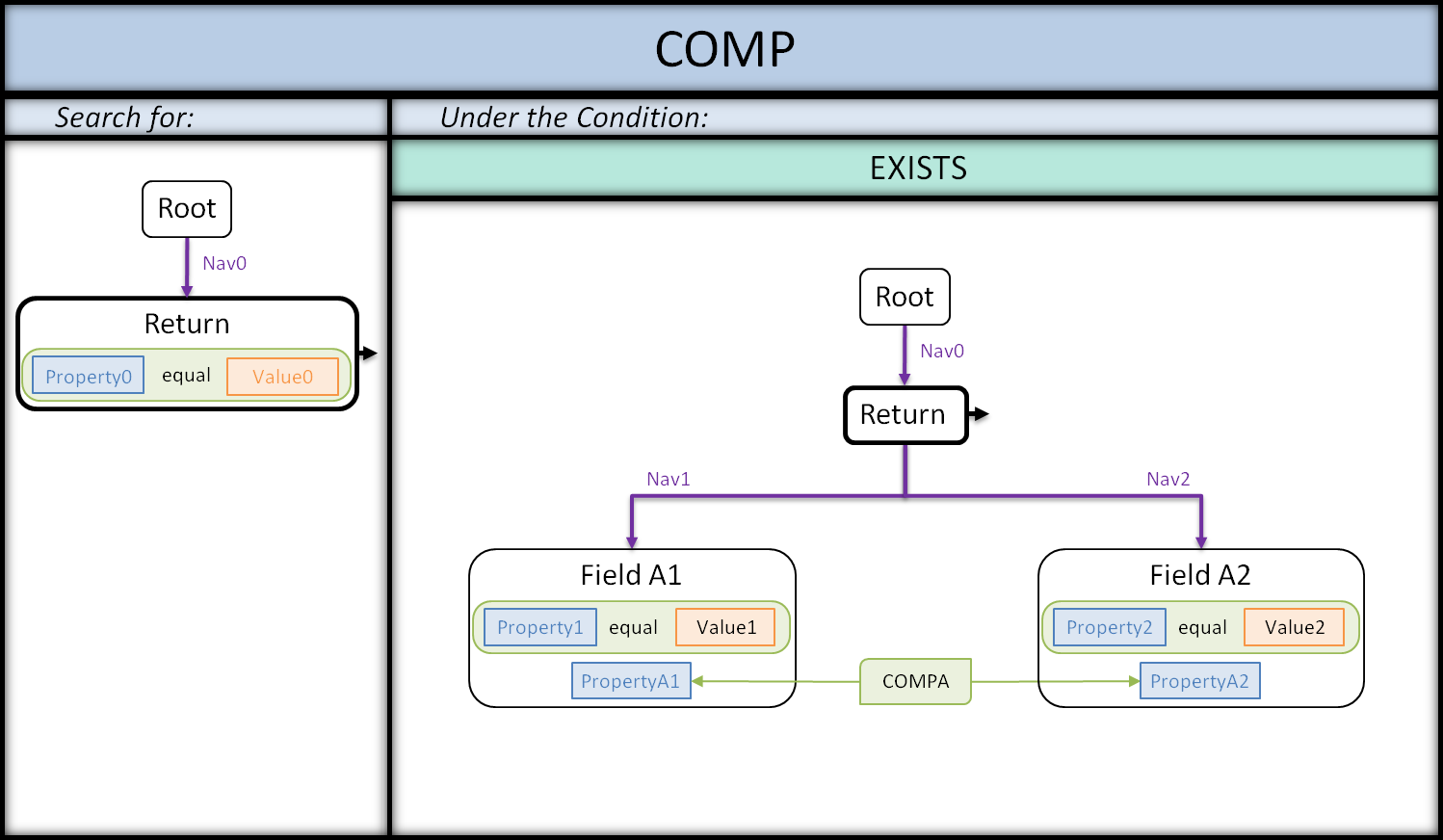}
	\caption{Abstract pattern COMP}
	\label{fig:comp_abs}
\end{figure}

\subsection{COMPVAL}
The COMPVAL patterns (Figures \ref{fig:compval_gen}-\ref{fig:compval2_abs}) compare a value in the data with a literal value via a comparison operator (COMP2 and COMP3, respectively).
\begin{figure}
	\centering
	\includegraphics[width=\linewidth]{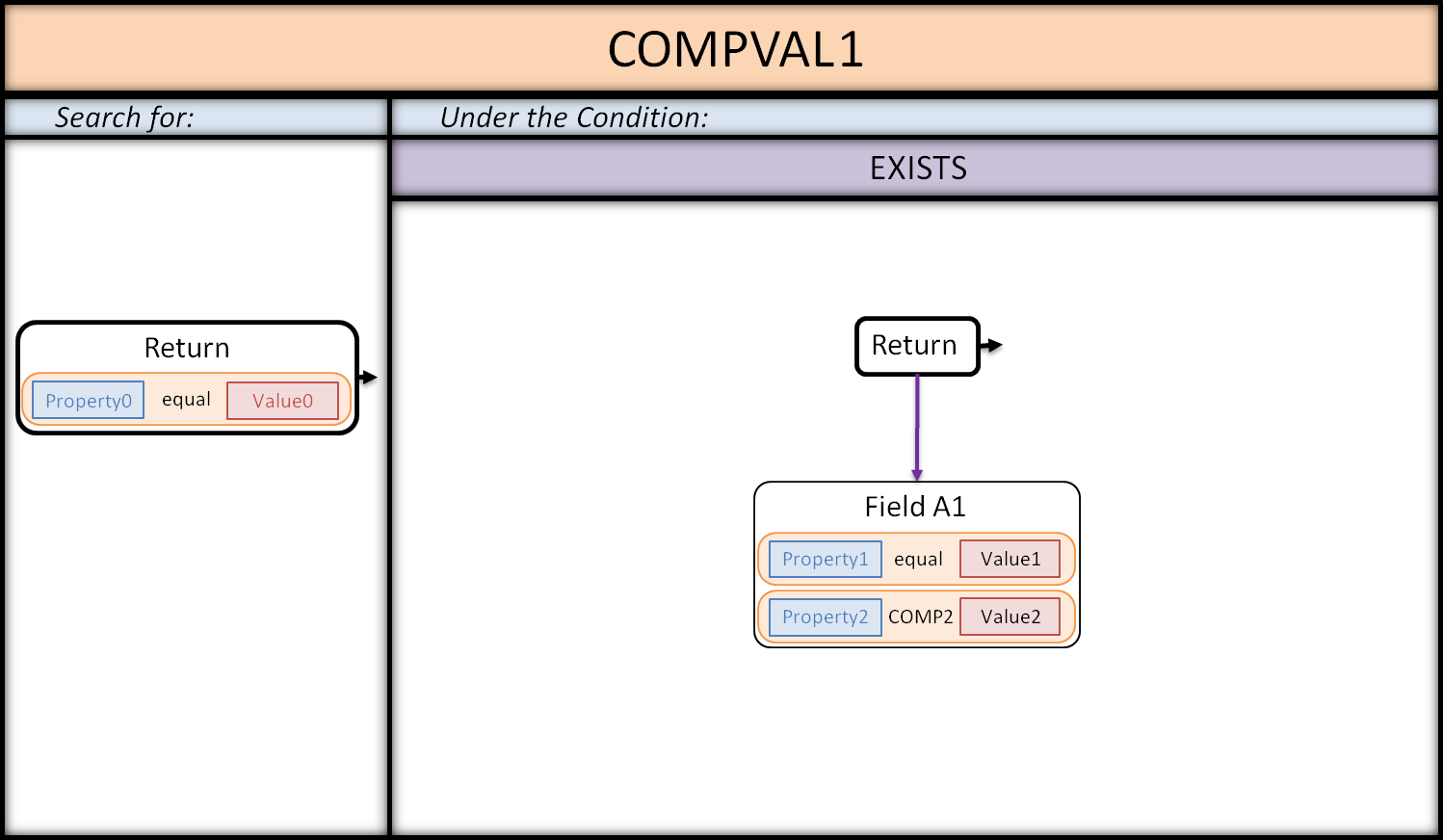}
	\caption{Generic pattern COMPVAL1}
	\label{fig:compval_gen}
\end{figure}
\begin{figure}
	\centering
	\includegraphics[width=\linewidth]{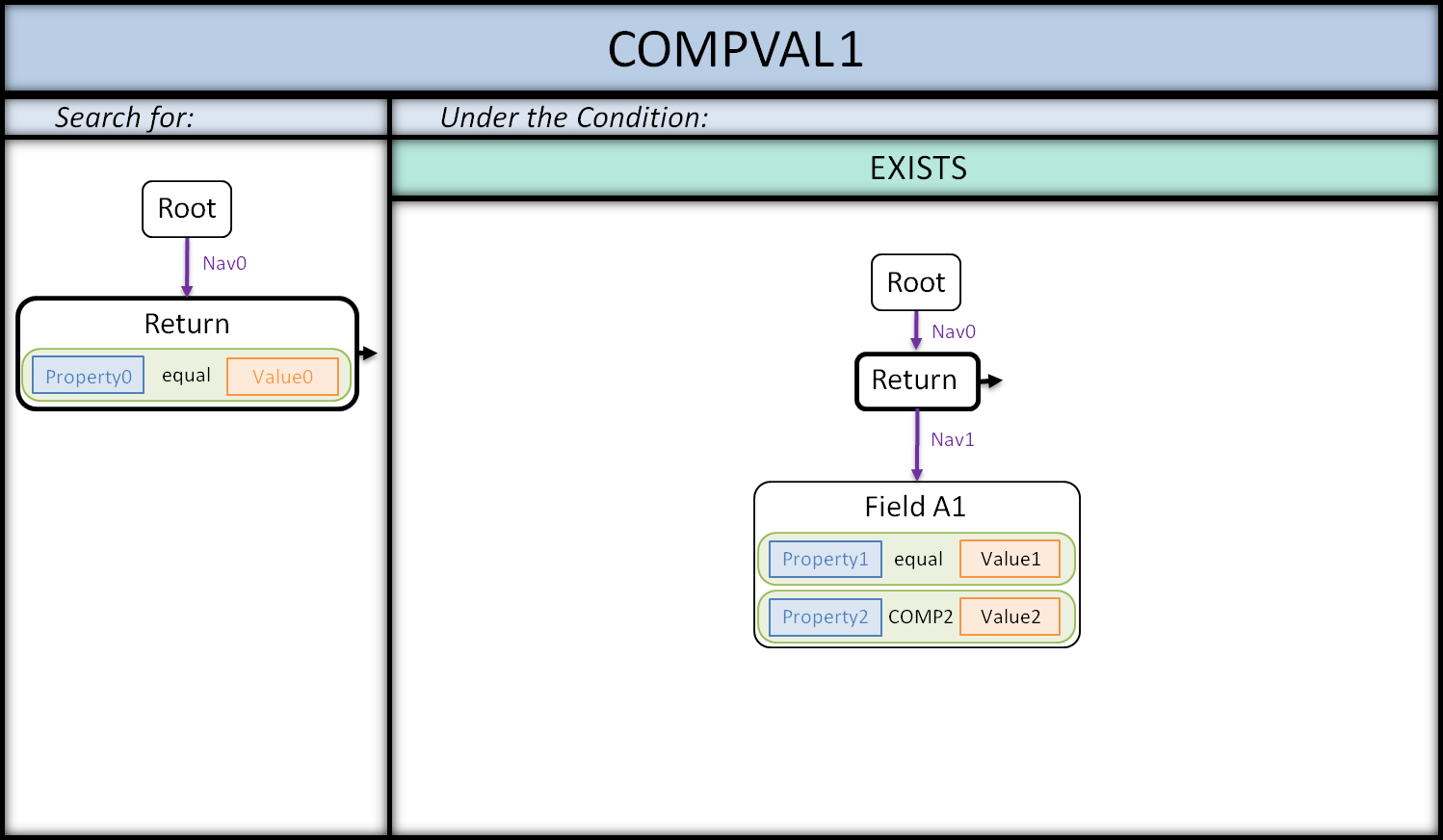}
	\caption{Abstract pattern COMPVAL1}
	\label{fig:compval_abs}
\end{figure}
\begin{figure}
	\centering
	\includegraphics[width=\linewidth]{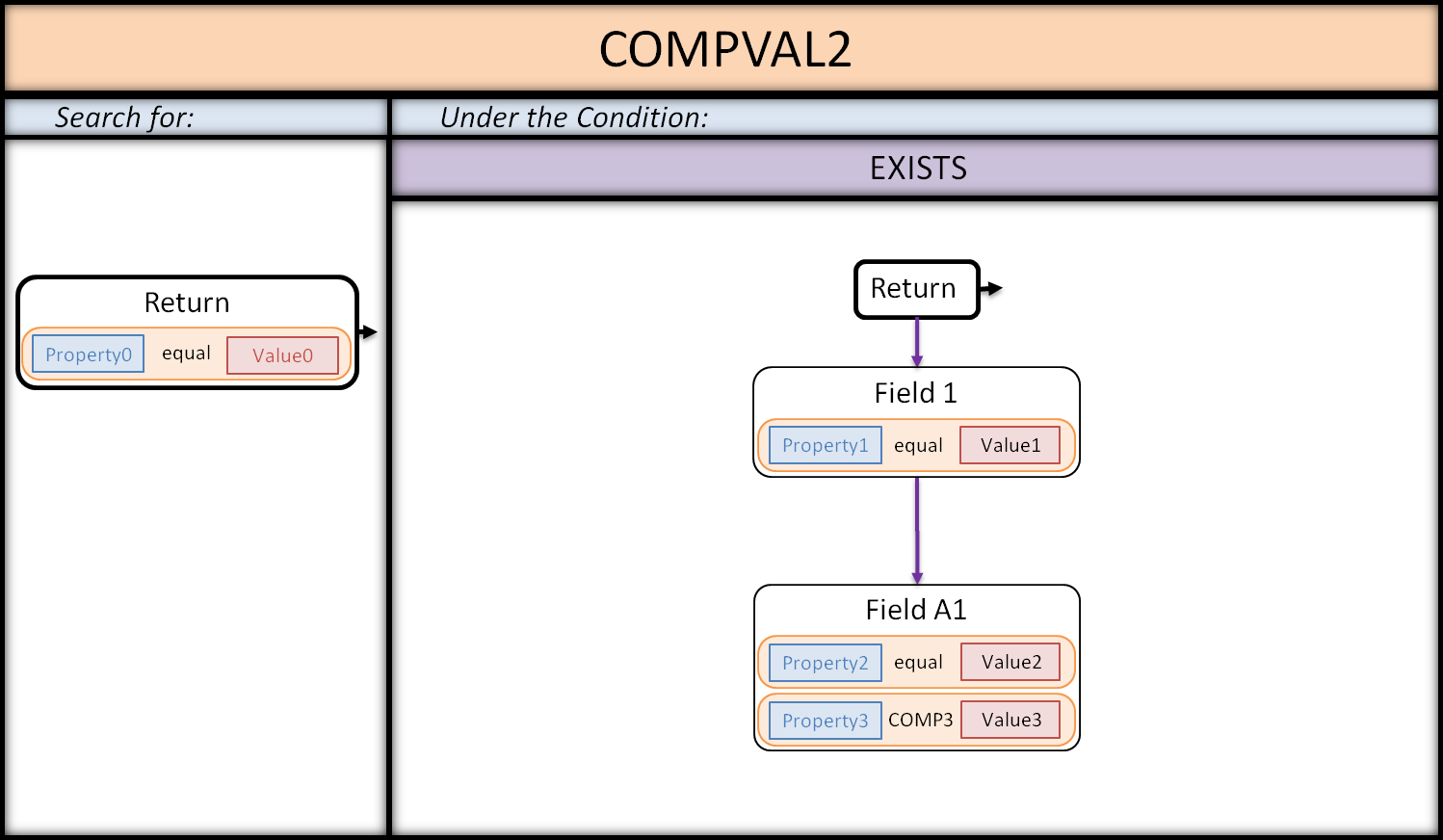}
	\caption{Generic pattern COMPVAL2}
	\label{fig:compval2_gen}
\end{figure}
\begin{figure}
	\centering
	\includegraphics[width=\linewidth]{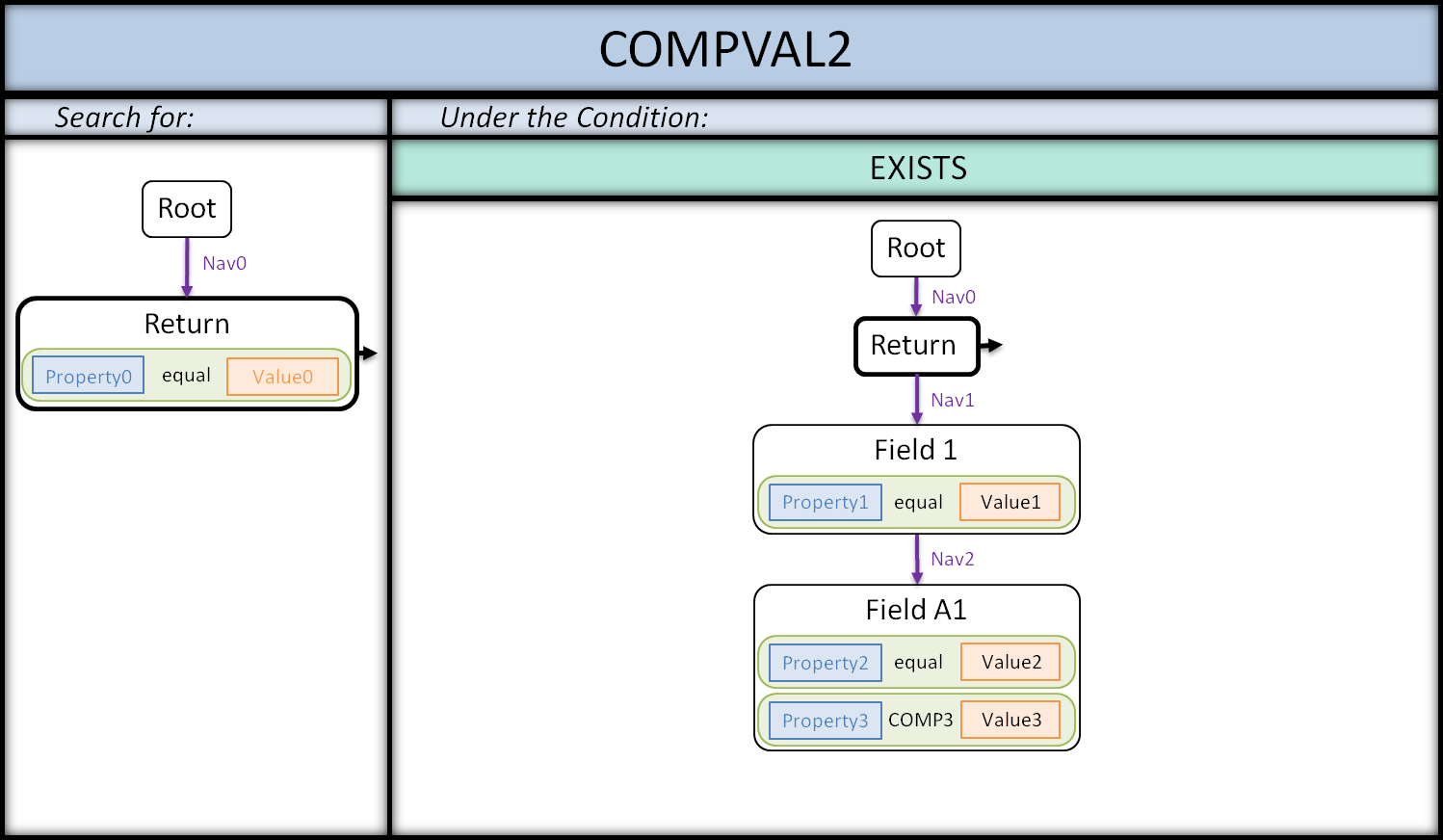}
	\caption{Abstract pattern COMPVAL2}
	\label{fig:compval2_abs}
\end{figure}

\subsection{REFINT}
The REFINT pattern (Figures \ref{fig:refint_gen}, \ref{fig:refint_abs}) detects referential integrity violations.
A reviewer pointed out that this pattern detects also cases in which the element called Field 1 does not have the PropertyA, which is actually not a referential integrity violation.
We will address this issue in near future.
The issue, however, does not affect our evaluation as in the chosen database the element Field 1 exists only if it also includes a reference to an element Field 2, thus has PropertyA. 
\begin{figure}
	\centering
	\includegraphics[width=\linewidth]{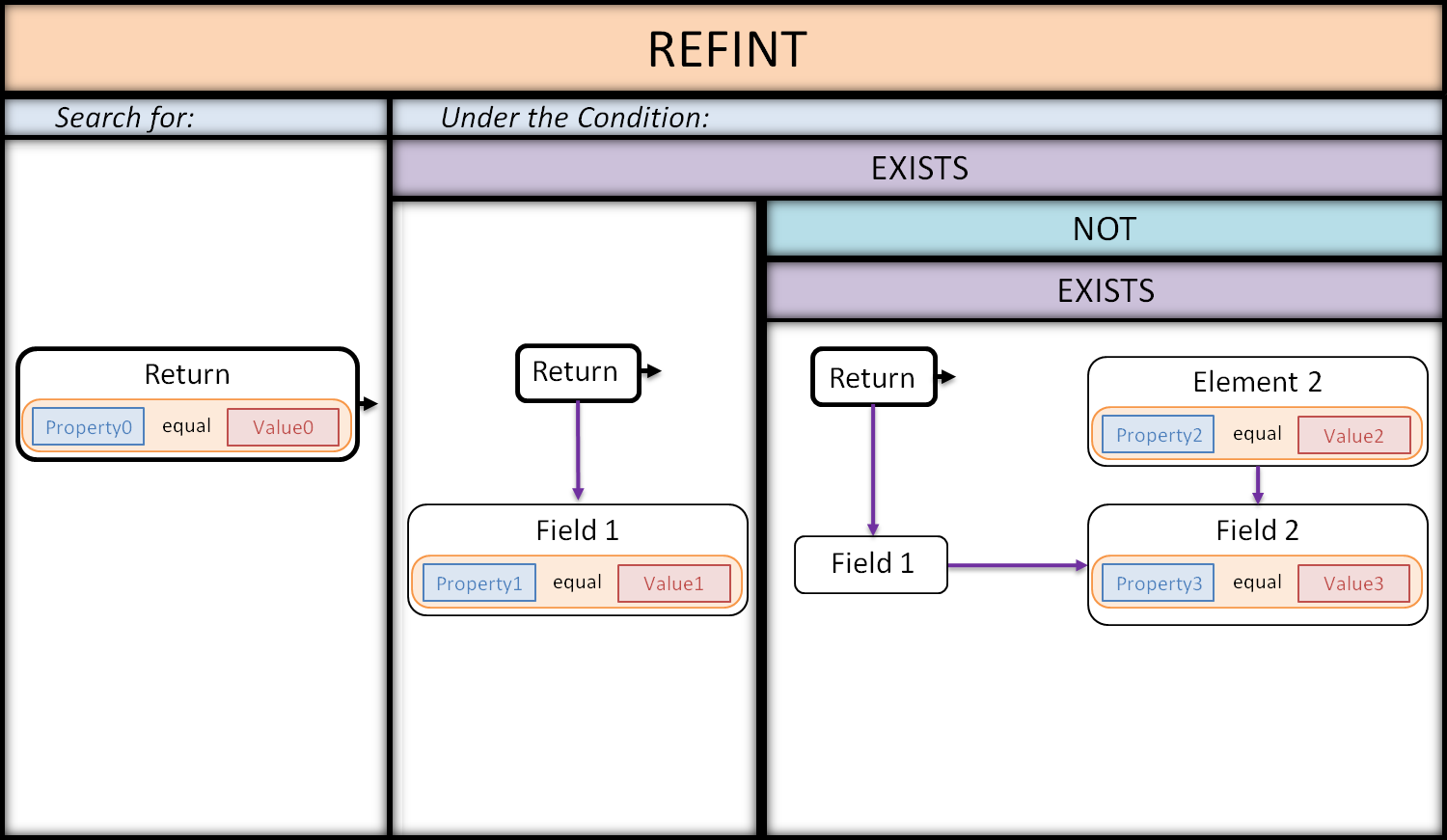}
	\caption{Generic pattern REFINT}
	\label{fig:refint_gen}
\end{figure}
\begin{figure}
	\centering
	\includegraphics[width=\linewidth]{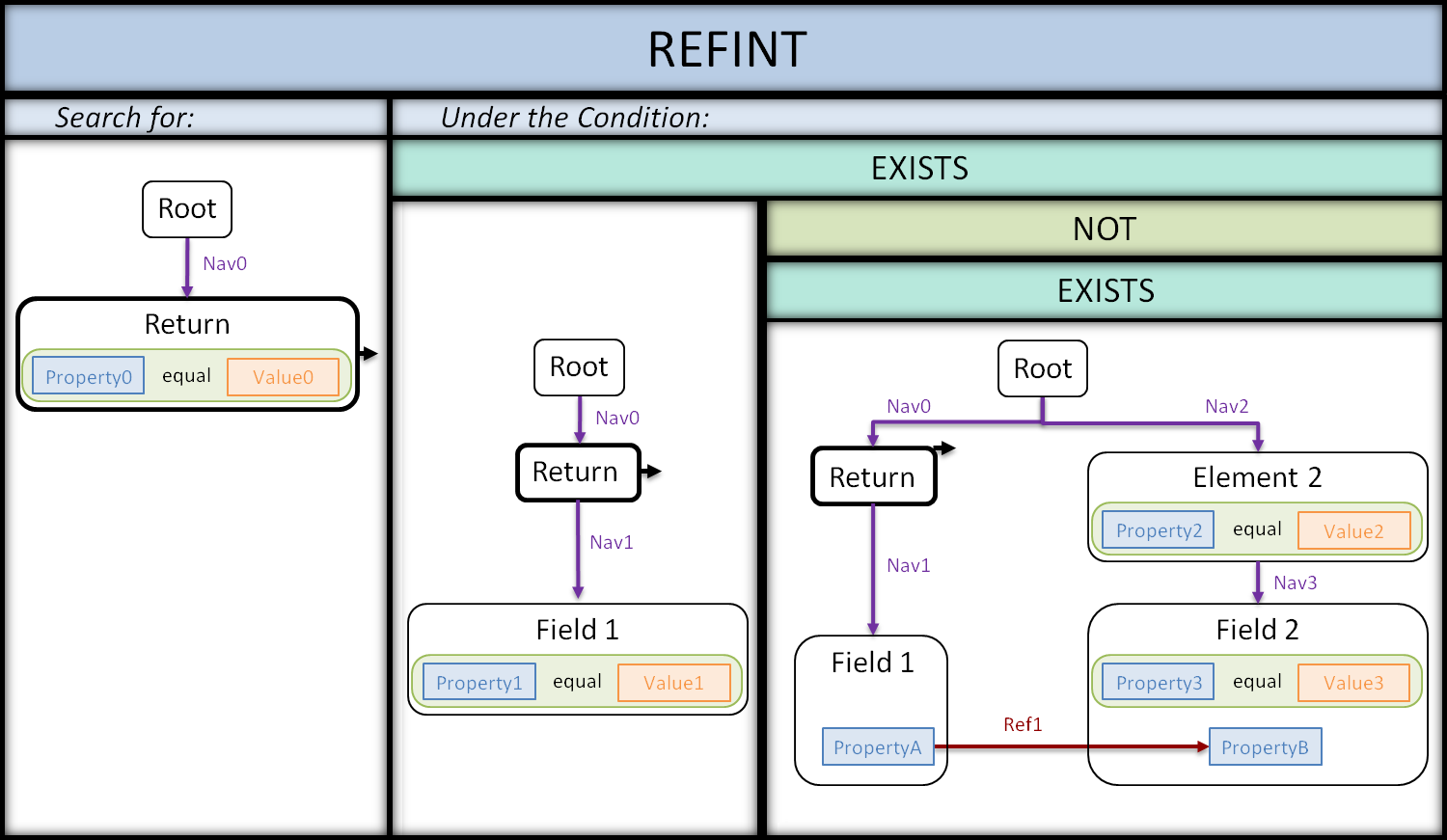}
	\caption{Abstract pattern REFINT}
	\label{fig:refint_abs}
\end{figure}

\subsection{CARD}
The CARD patterns (Figures \ref{fig:card_gen}-\ref{fig:card2_abs}) check whether a certain structure occurs more than once in the data.
The first CARD pattern is explained in detail in Section~\ref{subsec:examplepatterns}.
\begin{figure}
	\centering
	\includegraphics[width=\linewidth]{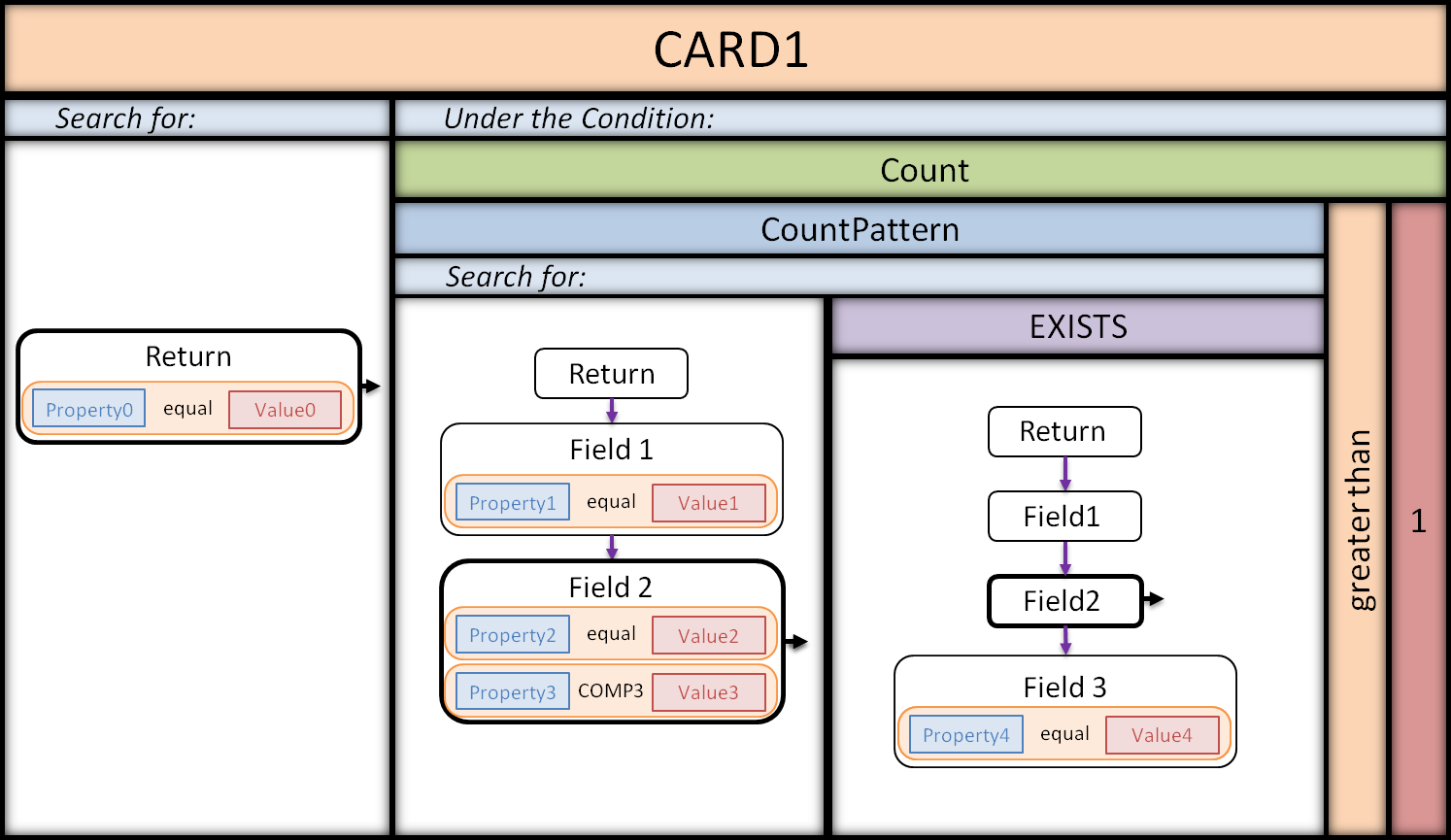}
	\caption{Generic pattern CARD1}
	\label{fig:card_gen}
\end{figure}
\begin{figure}
	\centering
	\includegraphics[width=\linewidth]{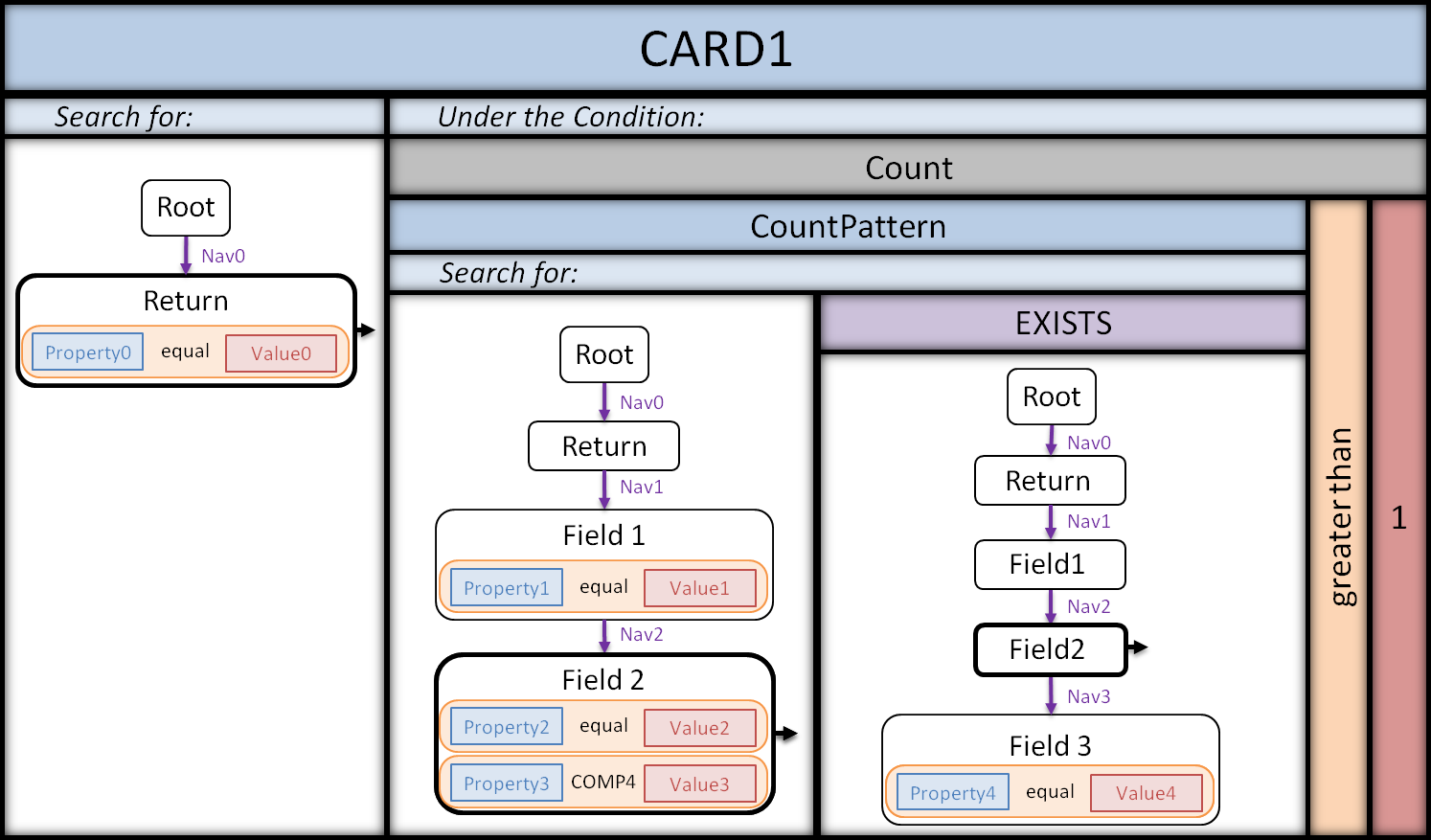}
	\caption{Abstract pattern CARD1}
	\label{fig:card_abs}
\end{figure}
\begin{figure}
	\centering
	\includegraphics[width=\linewidth]{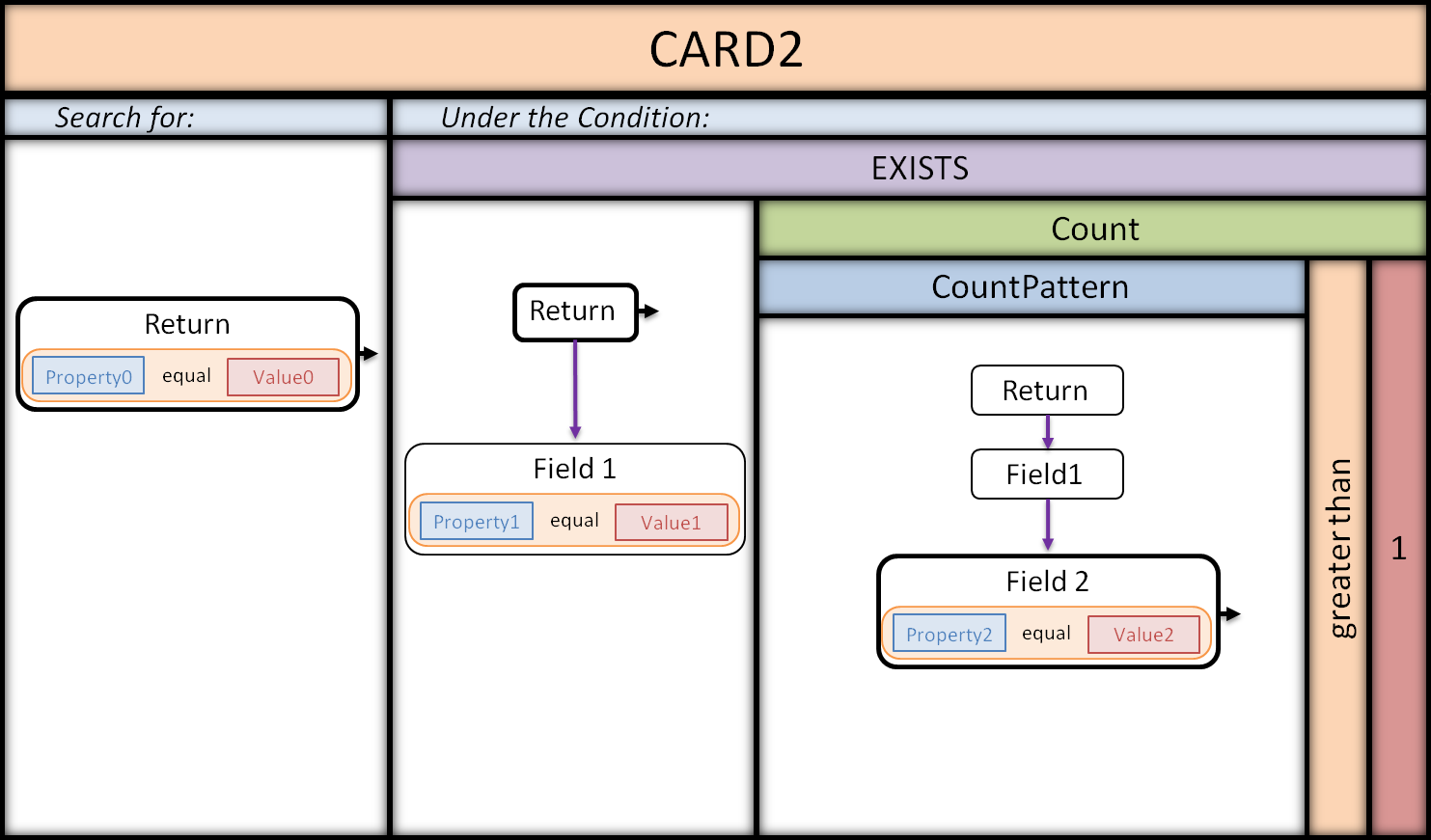}
	\caption{Generic pattern CARD2}
	\label{fig:card2_gen}
\end{figure}
\begin{figure}
	\centering
	\includegraphics[width=\linewidth]{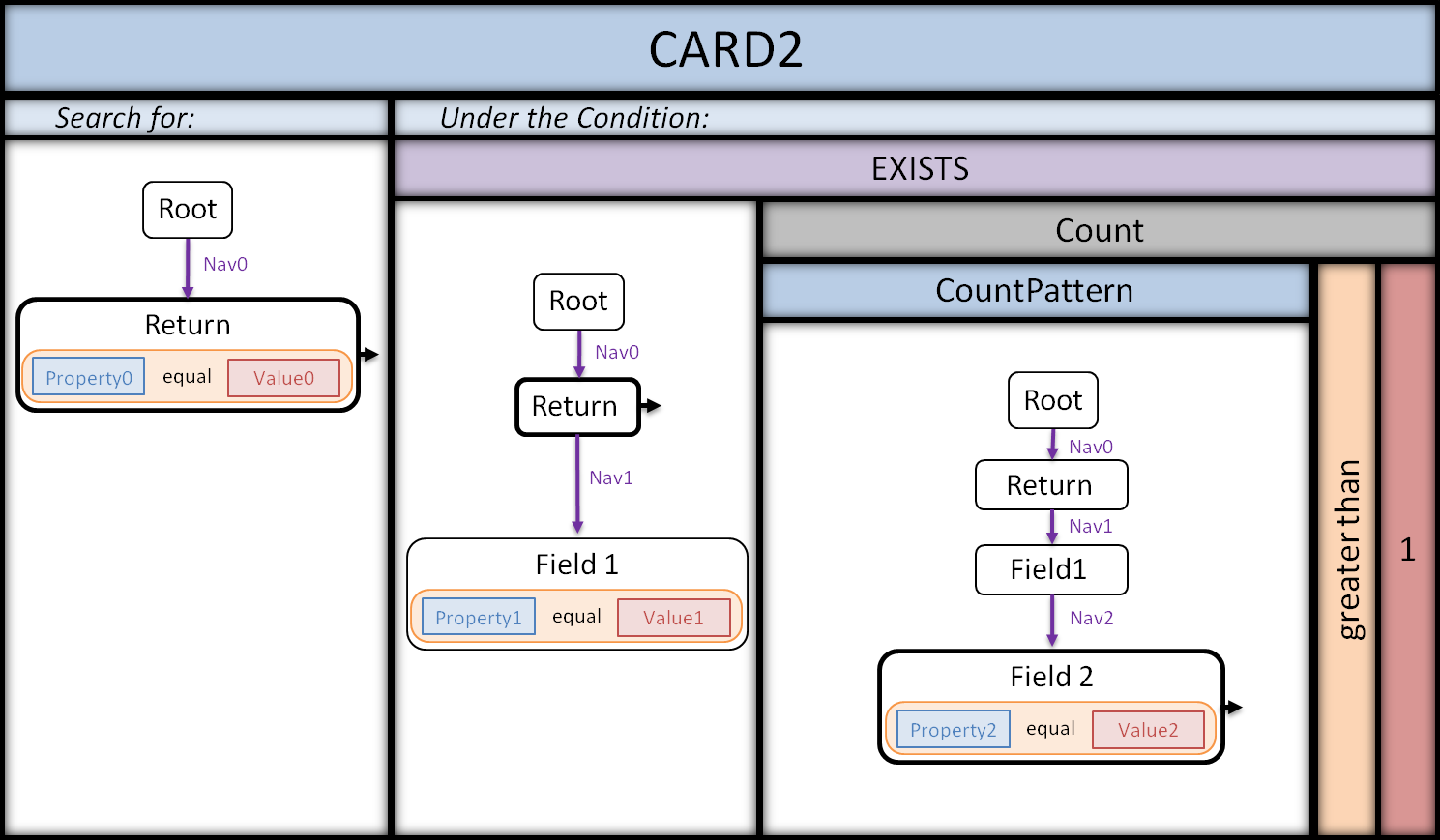}
	\caption{Abstract pattern CARD2}
	\label{fig:card2_abs}
\end{figure}

\subsection{FUNC}
The FUNC patterns (Figures \ref{fig:func_gen}, \ref{fig:func2_abs}) detect violations of functional dependencies.
The patterns check whether the values of a field in two data records do not match (Field B1 and B2), even though another (generally more specific) field (Field A1 and A2) has the same value in both records.
The first FUNC pattern is explained in detail in Section~\ref{subsec:examplepatterns}.
\begin{figure}
	\centering
	\includegraphics[width=\linewidth]{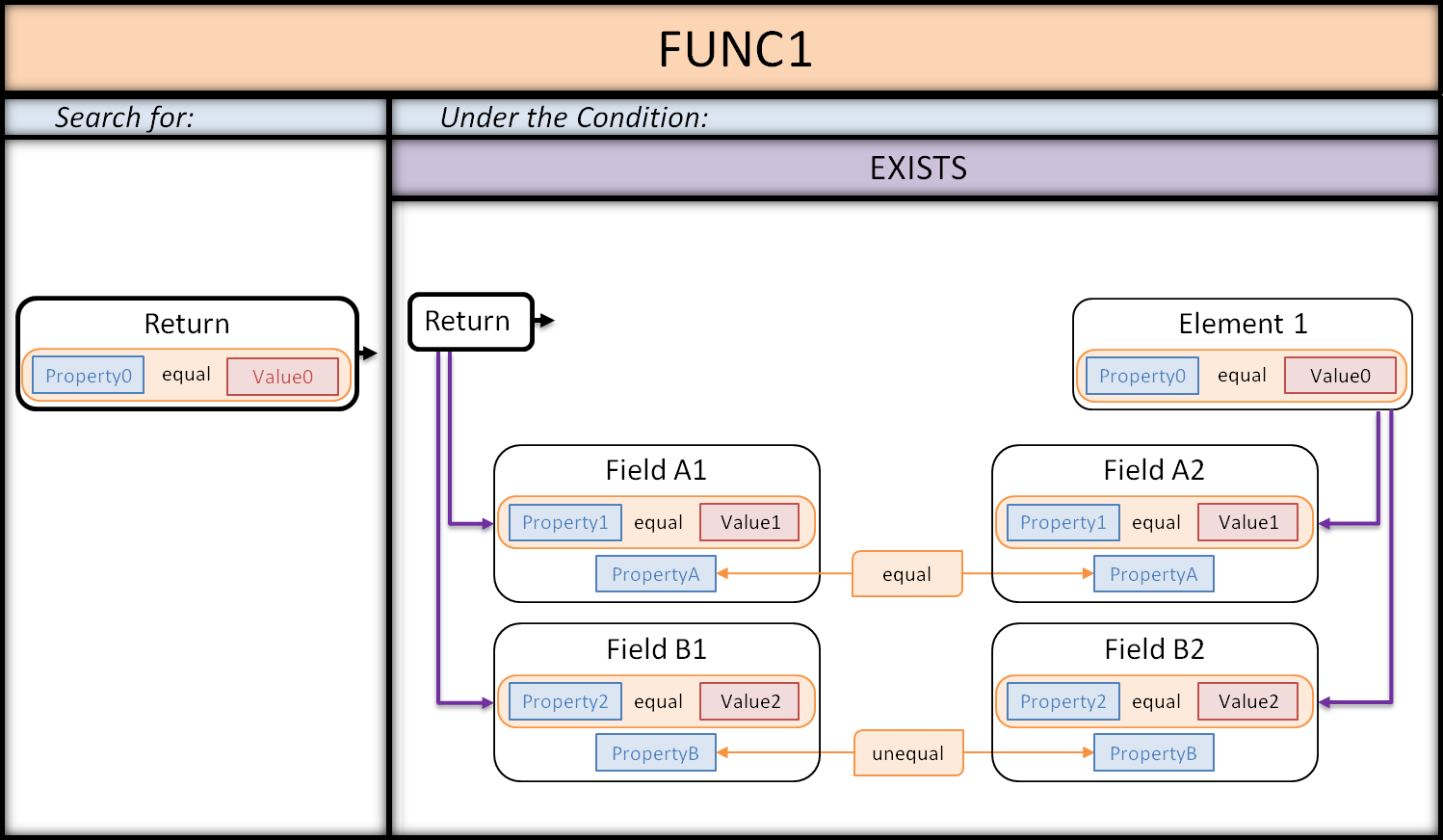}
	\caption{Generic pattern FUNC1}
	\label{fig:func_gen}
\end{figure}
\begin{figure}
	\centering
	\includegraphics[width=\linewidth]{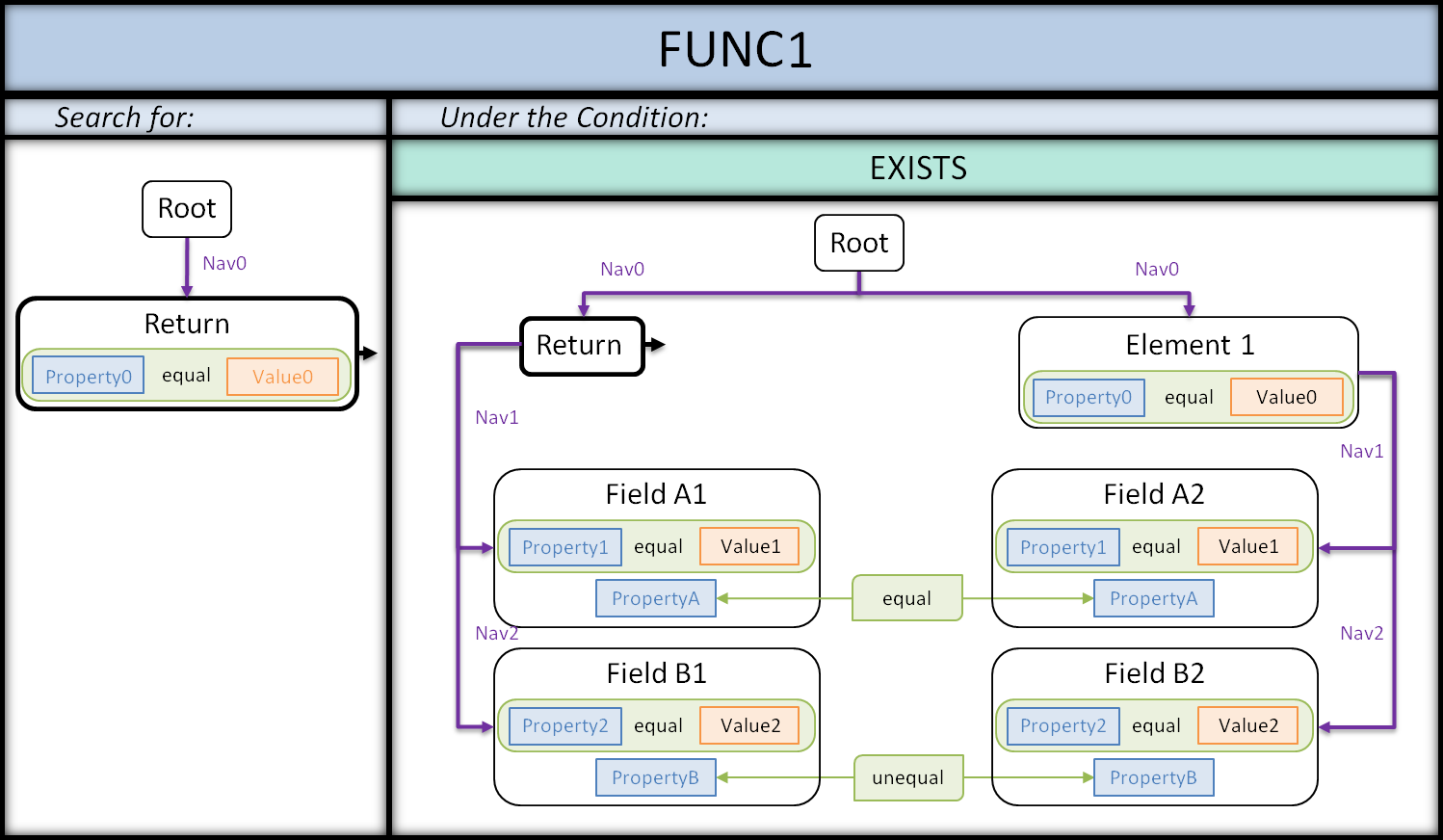}
	\caption{Abstract pattern FUNC1}
	\label{fig:func_abs}
\end{figure}
\begin{figure}
	\centering
	\includegraphics[width=\linewidth]{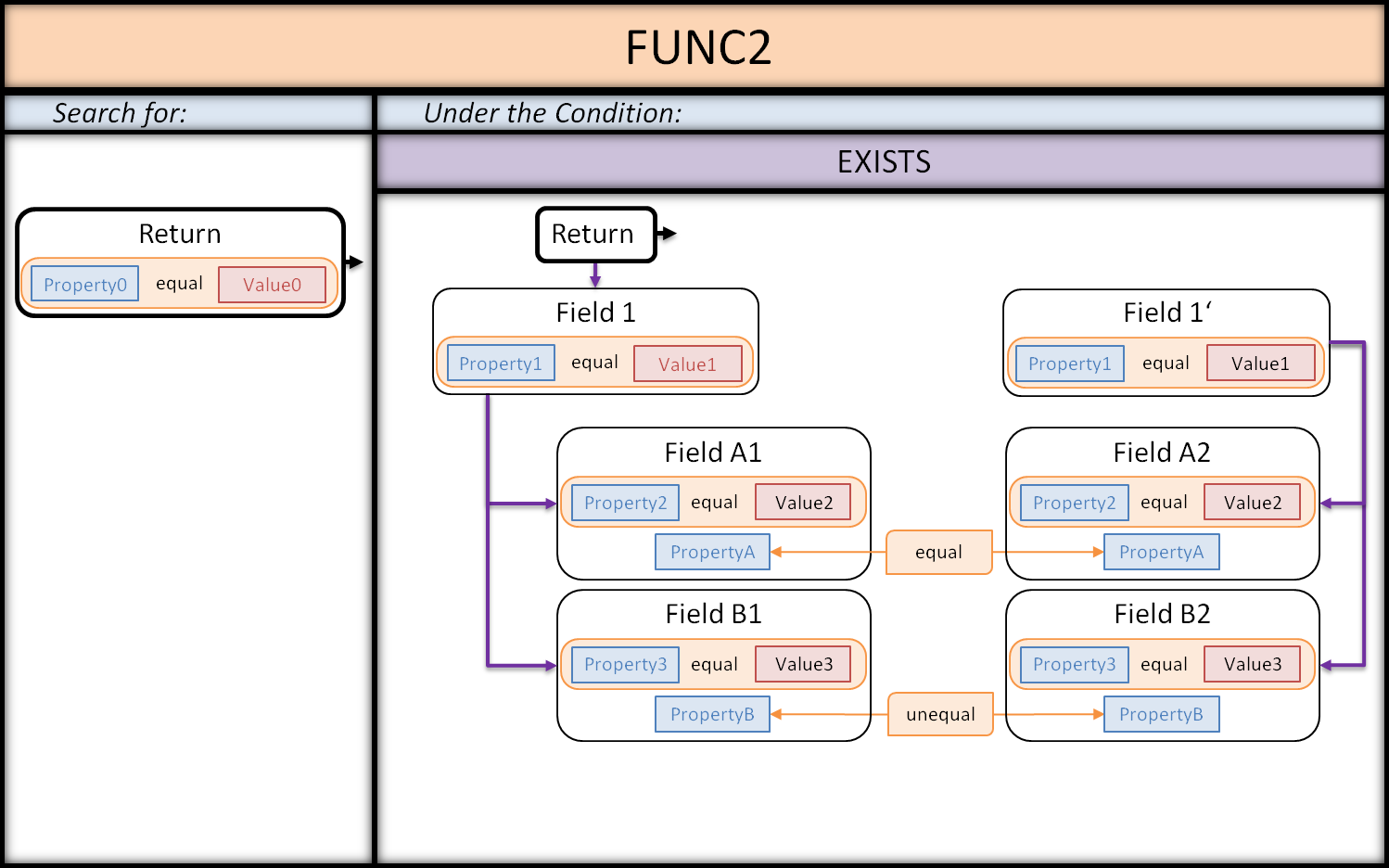}
	\caption{Generic pattern FUNC2}
	\label{fig:func2_gen}
\end{figure}
\begin{figure}
	\centering
	\includegraphics[width=\linewidth]{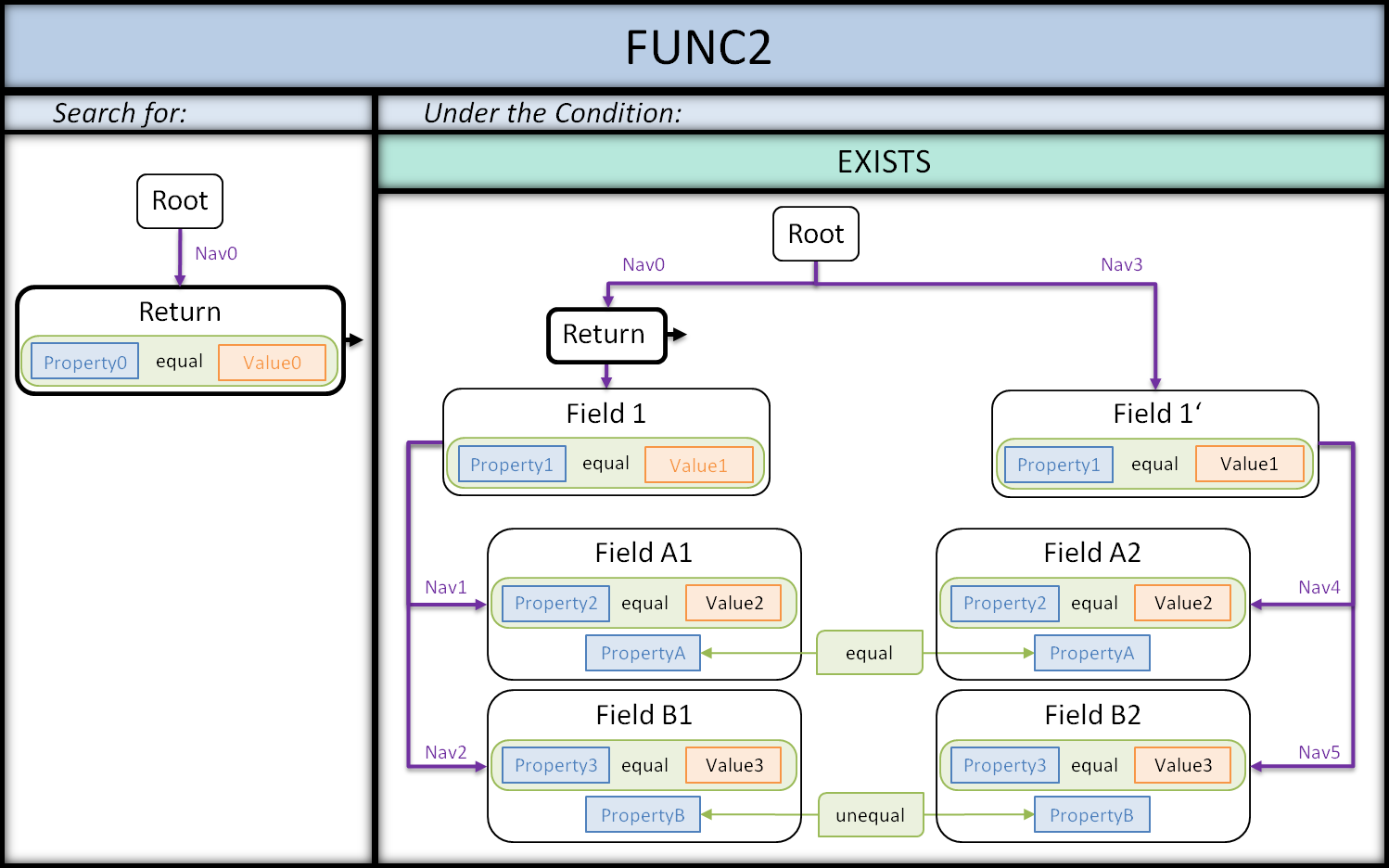}
	\caption{Abstract pattern FUNC2}
	\label{fig:func2_abs}
\end{figure}

\subsection{UNIQUE}
The UNIQUE patterns detect unique value violations across records (UNIQUE, Figures \ref{fig:unique_gen}, \ref{fig:unique_abs}) or within a record (UNIQUE2, Figures \ref{fig:unique2_gen}, \ref{fig:unique2_abs}).
\begin{figure}
	\centering
	\includegraphics[width=\linewidth]{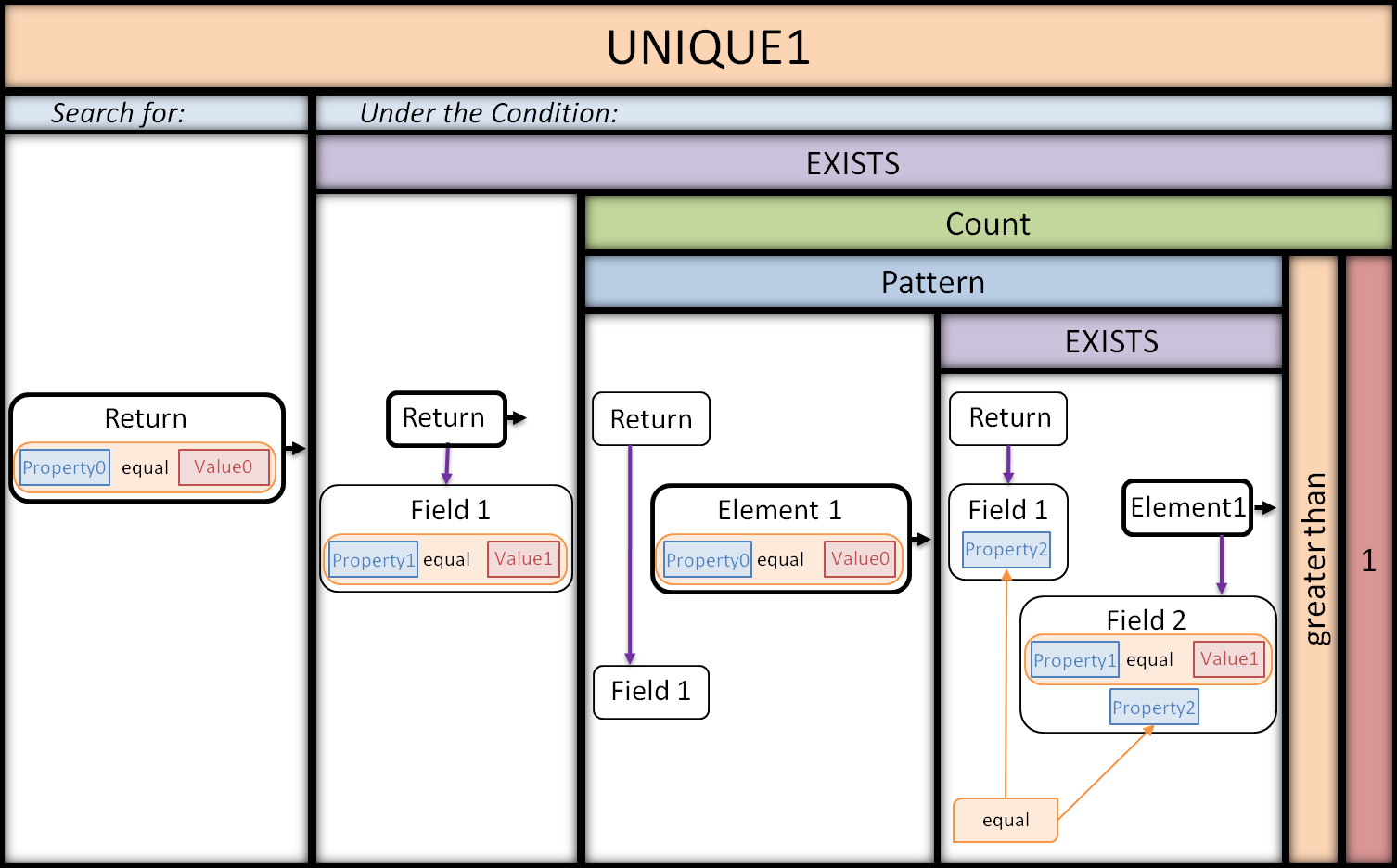}
	\caption{Generic pattern UNIQUE1}
	\label{fig:unique_gen}
\end{figure}
\begin{figure}
	\centering
	\includegraphics[width=\linewidth]{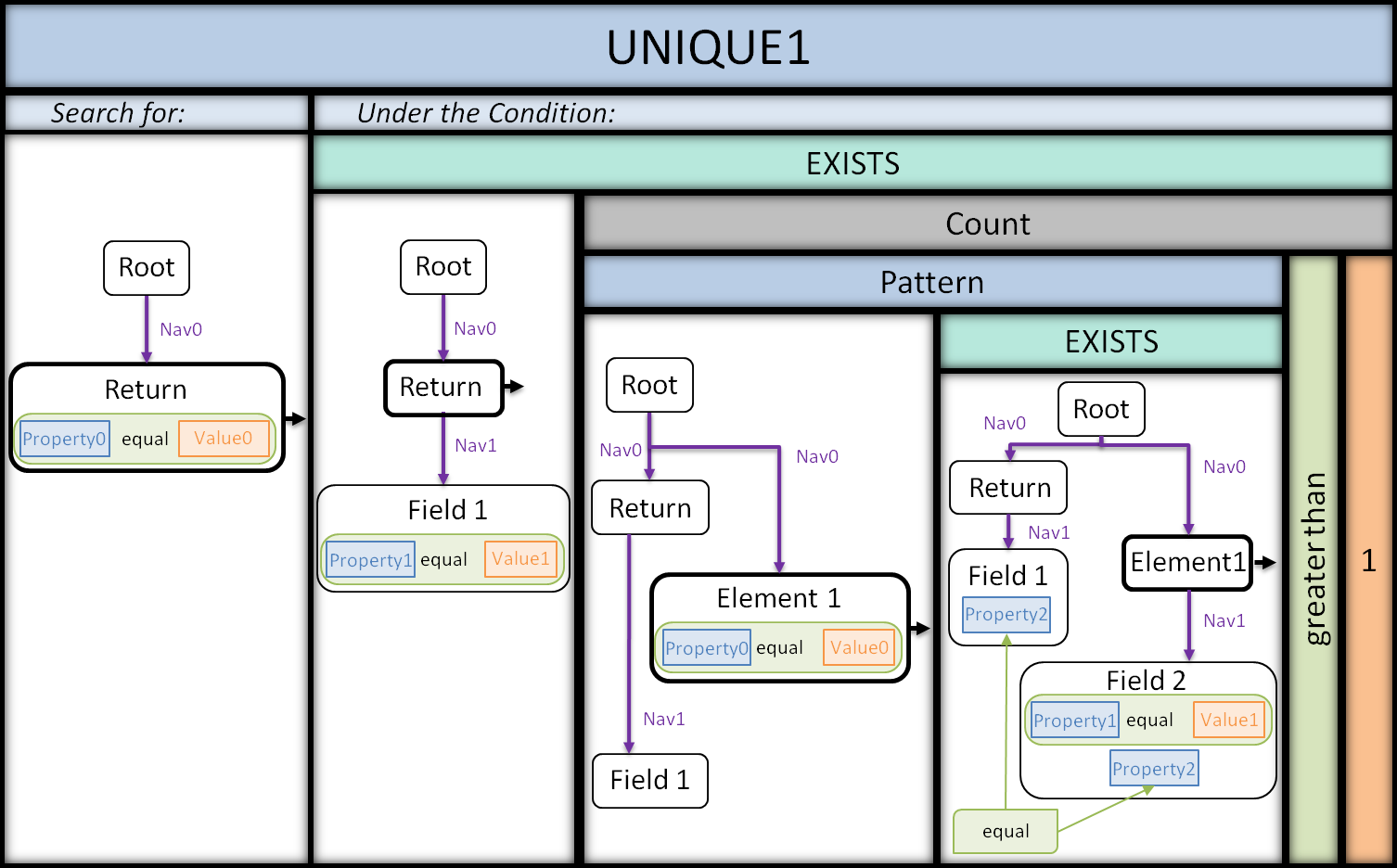}
	\caption{Abstract pattern UNIQUE1}
	\label{fig:unique_abs}
\end{figure}
\begin{figure}
	\centering
	\includegraphics[width=\linewidth]{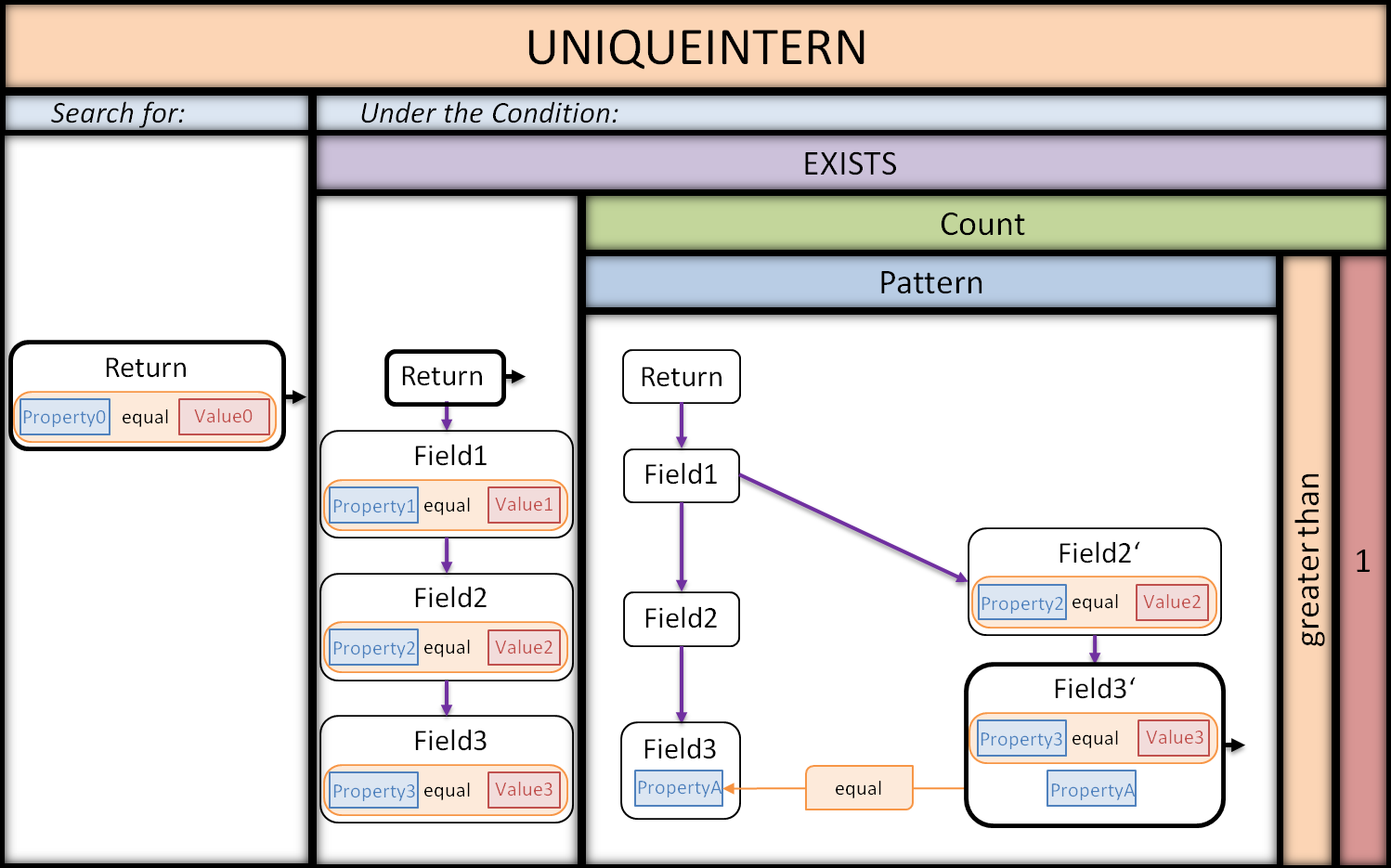}
	\caption{Generic pattern UNIQUE2}
	\label{fig:unique2_gen}
\end{figure}
\begin{figure}
	\centering
	\includegraphics[width=\linewidth]{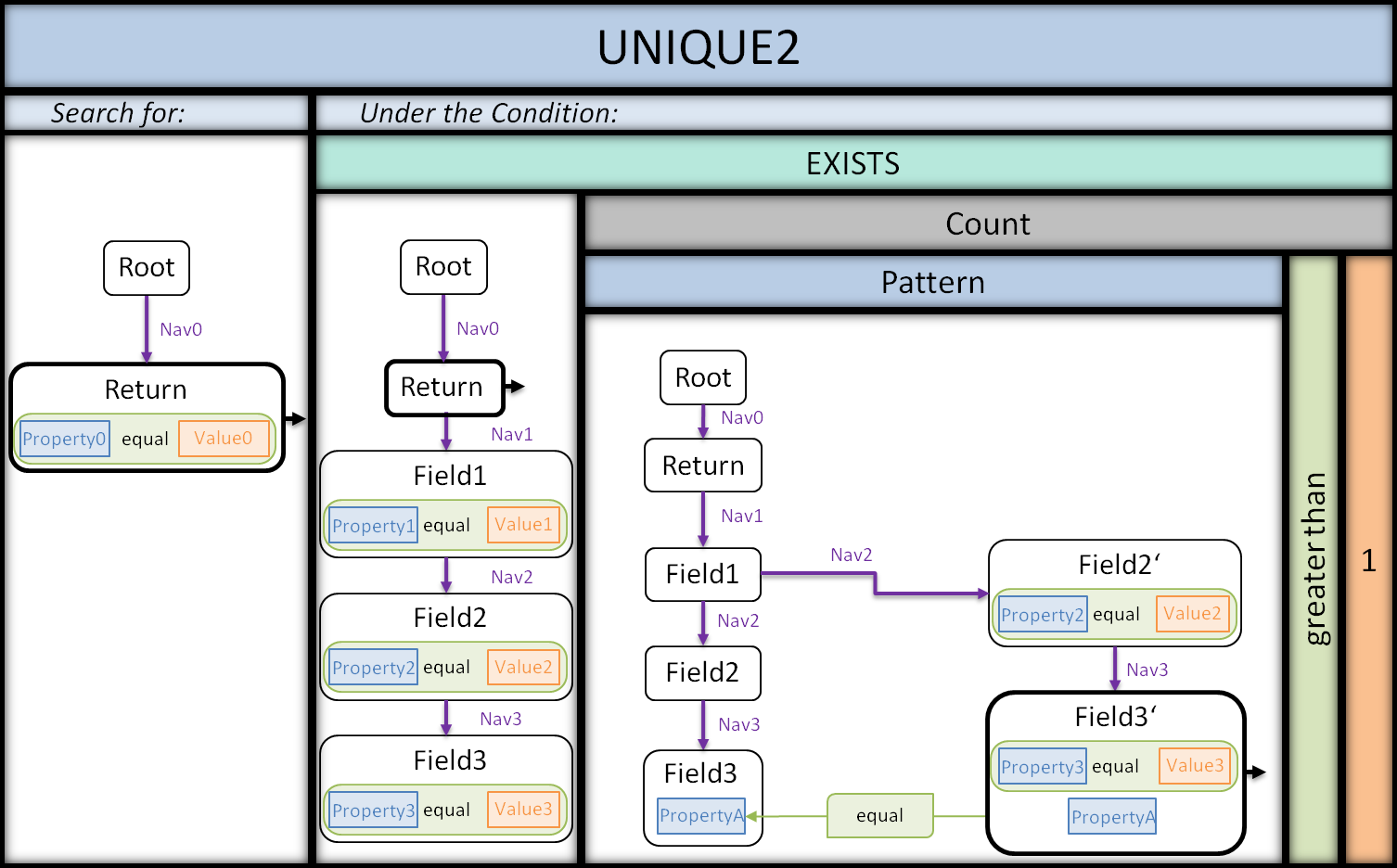}
	\caption{Abstract pattern UNIQUE2}
	\label{fig:unique2_abs}
\end{figure}

\subsection{MAND}
The MAND patterns (Figures \ref{fig:mand_gen}-\ref{fig:mand2_abs}) detect missing mandatory attributes.
The patterns match if a specific structure does not exist or if it exists but a specific included attribute is equal to one of several given dummy values.
\begin{figure}
	\centering
	\includegraphics[width=\linewidth]{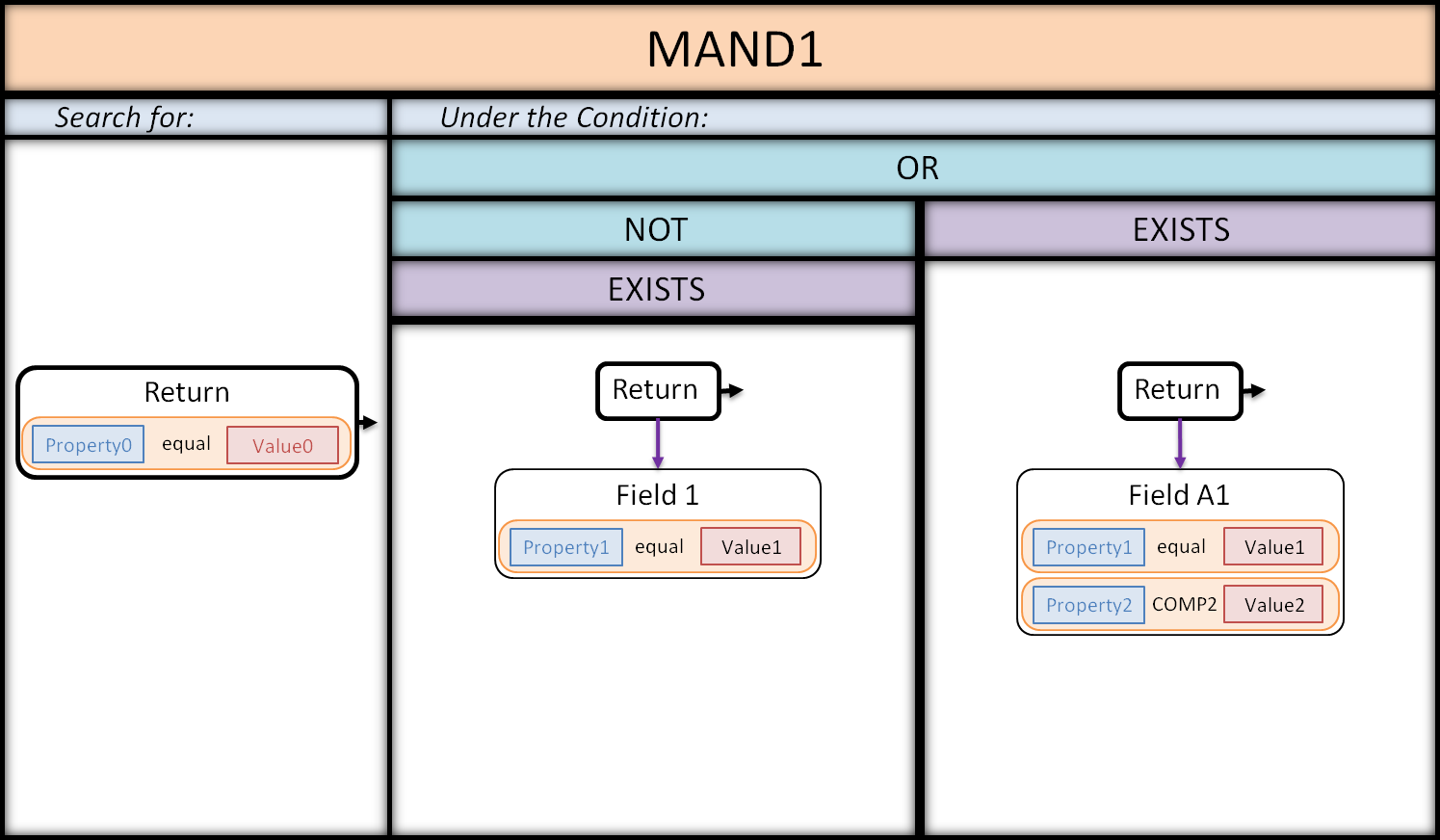}
	\caption{Generic pattern MAND1}
	\label{fig:mand_gen}
\end{figure}
\begin{figure}
	\centering
	\includegraphics[width=\linewidth]{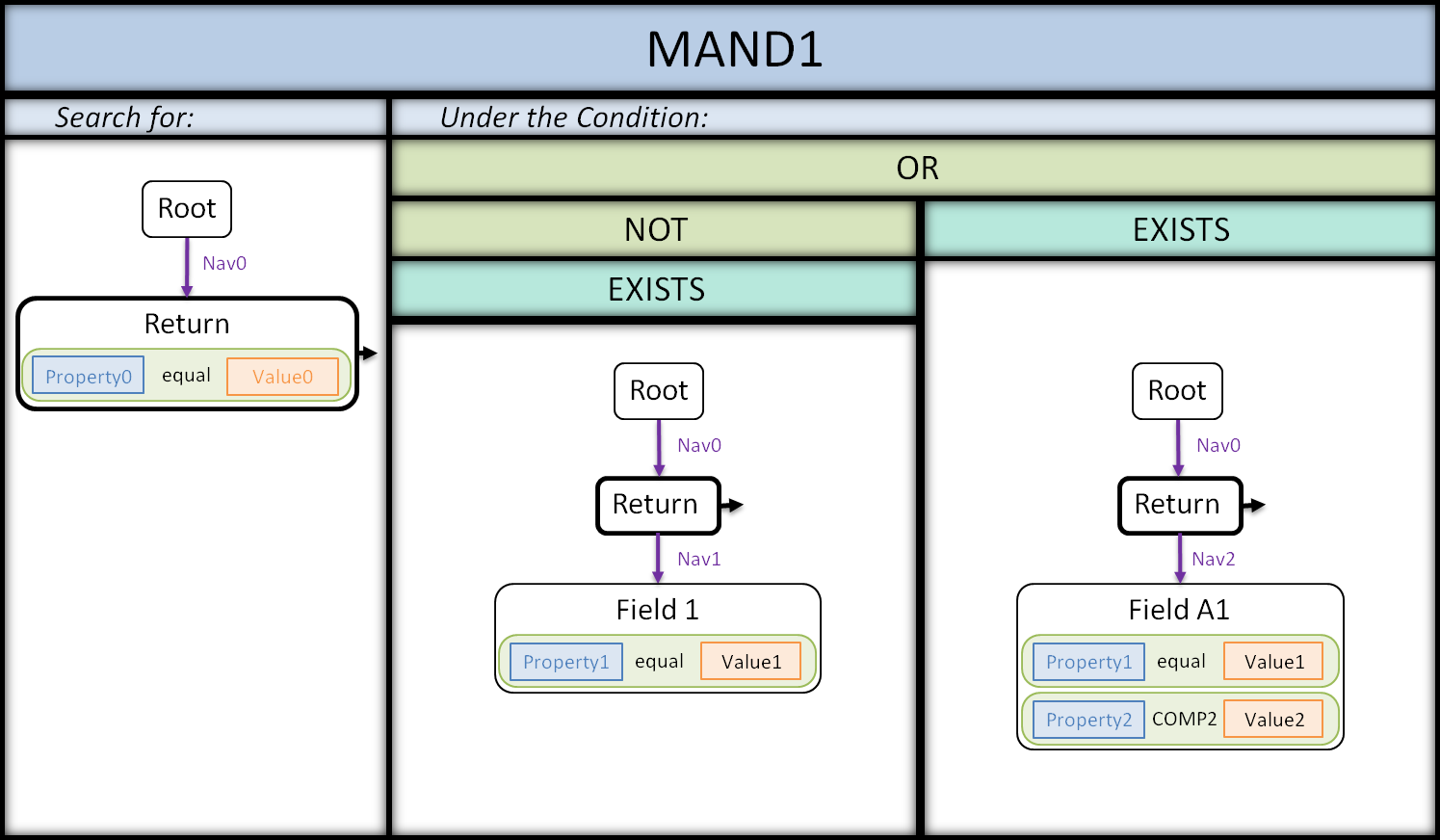}
	\caption{Abstract pattern MAND1}
	\label{fig:mand_abs}
\end{figure}
\begin{figure}
	\centering
	\includegraphics[width=\linewidth]{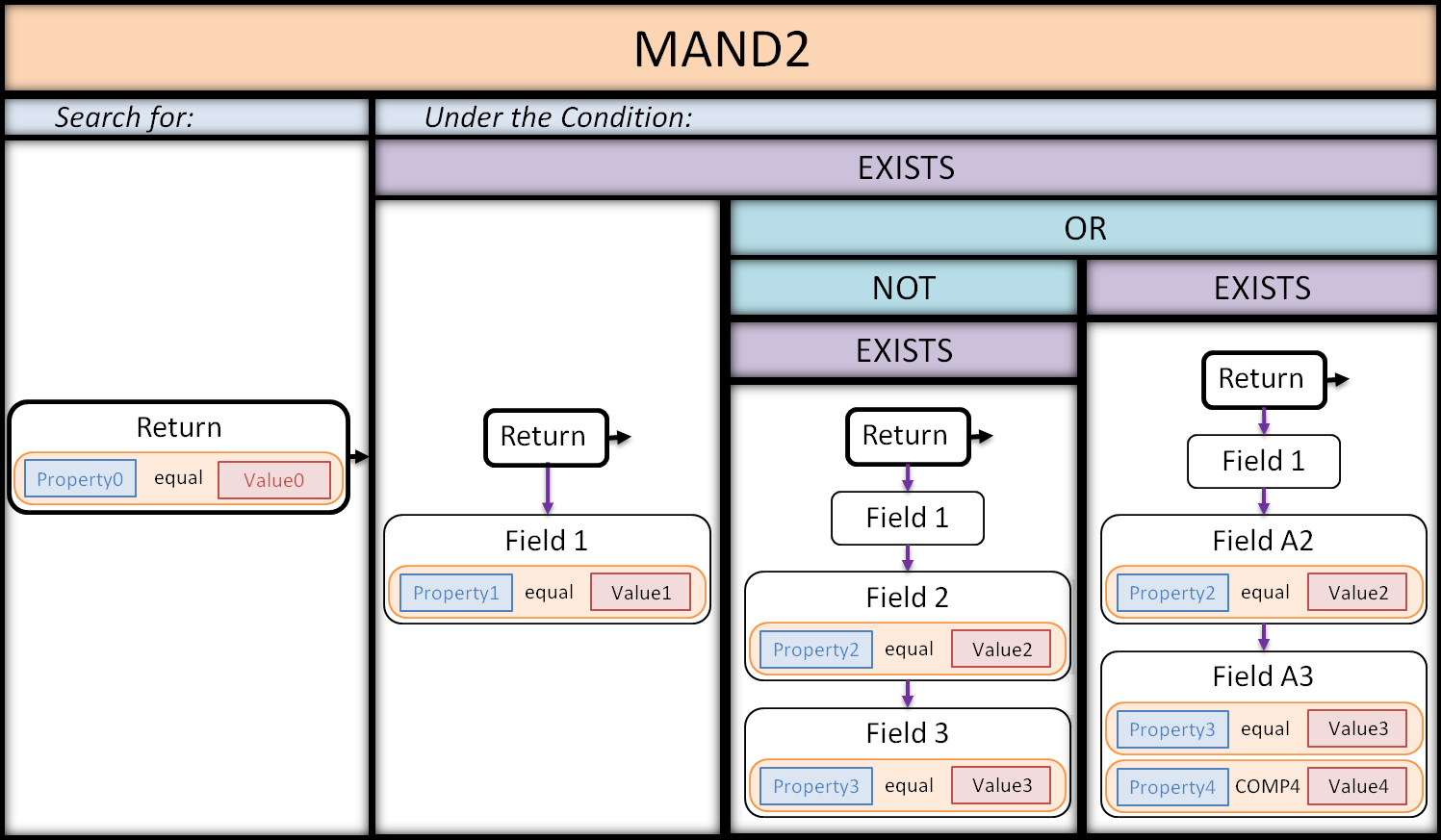}
	\caption{Generic pattern MAND2}
	\label{fig:mand2_gen}
\end{figure}
\begin{figure}
	\centering
	\includegraphics[width=\linewidth]{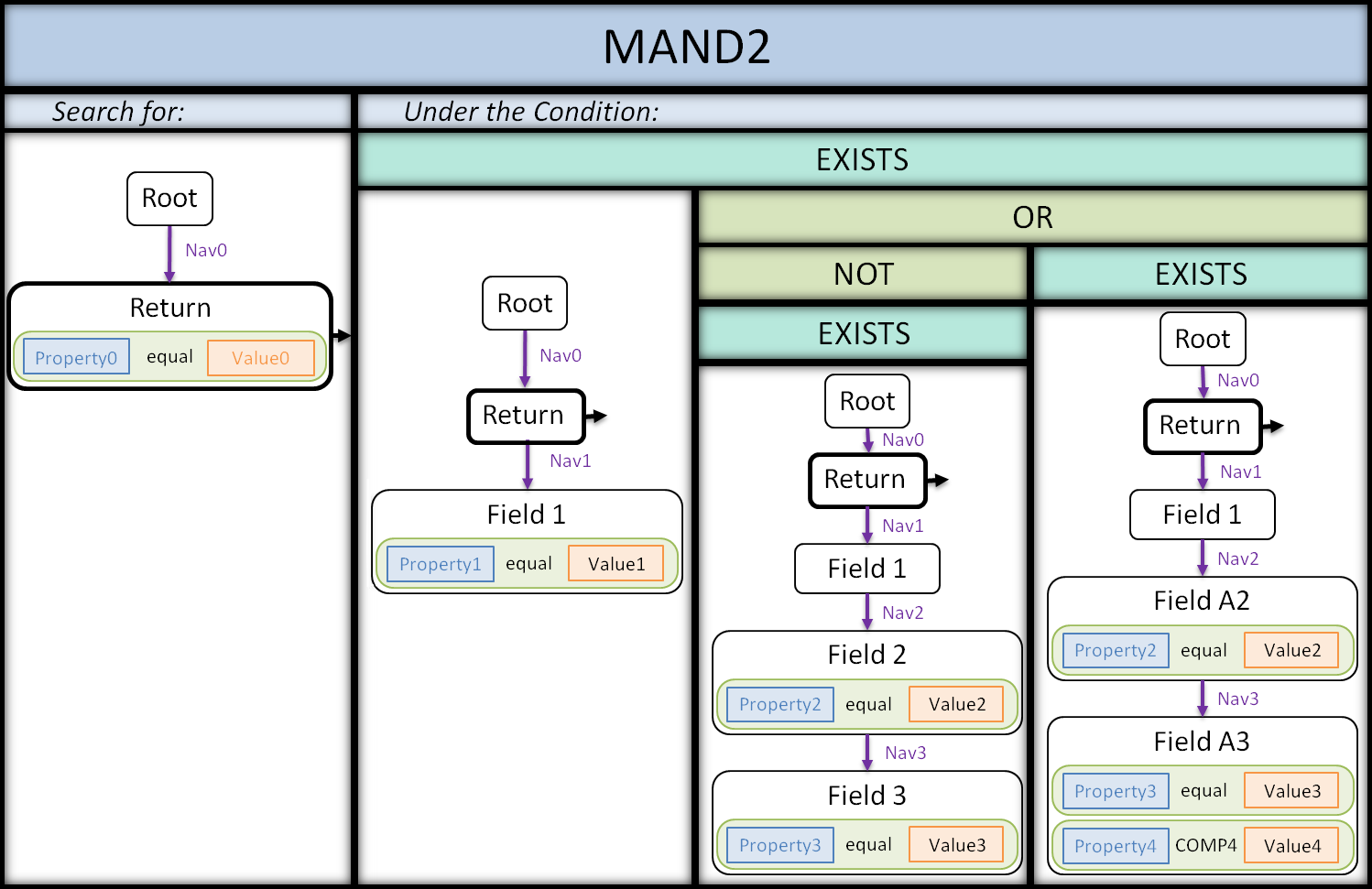}
	\caption{Abstract pattern MAND2}
	\label{fig:mand2_abs}
\end{figure}

\subsection{MANDSTRUC}
The MANDSTRUC patterns detect missing mandatory structures in the data.
MANDSTRUC1 (Figures \ref{fig:mandstruc_gen}, \ref{fig:mandstruc_abs}) detects data records, where Field 1 does not exist or its related Field 2 does not exist.
MANDSTRUC2 (Figures \ref{fig:mandstruc2_gen}, \ref{fig:mandstruc2_abs}) tests the existence of Field 2 under the precondition that the related Field 1 exists in the record.
\begin{figure}
	\centering
	\includegraphics[width=\linewidth]{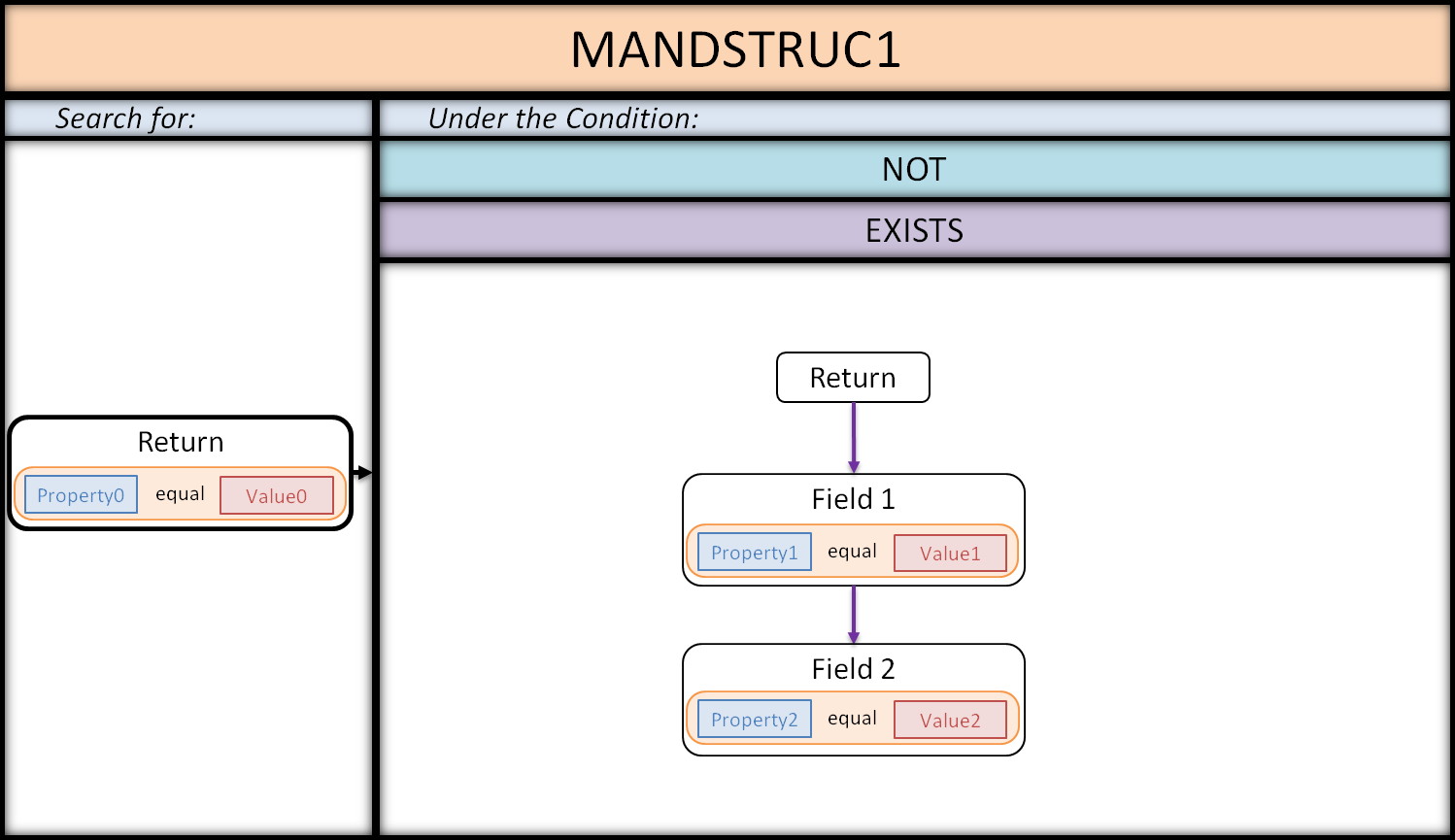}
	\caption{Generic pattern MANDSTRUC1}
	\label{fig:mandstruc_gen}
\end{figure}
\begin{figure}
	\centering
	\includegraphics[width=\linewidth]{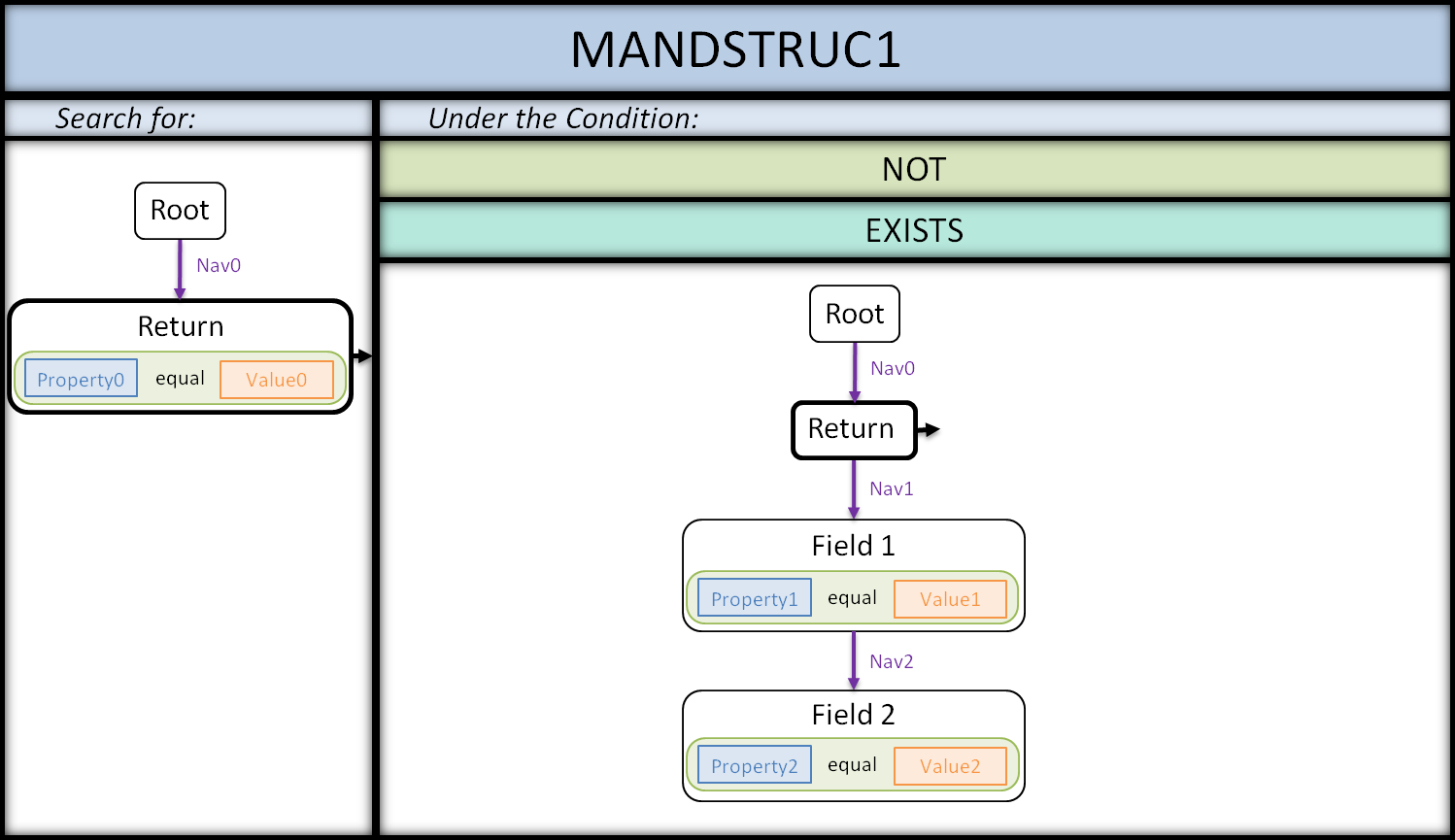}
	\caption{Abstract pattern MANDSTRUC1}
	\label{fig:mandstruc_abs}
\end{figure}
\begin{figure}
	\centering
	\includegraphics[width=\linewidth]{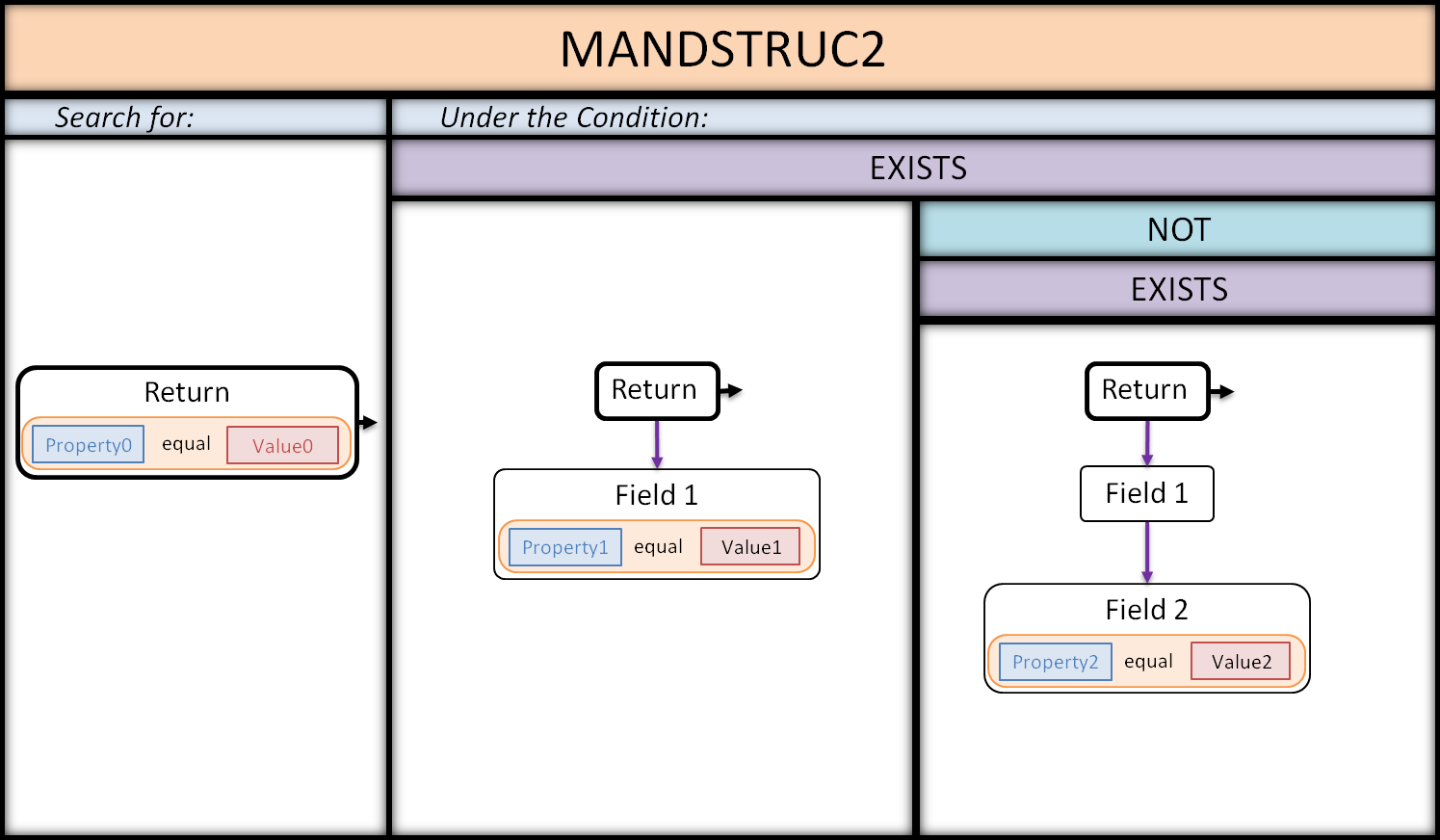}
	\caption{Generic pattern MANDSTRUC2}
	\label{fig:mandstruc2_gen}
\end{figure}
\begin{figure}
	\centering
	\includegraphics[width=\linewidth]{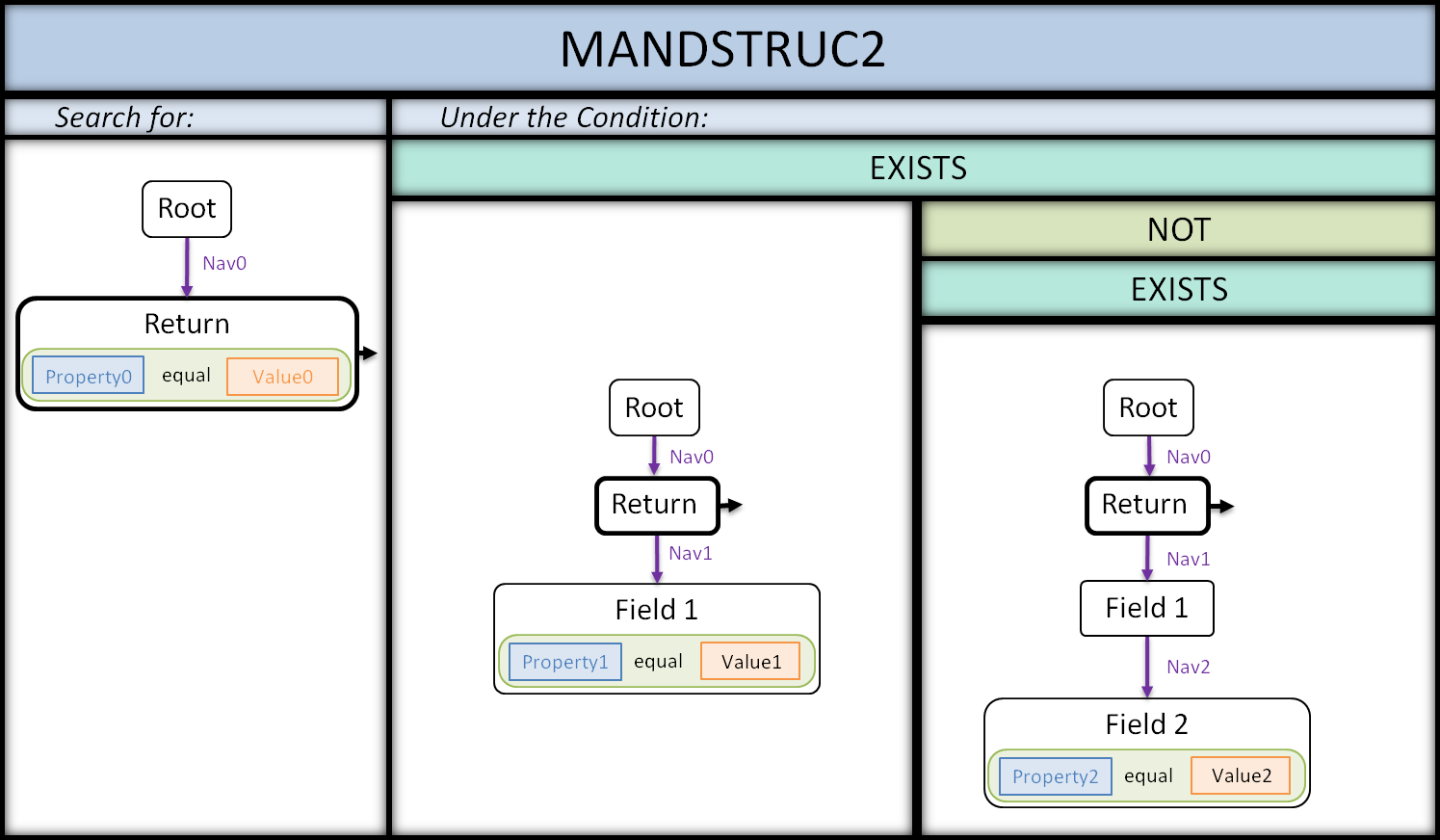}
	\caption{Abstract pattern MANDSTRUC2}
	\label{fig:mandstruc2_abs}
\end{figure}

\subsection{CONTREL}
The CONTREL pattern (Figures \ref{fig:contrel_gen}, \ref{fig:contrel_abs}) detects contradictory relationships of two records, more precisely violations of a symmetry constraint.
If Field 2 contains a reference to another record which contains a reference to the first record, the values of a specific field must satisfy a given comparison relation. 
Otherwise the relationships are contradictory.
\begin{figure}
	\centering
	\includegraphics[width=\linewidth]{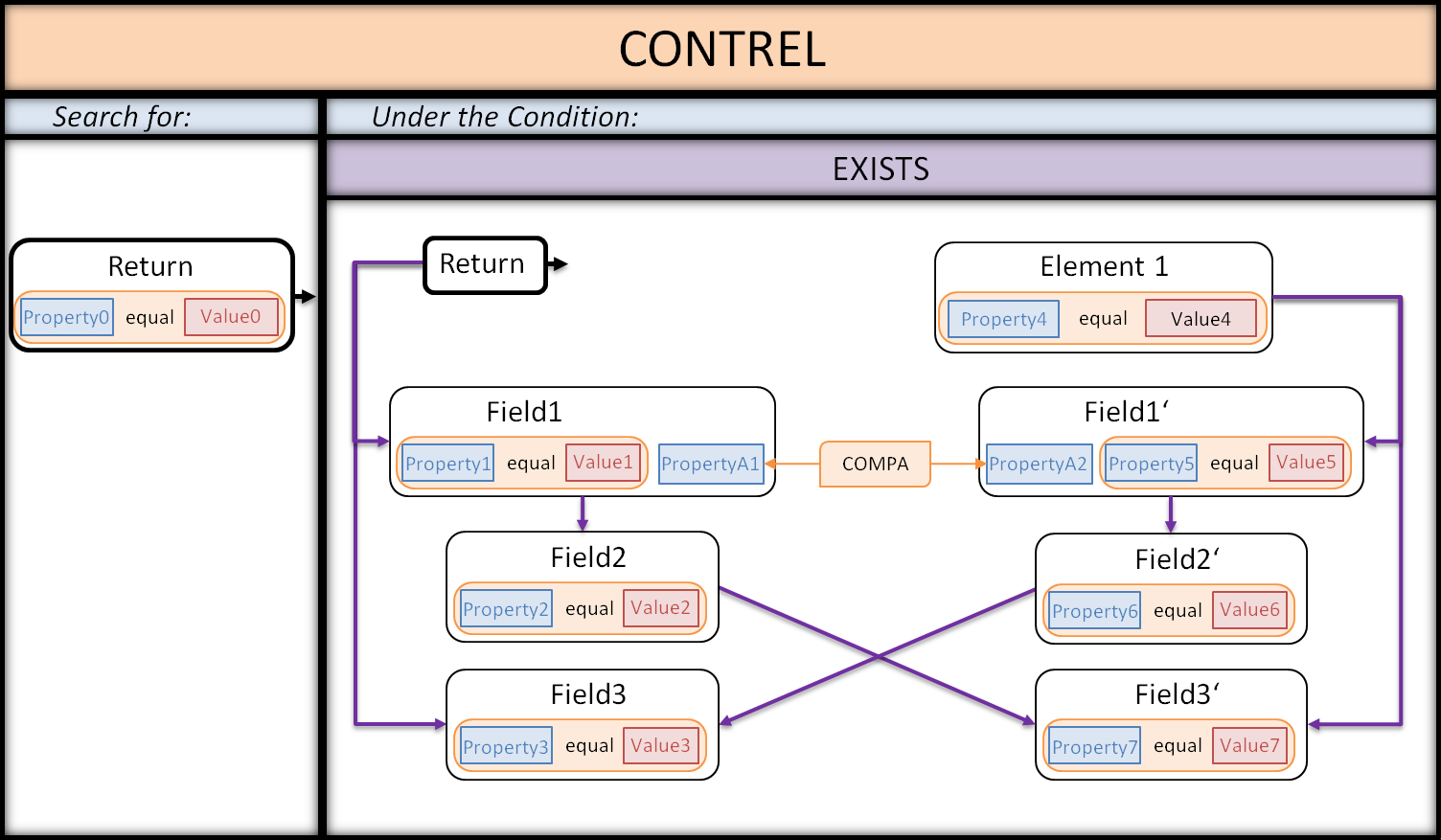}
	\caption{Generic pattern CONTREL}
	\label{fig:contrel_gen}
\end{figure}
\begin{figure}
	\centering
	\includegraphics[width=\linewidth]{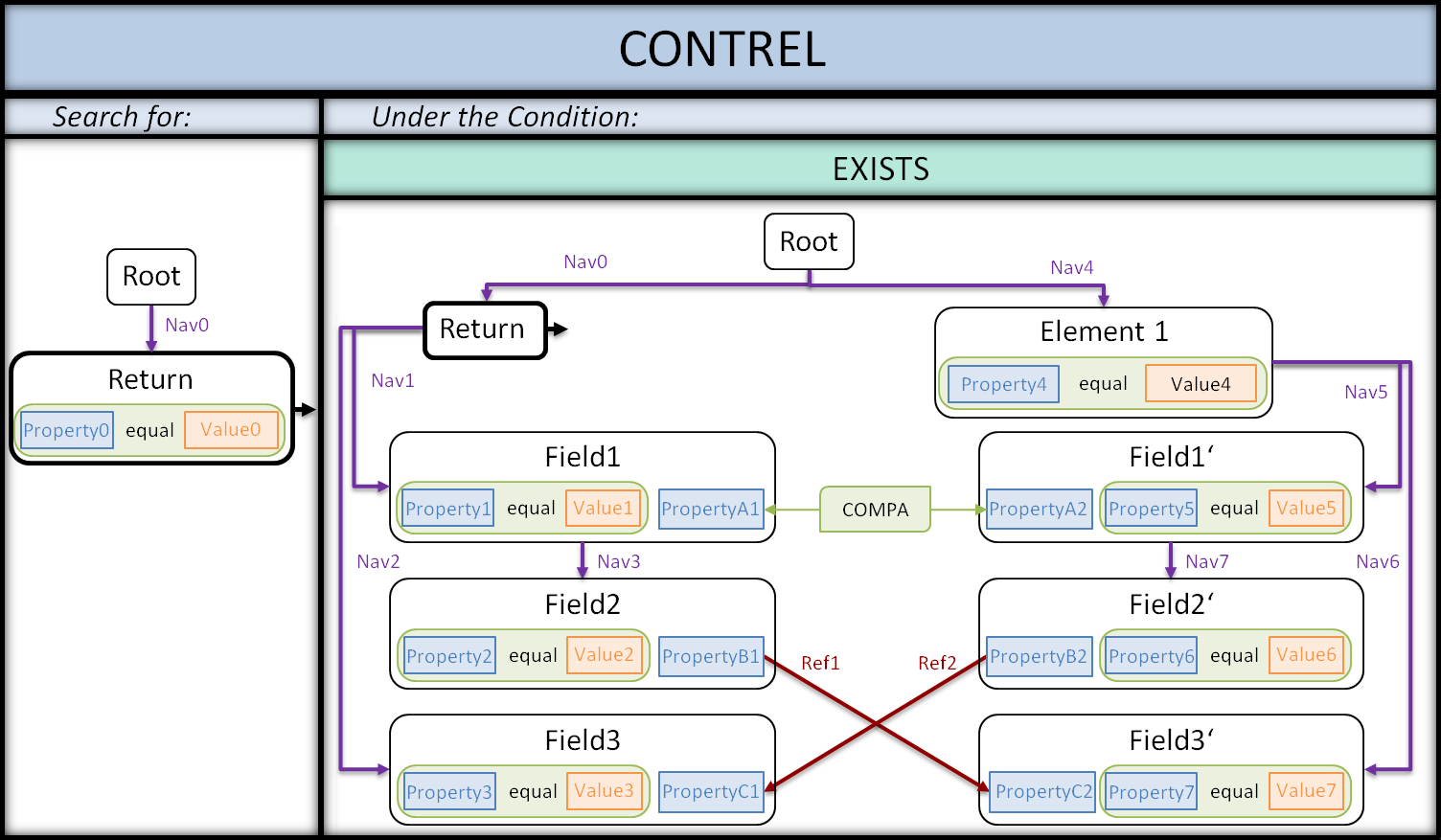}
	\caption{Abstract pattern CONTREL}
	\label{fig:contrel_abs}
\end{figure}

\subsection{EXDUP}
The EXDUP pattern (Figures \ref{fig:exdup_gen}, \ref{fig:exdup_abs}) detects exact duplicate records via a direct comparison between elements.
\begin{figure}
	\centering
	\includegraphics[width=\linewidth]{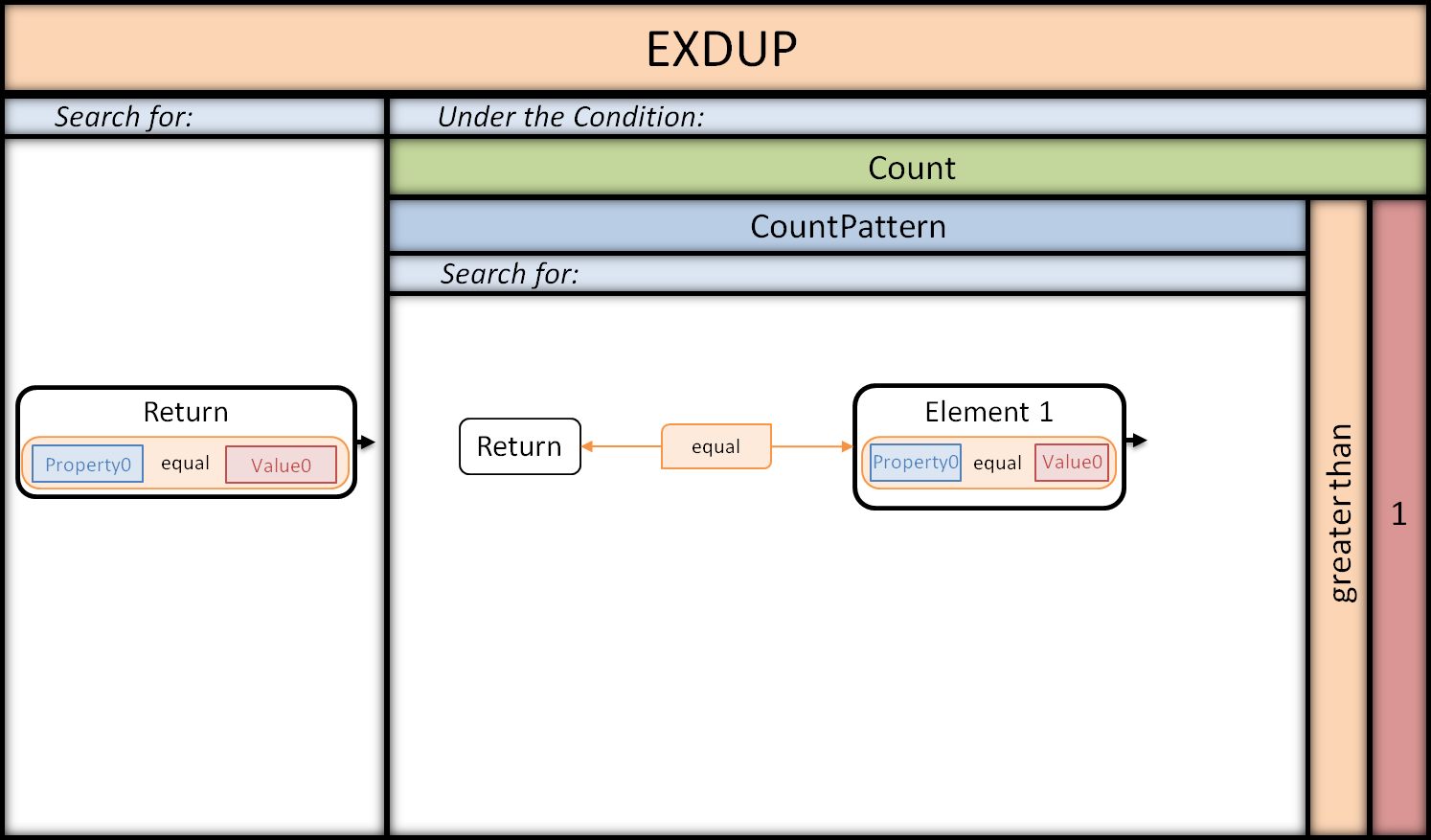}
	\caption{Generic pattern EXDUP}
	\label{fig:exdup_gen}
\end{figure}
\begin{figure}
	\centering
	\includegraphics[width=\linewidth]{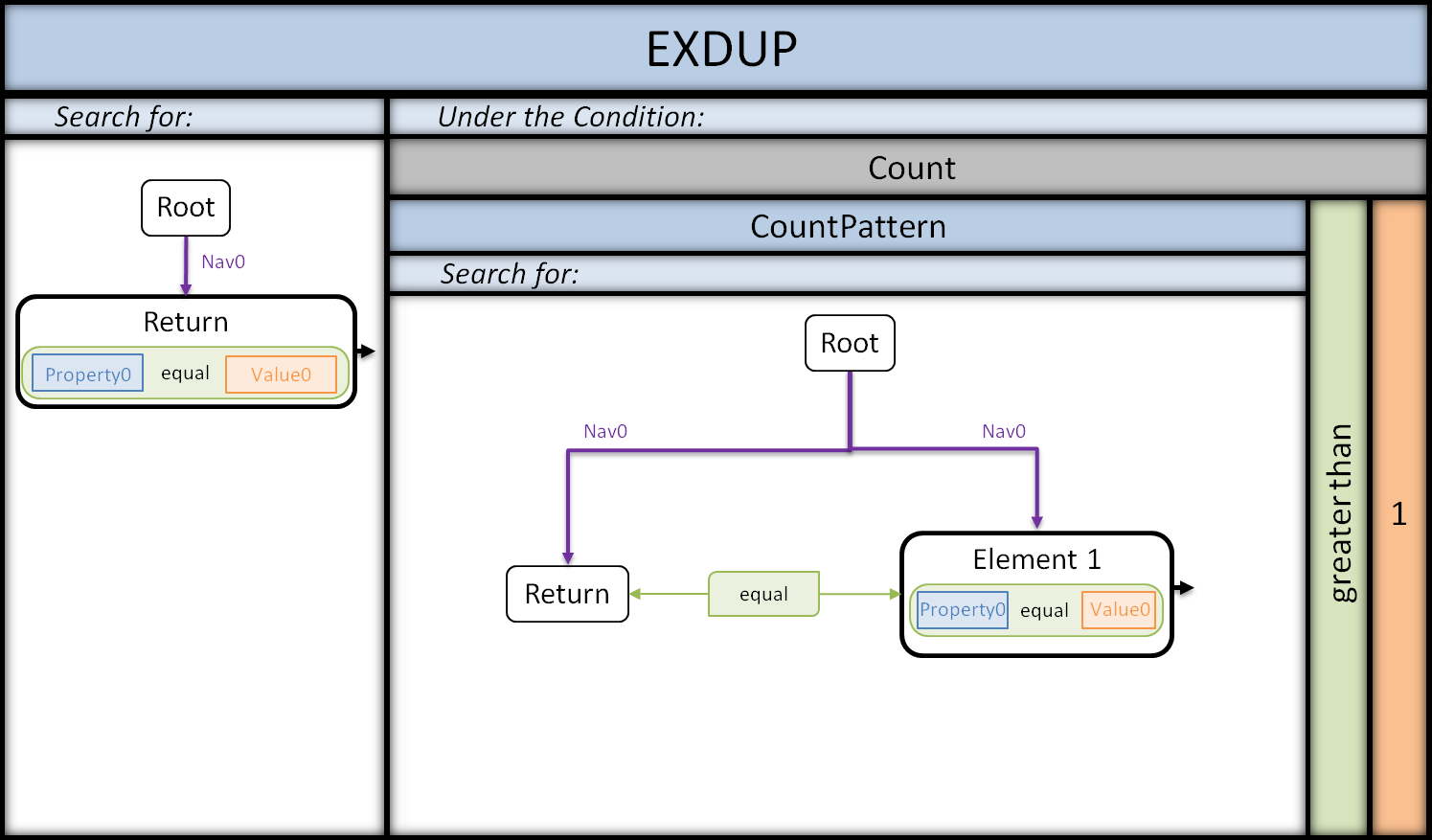}
	\caption{Abstract pattern EXDUP}
	\label{fig:exdup_abs}
\end{figure}

\subsection{APPDUP}
The APPDUP pattern (Figures \ref{fig:appdup_gen}, \ref{fig:appdup_abs}) detects approximate duplicate records by comparing three distinguishing attributes.
\begin{figure}
	\centering
	\includegraphics[width=\linewidth]{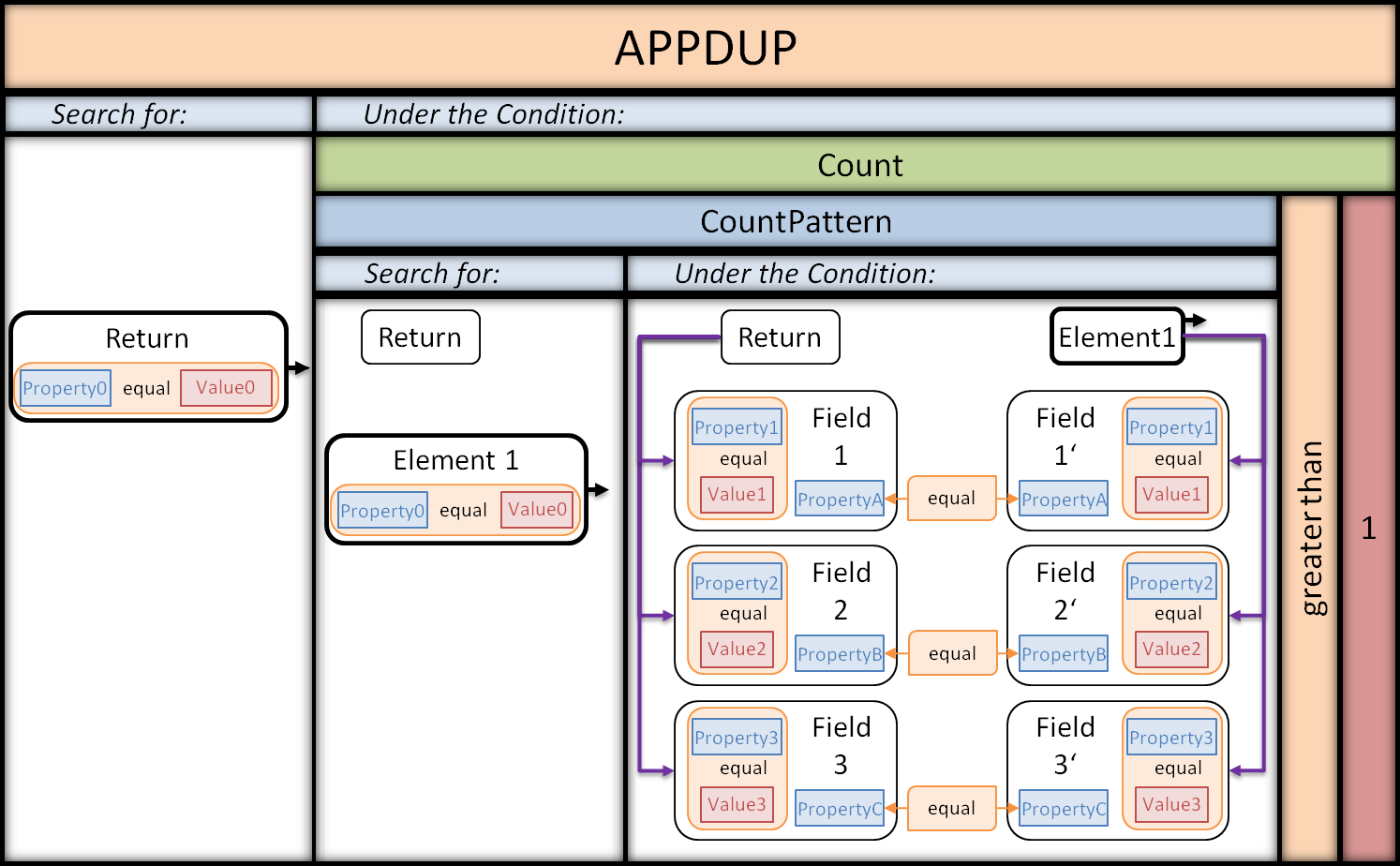}
	\caption{Generic pattern APPDUP}
	\label{fig:appdup_gen}
\end{figure}
\begin{figure}
	\centering
	\includegraphics[width=\linewidth]{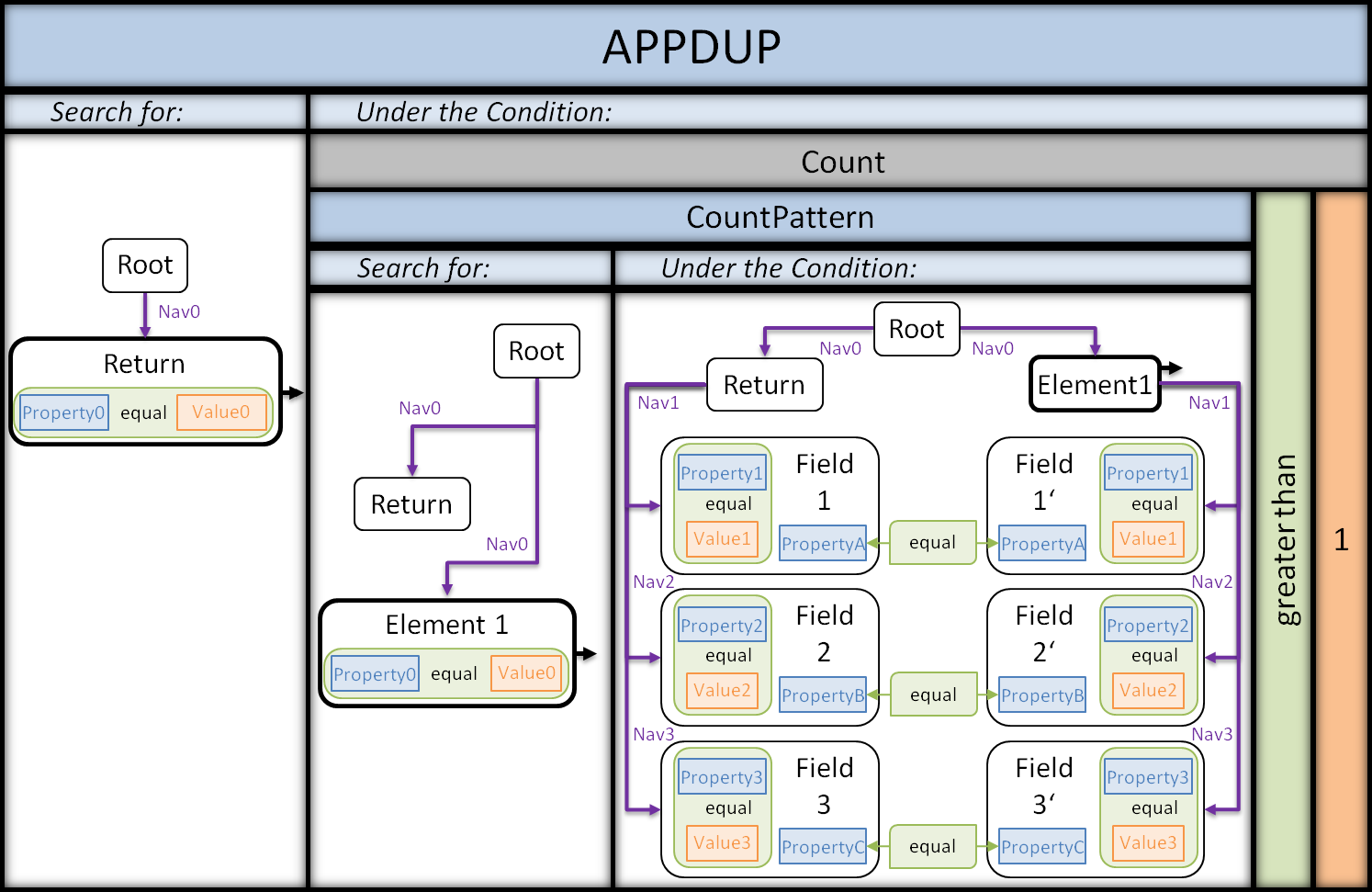}
	\caption{Abstract pattern APPDUP}
	\label{fig:appdup_abs}
\end{figure}

\clearpage
\section{Representative Quality Problems}\label{appendix:problems}
In the following, we will present the concrete quality problem instances for the MIDAS and LIDO data set that were covered by patterns in the course of the evaluation.
In some cases, we checked the LIDO data for the same concrete instance of the quality problem as the MIDAS data.
For each problem we will give a brief description, name the corresponding abstract pattern depicted in Section~\ref{appendix:generic}, give the runtime and list the parameter values that were necessary to concretise the corresponding XML-adapted abstract pattern for the MIDAS and LIDO format, respectively.
For each concretisation we specify properties, comparison operators, parameter values and relations.
If related properties, comparison operators and values shown in the generic diagram are not specified in the concretisation, this comparison is ignored.


\subsection{Illegal Values}
\subsubsection{Wrong Datatype}
\paragraph{MIDAS}
An XML element indicating the relation between an object and an artist contains an element of a wrong type. The contained elements should describe the artist, but that is not the case here. 

Runtime: 6653 ms
\begin{lstlisting}[label=lst:wrongdatatypem,caption={Concretisation of the pattern MATCH2},basicstyle=\ttfamily\footnotesize]
Nav0 = child3, Property0 = attribute "Type", Value0 = "obj"
Nav1 = child, Property1 = attribute "Type", Value1 = "ob30"
Nav2 = child, PropertyA = attribute "Value", ValueA = "^[12456789]"
\end{lstlisting}
\paragraph{LIDO}
A measurement value which should be given as a whole number or decimal fraction contains letters.

Runtime: 7206 ms
\begin{lstlisting}[label=lst:wrongdatatypel,caption={Concretisation of the pattern MATCH1},basicstyle=\ttfamily\footnotesize,language={[LaTeX]TeX}]
Nav0 = child2, Property0 = name, Value0 = "lido:lido"
Nav1 = child7, Property1 = name, Value1 = "lido:measurementValue"
PropertyA = data, ValueA = "[a-zA-ZüöäÜÖÄ]"
\end{lstlisting}
\subsubsection{Domain Violation}
\paragraph{MIDAS}
A set violation regarding an artist's gender.

Runtime: 770 ms
\begin{lstlisting}[label=lst:domainm,caption={Concretisation of the pattern COMPVAL1},basicstyle=\ttfamily\footnotesize]
Nav0 = child3, Property0 = attribute "Type", Value0 = "kue"
Nav1 = child, Property1 = attribute "Type", Value1 = "3140"
Property2 = attribute "Value", COMP2 = unequal, Value2 = ("m","f","unbekannt","m?","f?","?")
\end{lstlisting}
\paragraph{LIDO}
See above.

Runtime: 8021 ms
\begin{lstlisting}[label=lst:domainl,caption={Concretisation of the pattern COMPVAL1},basicstyle=\ttfamily\footnotesize]
Nav0 = child2, Property0 = name, Value0 = "lido:lido"
Nav1 = child8, Property1 = name, Value1 = "lido:genderActor"
Property2 = data, COMP2 = unequal, Value2 = ("male","männlich","weiblich","female","unknown","not applicable")
\end{lstlisting}

\subsection{Missing Data}
\subsubsection{Missing Values}
\paragraph{MIDAS}
The profession of an artist is not given.

Runtime: 1217 ms
\begin{lstlisting}[label=lst:missingvalm,caption={Concretisation of the pattern MAND1},basicstyle=\ttfamily\footnotesize]
Nav0 = child3, Property0 = attribute "Type", Value0 = "kue"
Nav1 = child, Property1 = attribute "Type", Value1 = "3162"
Property2 = attribute "Value", COMP2 = equal, Value2 = ""
\end{lstlisting}
\paragraph{LIDO}
An actor without a name specification.

Runtime: 8296 ms
\begin{lstlisting}[label=lst:missingvall,caption={Concretisation of the pattern MAND2},basicstyle=\ttfamily\footnotesize]
Nav0 = child2, Property0 = name, Value0 = "lido:lido"
Nav1 = child7, Property1 = name, Value1 = "lido:actor"
Nav2 = child, Property2 = name, Value2 = "lido:nameActorSet"
Nav3 = child, Property3 = name, Value3 = "lido:appellationValue"
Property4 = data, COMP2 = equal, Value4 = ""
\end{lstlisting}
\subsubsection{Missing References}
\paragraph{MIDAS}
An object for which no artist is specified.

Runtime: 6635 ms
\begin{lstlisting}[label=lst:missingrefm,caption={Concretisation of the pattern MANDSTRUC1},basicstyle=\ttfamily\footnotesize]
Nav0 = child3, Property0 = attribute "Type", Value0 = "obj"
Nav1 = child, Property1 = attribute "Type", Value1 = "ob30"
Nav2 = child, Property2 = attribute "Type", Value2 = "3100"
\end{lstlisting}
\paragraph{LIDO}
The role of an actor is not supplemented by a reference to a published controlled vocabulary.

Runtime: 6356 ms
\begin{lstlisting}[label=lst:missingrefl,caption={Concretisation of the pattern MANDSTRUC2},basicstyle=\ttfamily\footnotesize]
Nav0 = child2, Property0 = name, Value0 = "lido:lido"
Nav1 = child7, Property1 = name, Value1 = "lido:roleActor"
Nav2 = child, Property2 = name, Value2 = "lido:conceptID"
\end{lstlisting}
\subsubsection{Missing Records}
As explained in Section~\ref{subsec:expressiveness} this problem could not be covered.
\subsubsection{Dummy Values}
\paragraph{MIDAS}
The birthdate of an artist is one of: ``x'', ``y'', ``?'', ``unbekannt'' (German for ``unknown'').

Runtime: 926 ms
\begin{lstlisting}[label=lst:dummym,caption={Concretisation of the pattern COMPVAL1},basicstyle=\ttfamily\footnotesize]
Nav0 = child3, Property0 = attribute "Type", Value0 = "kue"
Nav1 = child, Property1 = attribute "Type", Value1 = "3270"
Property2 = attribute "Value", COMP2 = unequal, Value2 = ("x","y","?","unbekannt")
\end{lstlisting}
\paragraph{LIDO}
An appellation is stated to be one of: ``unbekannt'' (German for ``unknown''), empty string, ``?'', ``x'', ``unknown''.

Runtime: 9194 ms
\begin{lstlisting}[label=lst:dummyl,caption={Concretisation of the pattern COMPVAL1},basicstyle=\ttfamily\footnotesize]
Nav0 = child2, Property0 = name, Value0 = "lido:lido"
Nav1 = child5, Property1 = name, Value1 = "lido:appellationValue"
Property2 = data, COMP2 = unequal, Value2 = ("unbekannt","","?","x","unknown")
\end{lstlisting}

\subsection{Referential Integrity Violation}
\paragraph{MIDAS}
An object record containing a reference to a non-existent atelier record.

Runtime: 197244 ms
\begin{lstlisting}[label=lst:refintm,caption={Concretisation of the pattern REFINT},basicstyle=\ttfamily\footnotesize]
Nav0 = child3, Property0 = attribute "Type", COMP0 = equal, Value0 = "obj"
Nav1 = child2, Property1 = attribute "Type", COMP1 = equal, Value1 = "3600"
Nav2 = child3, Property2 = attribute "Type", COMP2 = equal, Value2 = "wer"
Nav3 = child, Property3 = attribute "Type", COMP3 = equal, Value3 = "3600"
PropertyA = attribute "Value", PropertyB = attribute "Value"
\end{lstlisting}
\paragraph{LIDO}
As explained in Section~\ref{subsec:application} this problem variant could not be covered for the LIDO data.

\subsection{Unique Value Violation}
\paragraph{MIDAS}
The atelier name is not unique even though it is used as an identifier.

Runtime: 90836 ms
\begin{lstlisting}[label=lst:uniquem,caption={Concretisation of the pattern UNIQUE1},basicstyle=\ttfamily\footnotesize]
Nav0 = child3, Property0 = attribute "Type", Value0 = "wer"
Nav1 = child, Property1 = attribute "Type", Value1 = "3600"
Property2 = attribute "Value"
\end{lstlisting}
\paragraph{LIDO}
A non-unique record ID.

Runtime: estimated 1.8 weeks (not finished)
\begin{lstlisting}[label=lst:uniquel,caption={Concretisation of the pattern UNIQUE1},basicstyle=\ttfamily\footnotesize]
Nav0 = child2, Property0 = name, Value0 = "lido:lido"
Nav1 = child, Property1 = name, Value1 = "lido:lidoRecID"
Property2 = data
\end{lstlisting}

\subsection{Violation of a Functional Dependency}
\paragraph{MIDAS}
Two atelier records with equal atelier names but different active years. The atelier name should be unique and thus determine the active years.

Runtime: 64299 ms
\begin{lstlisting}[label=lst:funcm,caption={Concretisation of the pattern FUNC1},basicstyle=\ttfamily\footnotesize]
Nav0 = child3, Property0 = attribute "Type", Value0 = "wer"
Nav1 = child, Property1 = attribute "Type", Value1 = "3600"
Nav2 = child, Property2 = attribute "Type", Value2 = "3680"
PropertyA = attribute "Value"
PropertyB = attribute "Value"
\end{lstlisting}
\paragraph{LIDO}
The functional dependency between a concept ID linking to a published controlled vocabulary and the corresponding name of the concept is violated for the description of a material or technique.

Runtime: 21474836 ms
\begin{lstlisting}[label=lst:funcl,caption={Concretisation of the pattern FUNC2},basicstyle=\ttfamily\footnotesize]
Nav0 = child2, Property0 = name, COMP0 = equal, Value0 = "lido:lido"
Nav1 = child7, Property1 = name, COMP1 = equal, Value1 = "lido:termMaterialsTech"
Nav2 = child, Property2 = name, COMP2 = equal, Value2 = "lido:conceptID"
Nav3 = child, Property3 = name, COMP3 = equal, Value3 = "lido:term"
PropertyA = data
PropertyB = data
\end{lstlisting}
\subsection{Contradictory Relationships}
\paragraph{MIDAS}
If an artist record references an atelier record and the atelier record references the artist record, both specified relation types must be equal, which is not the case here.

Runtime: 18957 ms
\begin{lstlisting}[label=lst:contrelm,caption={Concretisation of the pattern CONTREL},basicstyle=\ttfamily\footnotesize]
Nav0 = child3, Property0 = attribute "Type", Value0 = "kue"
Nav1 = child, Property1 = attribute "Type", Value1 = "ku35"
Nav2 = child, Property2 = attribute "Type", Value2 = "3600"
Nav3 = child, Property3 = attribute "Type", Value3 = "3100"
Nav4 = child3, Property4 = attribute "Type", Value4 = "wer"
Nav5 = child, Property5 = attribute "Type", Value5 = "we30"
Nav6 = child, Property6 = attribute "Type", Value6 = "3100"
Nav7 = child, Property7 = attribute "Type", Value7 = "3600"
PropertyA1 = attribute "Value", COMPA = unequal, PropertyA2 = attribute "Value"
PropertyB1 = attribute "Value", PropertyC2 = attribute "Value"
PropertyC1 = attribute "Value", PropertyB2 = attribute "Value"
\end{lstlisting}
\paragraph{LIDO}
As explained in Section~\ref{subsec:application} this problem variant could not be covered for the LIDO data.

\subsection{Imprecise Data}
\subsubsection{Alternative Possible Values}
\paragraph{MIDAS}
Multiple possible artists are listed for an artwork.

Runtime: 9813 ms
\begin{lstlisting}[label=lst:altm,caption={Concretisation of the pattern CARD1},basicstyle=\ttfamily\footnotesize]
Nav0 = child3, Property0 = attribute "Type", Value0 = "obj"
Nav1 = descendant-or-self, Property1 = name, Value1 = "h1:Block"
Nav2 = child, Property2 = attribute "Type", Value2 = "ob30"
Nav3 = child, Property3 = attribute "Value", COMP3 = equal, Value3 = "Herstellung"
Property4 = attribute "Type", Value4 = "ob30rl"
\end{lstlisting}
\paragraph{LIDO}
The place of an event is marked as ``alternative''.

Runtime: 5351 ms
\begin{lstlisting}[label=lst:altl,caption={Concretisation of the pattern COMPVAL1},basicstyle=\ttfamily\footnotesize]
Nav0 = child2, Property0 = name, Value0 = "lido:lido"
Nav1 = child5, Property1 = name, Value1 = "lido:eventPlace"
Property2 = attribute "lido:type", COMP2 = equal, Value2 = "alternative"
\end{lstlisting}

\subsubsection{Imprecise Numerical Values}
\paragraph{MIDAS}
An interval is given as the date of birth of an artist.

Runtime: 965 ms
\begin{lstlisting}[label=lst:imprm,caption={Concretisation of the pattern MATCH1},basicstyle=\ttfamily\footnotesize]
Nav0 = child3, Property0 = attribute "Type", Value0 = "kue"
Nav1 = child, Property1 = attribute "Type", Value1 = "3270"
PropertyA = attribute "Value", ValueA = "[0-9]/[0-9]"
\end{lstlisting}
\paragraph{LIDO}
The field for specifying the earliest possible date of something contains only the specification of a year.

Runtime: 4871 ms
\begin{lstlisting}[label=lst:imprl,caption={Concretisation of the pattern MATCH1},basicstyle=\ttfamily\footnotesize,language={[LaTeX]TeX}]
Nav0 = child2, Property0 = name, Value0 = "lido:lido"
Nav1 = child7, Property1 = name, Value1 = "lido:earliestDate"
PropertyA = data, ValueA = "^[0-9]{4}$"
\end{lstlisting}
\subsubsection{Abstract Terms}
\paragraph{MIDAS}
The type of the described artwork is specified as ``Objekt'' (German for ``object'').

Runtime: 8296 ms
\begin{lstlisting}[label=lst:abstrm,caption={Concretisation of the pattern COMPVAL1},basicstyle=\ttfamily\footnotesize]
Nav0 = child3, Property0 = attribute "Type", Value0 = "obj"
Nav1 = child, Property1 = attribute "Type", Value1 = "5230"
Property2 = attribute "Value", COMP2 = equal, Value2 = "Objekt"
\end{lstlisting}
\paragraph{LIDO}
See above.

Runtime: 5630 ms
\begin{lstlisting}[label=lst:abstrl,caption={Concretisation of the pattern COMPVAL2},basicstyle=\ttfamily\footnotesize]
Nav0 = child2, Property0 = name, Value0 = "lido:lido"
Nav1 = child4, Property1 = name, Value1 = "lido:objectWorkType"
Nav2 = child, Property2 = name, Value2 = "lido:term"
Property3 = data, COMP3 = equal, Value3 = "Objekt"
\end{lstlisting}
\subsubsection{Ambiguous Values}
\paragraph{MIDAS}
The type of the described artwork is specified as ``Schloss'' (German for ``castle'' and ``lock'').

Runtime: 7435 ms
\begin{lstlisting}[label=lst:ambm,caption={Concretisation of the pattern COMPVAL1},basicstyle=\ttfamily\footnotesize]
Nav0 = child3, Property0 = attribute "Type", Value0 = "obj"
Nav1 = child, Property1 = attribute "Type", Value1 = "5230"
Property2 = attribute "Value", COMP2 = equal, Value2 = "Schloss"
\end{lstlisting}
\paragraph{LIDO}
See above.

Runtime: 6011 ms
\begin{lstlisting}[label=lst:ambl,caption={Concretisation of the pattern COMPVAL2},basicstyle=\ttfamily\footnotesize]
Nav0 = child2, Property0 = name, Value0 = "lido:lido"
Nav1 = child4, Property1 = name, Value1 = "lido:objectWorkType"
Nav2 = child, Property2 = name, Value2 = "lido:term"
Property3 = data, COMP3 = equal, Value3 = "Schloss"
\end{lstlisting}
\subsubsection{Abbreviations}
\paragraph{MIDAS}
The name of an artist contains a dot.

Runtime: 1178 ms
\begin{lstlisting}[label=lst:abbrm,caption={Concretisation of the pattern MATCH1},basicstyle=\ttfamily\footnotesize]
Nav0 = child3, Property0 = attribute "Type", Value0 = "kue"
Nav1 = child, Property1 = attribute "Type", Value1 = "3100"
PropertyA = attribute "Value", ValueA = "\."
\end{lstlisting}
\paragraph{LIDO}
See above.

Runtime: 8729 ms
\begin{lstlisting}[label=lst:abbrl,caption={Concretisation of the pattern MATCH2},basicstyle=\ttfamily\footnotesize,language={[LaTeX]TeX}]
Nav0 = child2, Property0 = name, Value0 = "lido:lido"
Nav1 = child8, Property1 = name, Value1 = "lido:nameActorSet"
Nav2 = child, Property2 = name, COMP2 = equal, Value2 = "lido:appellationValue"
PropertyA = data, ValueA = "\."
\end{lstlisting}

\subsection{Misplaced Information}
\subsubsection{Misfielded Values}
\paragraph{MIDAS}
A field for specifying the kind of date given in another field contains only digits. This indicates that the date is given in the wrong field.

Runtime: 7445 ms
\begin{lstlisting}[label=lst:misfm,caption={Concretisation of the pattern MATCH1},basicstyle=\ttfamily\footnotesize]
Nav0 = child3, Property0 = attribute "Type", Value0 = "obj"
Nav1 = child, Property1 = attribute "Type", Value1 = "5060"
PropertyA = attribute "Value", ValueA = "^[0-9/]+$"
\end{lstlisting}
\paragraph{LIDO}
The measurement unit field contains digits probably representing the actual measurement value which, however, should be given in another field.

Runtime: 7424 ms
\begin{lstlisting}[label=lst:misfl,caption={Concretisation of the pattern MATCH1},basicstyle=\ttfamily\footnotesize,language={[LaTeX]TeX}]
Nav0 = child2, Property0 = name, Value0 = "lido:lido"
Nav1 = child7, Property1 = name, Value1 = "lido:measurementUnit"
PropertyA = data, ValueA = "[0-9]"
\end{lstlisting}
\subsubsection{Extraneous Data}
\paragraph{MIDAS}
The field for specifying the dating of an artwork contains more than 10 letters. This indicates that more than the expected information is given.

Runtime: 10235 ms
\begin{lstlisting}[label=lst:extram,caption={Concretisation of the pattern MATCH1},basicstyle=\ttfamily\footnotesize]
Nav0 = child3, Property0 = attribute "Type", Value0 = "obj"
Nav1 = child, Property1 = attribute "Type", Value1 = "5064"
PropertyA = attribute "Value", ValueA = "[a-zA-Z ]{10}"
\end{lstlisting}
\paragraph{LIDO}
The political entity is additionally given in a field for specifying the name of a geographic place. It follows after a comma.

Runtime: 6453 ms
\begin{lstlisting}[label=lst:extral,caption={Concretisation of the pattern MATCH1},basicstyle=\ttfamily\footnotesize,language={[LaTeX]TeX}]
Nav0 = child2, Property0 = name, Value0 = "lido:lido"
Nav1 = child7, Property1 = name, Value1 = "lido:namePlaceSet"
PropertyA = data, ValueA = ","
\end{lstlisting}

\subsection{Redundant Data}
\subsubsection{Exact Duplicate Records}
\paragraph{MIDAS}
Exact duplicate atelier records.

Runtime: 39046 ms
\begin{lstlisting}[label=lst:exactm,caption={Concretisation of the pattern EXDUP},basicstyle=\ttfamily\footnotesize]
Nav0 = child3, Property0 = attribute "Type", Value0 = "wer"
\end{lstlisting}
\paragraph{LIDO}
Exact duplicate object records.

Runtime: estimated 4.1 weeks (not finished)
\begin{lstlisting}[label=lst:exactl,caption={Concretisation of the pattern EXDUP},basicstyle=\ttfamily\footnotesize,language={[LaTeX]TeX}]
Nav0 = child2, Property0 = name, COMP0 = equal, Value0 = "lido:lido"
\end{lstlisting}
\subsubsection{Approximate Duplicate Records}
\paragraph{MIDAS}
Multiple atelier records with equal location, type and active time may be approximate duplicates.

Runtime: 121129 ms
\begin{lstlisting}[label=lst:approxm,caption={Concretisation of the pattern APPDUP},basicstyle=\ttfamily\footnotesize]
Nav0 = child3, Property0 = attribute "Type", Value0 = "wer"
Nav1 = child, Property1 = attribute "Type", Value1 = "3560"
Nav2 = child, Property2 = attribute "Type", Value2 = "3580"
Nav3 = child, Property3 = attribute "Type", Value3 = "3680"
\end{lstlisting}
\paragraph{LIDO}
Multiple records reference the same unique, published identification of the described object. This indicates duplicate records describing the same object.

Runtime: 4330299 ms
\begin{lstlisting}[label=lst:approxl,caption={Concretisation of the pattern UNIQUE1},basicstyle=\ttfamily\footnotesize]
Nav0 = child2, Property0 = name, Value0 = "lido:lido"
Nav1 = child, Property1 = name, Value1 = "lido:objectPublishedID"
Property2 = data
\end{lstlisting}
\subsubsection{Information Placed in Multiple Locations}
\paragraph{MIDAS}
The given first and middle names of an artist are equal.

Runtime: 1380 ms
\begin{lstlisting}[label=lst:redm,caption={Concretisation of the pattern COMP},basicstyle=\ttfamily\footnotesize]
Nav0 = child3, Property0 = attribute "Type", COMP0 = equal, Value0 = "kue"
Nav1 = child, Property1 = attribute "Type", COMP1 = equal, Value1 = "3100"
Nav2 = child, Property2 = attribute "Type", COMP2 = equal, Value2 = "3105"
PropertyA = attribute "Value", COMPA = equal, PropertyB = attribute "Value"
\end{lstlisting}
\paragraph{LIDO}
The same name is given multiple times in one actor element.

Runtime: 10029 ms
\begin{lstlisting}[label=lst:redl,caption={Concretisation of the pattern UNIQUE2},basicstyle=\ttfamily\footnotesize]
Nav0 = child2, Property0 = name, Value0 = "lido:lido"
Nav1 = child7, Property1 = name, Value1 = "lido:actor"
Nav2 = child, Property2 = name, Value2 = "lido:nameActorSet"
Nav3 = child, Property3 = name, Value3 = "lido:appellationValue"
PropertyA = data
\end{lstlisting}

\subsection{Heterogeneous Data}
\subsubsection{Heterogeneous Measure Units}
\paragraph{MIDAS}
The size of an artwork must be given as height times width in cm without stating the measure unit explicitly. The pattern detects violations.

Runtime: 6445 ms
\begin{lstlisting}[label=lst:hetmm,caption={Concretisation of the pattern MATCH1},basicstyle=\ttfamily\footnotesize]
Nav0 = child3, Property0 = attribute "Type", Value0 = "obj"
Nav1 = child, Property1 = attribute "Type", Value1 = "5360"
PropertyA = attribute "Value", ValueA = "^[0-9]+(,[0-9]+)?( x [0-9]+(,[0-9]+)?)? (m|mm)( \([a-zA-ZäüöÄÜÖ ]+\))?$"
\end{lstlisting}
\paragraph{LIDO}
LIDO explicitly allows heterogeneous measure units. This is why we do not consider heterogeneous measure units to be a problem in the LIDO data and did not create a corresponding pattern.
\subsubsection{Heterogeneous Value Representations}
\paragraph{MIDAS}
The type of the described artwork is specified as ``Print'' even though the equivalent German word ``Druck'' is preferred.

Runtime: 7564 ms
\begin{lstlisting}[label=lst:hetvalm,caption={Concretisation of the pattern COMPVAL1},basicstyle=\ttfamily\footnotesize]
Nav0 = child3, Property0 = attribute "Type", Value0 = "obj"
Nav1 = child, Property1 = attribute "Type", Value1 = "5230"
Property2 = attribute "Value", COMP2 = equal, Value2 = "Print"
\end{lstlisting}
\paragraph{LIDO}
See above.

Runtime: 5652 ms
\begin{lstlisting}[label=lst:hetvall,caption={Concretisation of the pattern COMPVAL2},basicstyle=\ttfamily\footnotesize]
Nav0 = child2, Property0 = name, Value0 = "lido:lido"
Nav1 = child4, Property1 = name, Value1 = "lido:objectWorkType"
Nav2 = child, Property2 = name, Value2 = "lido:term"
Property3 = data, COMP3 = equal, Value3 = "Print"
\end{lstlisting}
\subsubsection{Heterogeneous Structural Representations}
\paragraph{MIDAS}
If no artist can be specified for an artwork, then the XML element indicating a relation to an artist must be omitted in the records describing the artwork. The pattern detects violations.

Runtime: 6503 ms
\begin{lstlisting}[label=lst:hetstrucm,caption={Concretisation of the pattern MANDSTRUC2},basicstyle=\ttfamily\footnotesize]
Nav0 = child3, Property0 = attribute "Type", Value0 = "obj"
Nav1 = child, Property1 = attribute "Type", Value1 = "ob30"
Nav2 = child, Property2 = attribute "Type", Value2 = "3100"
\end{lstlisting}
\paragraph{LIDO}
An element for specifying an actor's name contains multiple names instead of the actor record containing multiple of such elements.

Runtime: 7809 ms
\begin{lstlisting}[label=lst:hetstrucl,caption={Concretisation of the pattern CARD2},basicstyle=\ttfamily\footnotesize]
Nav0 = child2, Property0 = name, Value0 = "lido:lido"
Nav1 = child8, Property1 = name, Value1 = "lido:nameActorSet"
Nav2 = child, Property2 = name, Value2 = "lido:appellationValue"
\end{lstlisting}
\subsection{Misspellings}
As explained in Section~\ref{subsec:expressiveness} this problem could not be covered.
\subsection{Semantically Incorrect Data}
\subsubsection{False Values}
As explained in Section~\ref{subsec:expressiveness} this problem variant could not be covered.
\subsubsection{False References}
As explained in Section~\ref{subsec:expressiveness} this problem variant could not be covered.
\subsubsection{Doubtful Data}
\paragraph{MIDAS}
The given birth date of an artist is followed by a question mark indicating uncertainty.

Runtime: 1592 ms
\begin{lstlisting}[label=lst:doubtm,caption={Concretisation of the pattern MATCH1},basicstyle=\ttfamily\footnotesize]
Nav0 = child3, Property0 = attribute "Type", Value0 = "kue"
Nav1 = child, Property1 = attribute "Type", Value1 = "3270"
PropertyA = attribute "Value", ValueA = "\?$"
\end{lstlisting}
\paragraph{LIDO}
A given appellation value is followed by a question mark indicating uncertainty.

Runtime: 10046 ms
\begin{lstlisting}[label=lst:doubtl,caption={Concretisation of the pattern MATCH1},basicstyle=\ttfamily\footnotesize,language={[LaTeX]TeX}]
Nav0 = child2, Property0 = name, Value0 = "lido:lido"
Nav1 = descendant, Property1 = name, Value1 = "lido:appellationValue"
PropertyA = data, ValueA = "\?$"
\end{lstlisting}

}{
\clearpage
\appendix

\section{Further Example}
\label{appendix:example}
To get a better understanding of our metamodel, we will present one further example pattern in the following.
First we modify the running example.
In the version depicted in Listing~\ref{lst:runningexamplelong} \lstinline|building| elements may contain a
\lstinline|creator| element containing a reference to its associated architect.

\begin{lstlisting}[language=XML,caption={Running example: data describing paintings, buildings and artists which includes 3 of the quality problems listed in Table \ref{table:problems}.},label={lst:runningexamplelong},basicstyle=\ttfamily\footnotesize,float]
<data>
	<building id="1">
		<name>Empire State Building</name>
		<city>New York City</city>
		<country>USA</country>	
		<creator ref="3"/>
	</building>		
	<building id="2">
		<name>Chrysler Building</name>
		<city>New York City</city>
		<country>unknown</country>
		<creator ref="4"/>	
	</building>	
	<architect id="3">
		<name>William F. Lamb</name>
		<birthyear>1883</birthyear>
		<birthyear>1884</birthyear>
	</architect>	
</data>
\end{lstlisting}

Note, that in this example we have no \lstinline|architect| with \lstinline|ID 4|, however such a data record is referenced by the \lstinline|building| with \lstinline|ID 2|.
Hence, this is a referential integrity violation.
Such problems usually do not arise during data creation as this is controlled by database management systems.
Instead, they often stem from deletions or transfer of parts of the database.

The \emph{abstract pattern} \texttt{REFINT} shown in Fig.~\ref{fig:refint} detects such referential integrity violations. 
The right part of this pattern specifies the condition that determines whether referential integrity has been violated. 
The return element has to contain another element (called Element 1, identified by Property1 and Value1).
Its PropertyA1 refers to some PropertyA2 of an Element 2 (identified by Property2 and Value2) which is not contained in the root element as indicated and thus does not exist.  
\begin{figure}	
	\centering
	\spaceabove
	\includegraphics[width=\linewidth]{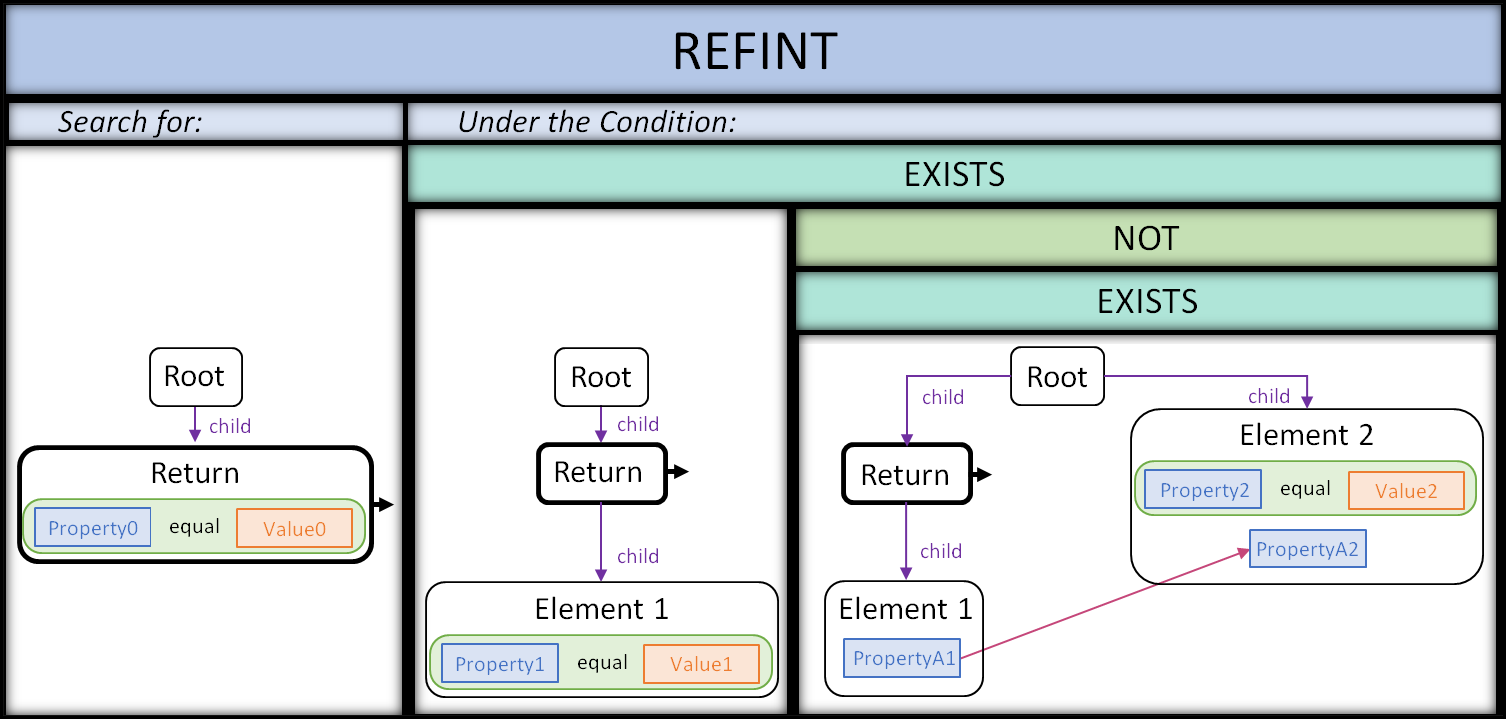}
	\caption{Abstract pattern REFINT for detecting referential integrity violations; XML-adaption}
	\label{fig:refint}
	\spacebelow
\end{figure}

Note that Fig.~\ref{fig:refint} shows three graphs which are included into each other from left to right. 
Pattern elements are bound from left to right.
This allows expressing relations and conditions between elements already existing in a previous graph (to the left) and new elements that are added to the pattern in this graph.
To detect the referential integrity violation in the example data in Listing~\ref{lst:runningexamplelong}, the parameters need to be specified as shown in Listing~\ref{lst:refint}.
\begin{lstlisting}[label=lst:refint,caption={Example concretisation of the pattern REFINT},basicstyle=\ttfamily\footnotesize]
Property0 = name, Value0 = "building"
Property1 = name, Value1 = "creator"
Property2 = name, Value2 = "architect"
PropertyA1 = attribute "ref", PropertyA2 = attribute "id"
\end{lstlisting}
\emph{The concrete REFINT pattern detects all \lstinline|building| elements whose \lstinline|creator| element contains a reference to a non-existent \lstinline|architect| element.}
When applied to the example data depicted in Listing~\ref{lst:runningexamplelong}, the pattern returns the \lstinline|building| with \lstinline|ID 2|.

A reviewer pointed out that this pattern detects also cases in which Element 1 does not have the PropertyA1, which is actually not a referential integrity violation.
We will address this issue in near future.
The issue, however, does not affect our evaluation as in the chosen database Element 1 exists only if it also includes a reference to an element Element 2, thus has PropertyA1.

\clearpage

\section{Generated Queries}\label{appendix:queries}

The following listings show the XQuery expressions generated by our tool for the concrete example pattern \texttt{CARD} presented in Section~\ref{subsec:examplepatterns} and \texttt{REFINT} presented in Section~\ref{appendix:example} of the appendix.

\begin{lstlisting}[label=lst:queryCard,caption={XQuery for the concrete CARD pattern presented in Section~\ref{subsec:examplepatterns}},language=XQuery]
for $var1 in /child::*[./name()="artist"]
where count(
 for $var2 in /child::*[./name()="name"]
 where true()
 return $var2
) > 1.0
return $var1
\end{lstlisting}

\begin{lstlisting}[label=lst:queryRefint,caption={XQuery for the concrete REFINT pattern presented in Section~\ref{appendix:example}},language=XQuery,basicstyle=\ttfamily\footnotesize]
for $var1 in /child::*[./name()="building"]
where some $var2 in $var1/child::*[./name()="creator"]
 satisfies not(
  some $var3 in /child::*[./name()="artist"]
     [$var2/@ref=./@id]
   satisfies true()
 )
return $var25
\end{lstlisting}


\clearpage

\clearpage
}

\end{document}